\documentclass[reqno, twocolumn, 10pt]{wlscirep}
\usepackage[utf8]{inputenc}
\usepackage[T1]{fontenc}
\usepackage{quotes}
\usepackage{sidecap}
\usepackage{comment}
\sidecaptionvpos{figure*}{t}

\usepackage{graphicx}
\usepackage{multirow}
\usepackage{amsmath,amssymb,amsfonts}
\usepackage{amsthm}
\usepackage{mathrsfs}
\usepackage[title]{appendix}
\usepackage{textcomp}
\usepackage{manyfoot}
\usepackage{booktabs}
\usepackage{algorithm}
\usepackage{algorithmicx}
\usepackage{algpseudocode}
\usepackage{listings}
\usepackage{placeins}
\usepackage{lmodern}
\usepackage{fix-cm}
\usepackage{color}
\usepackage{subcaption}
\usepackage{bbding}
\usepackage{float}
\usepackage{hyperref}
\usepackage{url}

\newcommand{\hypergraph}{\texttt{H3}}
\newcommand{\contagion}{\texttt{naSI}}

\title{Effects of higher-order interactions and homophily on information access inequality}

\author[1,2,*]{Moritz Laber}
\author[3,*]{Samantha Dies}
\author[1,*]{Joseph Ehlert}
\author[1,$\sharp$]{Brennan Klein}
\author[1,3,4]{Tina Eliassi-Rad}

\affil[1]{Network Science Institute, Northeastern University, 177 Huntington Ave., Boston, MA 02115, USA}

\affil[2]{Complexity Science Hub Vienna, Metternichgasse 8, Vienna, 1030 Austria}

\affil[3]{Khoury College of Computer Sciences, Northeastern University, 440 Huntington Ave, Boston, MA 02115, USA}

\affil[4]{Santa Fe Institute, 1399 Hyde Park Road, Santa Fe, NM 87501, USA}

\affil[*]{\textit{These authors contributed equally.}}
\affil[$\sharp$]{\href{mailto:b.klein@northeastern.edu}{b.klein@northeastern.edu}}

\begin{abstract}
The spread of information through socio-technical systems determines which individuals are the first to gain access to opportunities and insights.
Yet, the pathways through which information flows can be skewed, leading to systematic differences in access across social groups. These inequalities remain poorly characterized in settings involving nonlinear social contagion and higher-order interactions that exhibit homophily. 
We introduce a generative model for hypergraphs with hyperedge homophily, a hyperedge size-dependent property, and tunable degree distribution, called the \hypergraph~model, along with a model for nonlinear social contagion that incorporates asymmetric transmission between in-group and out-group nodes.
Using stochastic simulations of a social contagion process on hypergraphs from the \hypergraph~model and diverse real-world datasets, we show that the interaction between social contagion dynamics and hyperedge homophily---an effect unique to higher-order networks due to its dependence on hyperedge size---can critically shape group-level differences in information access.
By emphasizing how hyperedge homophily shapes interaction patterns, our findings underscore the need to rethink socio-technical system design through a higher-order perspective and suggest that dynamics-informed, targeted interventions at specific hyperedge sizes, embedded in a platform architecture, offer a powerful lever for reducing inequality.
\end{abstract}

\begin{document}

\flushbottom
\maketitle
\thispagestyle{empty}

\section{Introduction}\label{sec:introduction}
Information is a crucial resource in modern societies~\cite{capurro2003_ConceptInformation}. Access to information---whether about professional opportunities~\cite{ibarra1995_RaceOpportunityDiversity, burt1998_GenderSocialCapital, lambrecht2019_AlgorithmicBiasEmpirical,rajkumar2022_CausalTestStrength}, research insights~\cite{jadidi2018_GenderDisparitiesScience,bachmann2024_CumulativeAdvantageBrokerage,zappala2025_GenderDisparitiesDissemination}, or even the latest news and trends~\cite{shifman2013_MemesDigitalCulture, bakshy2012_RoleSocialNetworks, meng2025_SpreadingDynamicsInformation}---can shape an individual's social capital~\cite{burt2004_StructuralHolesGood, burt2000_NetworkStructureSocial,lin2001_BuildingNetworkTheory} and, consequently, their chances of success. As a result, issues of information access inequality~\cite{wang2022_InformationAccessEquality,stoica2020_SeedingNetworkInfluence,fish2019_GapsInformationAccess} and the fairness of the socio-technical systems~\cite{abebe2020_RolesComputingSocial,minow2021_EqualityVsEquity,barocas2023_FairnessMachineLearning} that govern it are unavoidable. These issues are particularly relevant for systems such as social media platforms, online marketplaces, and scientific publishing ecosystems. In this work, we focus on two aspects of information access that add nuance to these discussions. First, information spreads through social networks, meaning that the structure of social ties plays a fundamental role in determining who receives information~\cite{burt2000_NetworkStructureSocial,bakshy2012_RoleSocialNetworks,meng2025_SpreadingDynamicsInformation}. Second, the utility of information is time-sensitive since early access often confers a competitive advantage~\cite{ali2023_FairnessTimeCriticalInfluence, li2024_AgeInformationDiffusion}. In particular, structural factors such as homophily and the differences between pairwise and higher-order interactions shape who gains access to information and when, often reinforcing systemic advantages for certain groups.

Consider a new job posting on a social media platform. Information about the opportunity propagates through news feeds, direct messages, and group chats. 
Because social ties operate as conduits for the flow of information, awareness cascades outward: friends of early viewers hear next, then friends‑of‑friends, and so on. This spread lets well‑connected applicants reach the posting sooner than the peripheral applicants, creating achievement gaps. When social connections are shaped by homophily, cascades travel disproportionately within a single group. This can skew opportunities, underscoring the need to understand how network ties and diffusion mechanisms shape information access.

Network structure plays an important role in social contagion, and its influence on the contagion process has been extensively studied~\cite{kiss2017_MathematicsEpidemicsNetworks, lehmann2018_SpreadingSocialSystems}. In the case of networks with pairwise interactions, the outcome of a social contagion process depends on homophily and community structure~\cite{gleeson2008_CascadesCorrelatedModular,hebertdufresne2010_PropagationDynamicsNetworks,davis2020_PhaseTransitionsInformation,patwardhan2023_EpidemicSpreadingGroupStructured}, clustering~\cite{miller2009_PercolationEpidemicsRandom, keating2022_MultitypeBranchingProcess}, degree heterogeneity~\cite{pastor-satorras2001_EpidemicSpreadingScaleFree, allard2023_RoleDirectionalityHeterogeneity}, and degree assortativity~\cite{boguna2003_AbsenceEpidemicThreshold, gleeson2008_CascadesCorrelatedModular,dodds2009_AnalysisThresholdModel}. Likewise, contagion dynamics have been characterized through distinctions between simple~\cite{kermack1997_ContributionMathematicalTheory, maki1973_MathematicalModelsApplications, daley1964_EpidemicsRumours,
pastor-satorras2001_EpidemicSpreadingScaleFree} and complex contagion~\cite{granovetter1978_ThresholdModelsCollective,watts2002_SimpleModelGlobal,centola2007_ComplexContagionsWeakness}, as well as through many other models~\cite{castellano2009_StatisticalPhysicsSocial,olsson2024_AnalogiesModelingBelief}.

The interplay between network structure and contagion dynamics has also been examined through the lens of fairness, demonstrating how homophily~\cite{coleman1958_RelationalAnalysisStudy,mcpherson2001_BirdsFeatherHomophily,park2007_DistributionNodeCharacteristics,karimi2023_InadequacyNominalAssortativity} influences the ability of minority groups to access~\cite{jalali2020_InformationUnfairnessSocial,wang2022_InformationAccessEquality,zappala2025_GenderDisparitiesDissemination} and disseminate~\cite{furutani2023_AnalysisHomophilyEffects,zappala2025_GenderDisparitiesDissemination}  information. Researchers have explored strategies to improve fairness in information spread, often focusing on seeding techniques designed to enhance overall outreach~\cite{fish2019_GapsInformationAccess,stoica2020_SeedingNetworkInfluence,anwar2021_BalancedInfluenceMaximization} or related objectives~\cite{farnad2020_UnifyingFrameworkFairnessAware,chowdhary2025_FairnessSocialInfluence}. Only recently have algorithms accounting for the time-sensitive nature of information been proposed~\cite{ali2023_FairnessTimeCriticalInfluence}.

While much of this research has focused on pairwise interactions in networks, many real-world socio-technological systems involve higher-order, group-based interactions that traditional graph models do not explicitly capture. Higher-order networks, formalized as simplicial complexes or hypergraphs, provide a more expressive framework for modeling these systems since they can represent both pairwise and group interactions.~\cite{battiston2020_NetworksPairwiseInteractions, torres2021_WhyHowWhen, bick2023_WhatAreHigherOrder, lee2025_SurveyHypergraphMining}. Compared to simplicial complexes, which require the existence of all possible lower-order interactions among nodes that participate in a given higher-order interaction, hypergraphs do not impose restrictions on the structure of interactions. Thus, hypergraphs naturally differentiate between different modes of information spread, such as direct messages between individuals and group chats involving multiple participants on online social networks.

Accurately modeling these processes on hypergraphs requires generative hypergraph models that capture key structural properties, especially hyperedge homophily, which depends on hyperedge size, while remaining computationally feasible. However, existing models often face trade-offs between realism and tractability. Stochastic block models permit analytically tractable graph likelihoods; however, generating hypergraphs with large numbers of nodes quickly becomes computationally infeasible~\cite{chodrow2021_GenerativeHypergraphClustering, contisciani2022_InferenceHyperedgesOverlapping, ruggeri2023_CommunityDetectionLarge, ruggeri2024_FrameworkGenerateHypergraphs}. Markov Chain Monte Carlo methods generate randomized hypergraphs that preserve certain structural properties, provided there is an initial hypergraph. Yet, uniform sampling from all hypergraphs with a given set of properties is only possible in simple cases, e.g., for degree sequence~\cite{chodrow2020_ConfigurationModelsRandom,chodrow2020_AnnotatedHypergraphsModels,preti2024_HigherOrderNullModels,ruggeri2024_FrameworkGenerateHypergraphs}. Alternatively, growing hypergraph models offer computationally efficient approaches for generating hypergraphs from scratch but often lack the analytical tractability to determine the probability of hypergraphs with specific properties~\cite{do2020_StructuralPatternsGenerative, barthelemy2022_ClassModelsRandom, papachristou2022_CoreperipheryModelsHypergraphs, pister2024_StochasticBlockHypergraph}. Finally, hypergraph models tailored towards specific applications, such as face-to-face interactions, have been proposed, yet they are context-specific~\cite{gallo2024_HigherOrderModelingFacetoFace, akcakir2024_ExploringInterplayIndividualTraits}.

Structure alone, however, does not determine how information spreads. A comprehensive understanding of information access inequality requires not only realistic models with group structure, but also explicit models of the dynamics that govern information flow. Incorporating higher-order interactions into contagion models provides a more realistic framework for capturing social dynamics, particularly in the context of information spread~\cite{majhi2022_DynamicsHigherorderNetworks, ferrazdearruda2024_ContagionDynamicsHigherorder}. These interactions can account for social reinforcement~\cite{iacopini2019_SimplicialModelsSocial, stonge2022_InfluentialGroupsSeeding} and inhibition~\cite{landry2020_EffectHeterogeneityHypergraph, burgio2023_SpreadingDynamicsNetworks}, often influenced by group membership. Higher-order effects can also give rise to multi-stability~\cite{iacopini2019_SimplicialModelsSocial, ferrazdearruda2020_SocialContagionModels, kiss2023_InsightsExactSocial, ferrazdearruda2023_MultistabilityIntermittencyHybrid} and localization~\cite{stonge2022_InfluentialGroupsSeeding, mancastroppa2023_HypercoresPromoteLocalization}, qualitatively altering the dynamics of information spread compared to pairwise networks. Recent work further shows that hypergraphs can capture mixed hyperedge homophily patterns---where the amount of hyperedge homophily varies with interaction (i.e., hyperedge) size---revealing a richer structural vocabulary than traditional graphs~\cite{veldt2023_CombinatorialCharacterizationsImpossibilities, sarker2024_HigherorderHomophilySimplicial, rizi2024_homophilywithinacross}. However, despite advances, the question of who gains access to information and when, particularly under structural bias, remains largely unexplored.

Here, we study how higher-order homophily in hypergraphs influences fairness under potentially group-biased social contagion. First, we introduce the Hypergraphs with Hyperedge Homophily (\hypergraph), a computationally efficient generative framework that produces hypergraphs with a tunable degree distribution and controllable levels of homophily as a function of hyperedge size and a tunable degree distribution.
Second, we extend prior hypergraph contagion models~\cite{stonge2022_InfluentialGroupsSeeding} to systematically study information access inequality by incorporating group-specific asymmetric transmission. While our primary focus is on time-critical information access, our approach also offers a foundation for examining dynamics in domains where structural bias and nonlinear contagion interact---including innovation diffusion, misinformation spread, public health campaigns, and collaborative knowledge production.


\begin{figure*}[t!]
    \centering
    \includegraphics[width=0.825\textwidth]{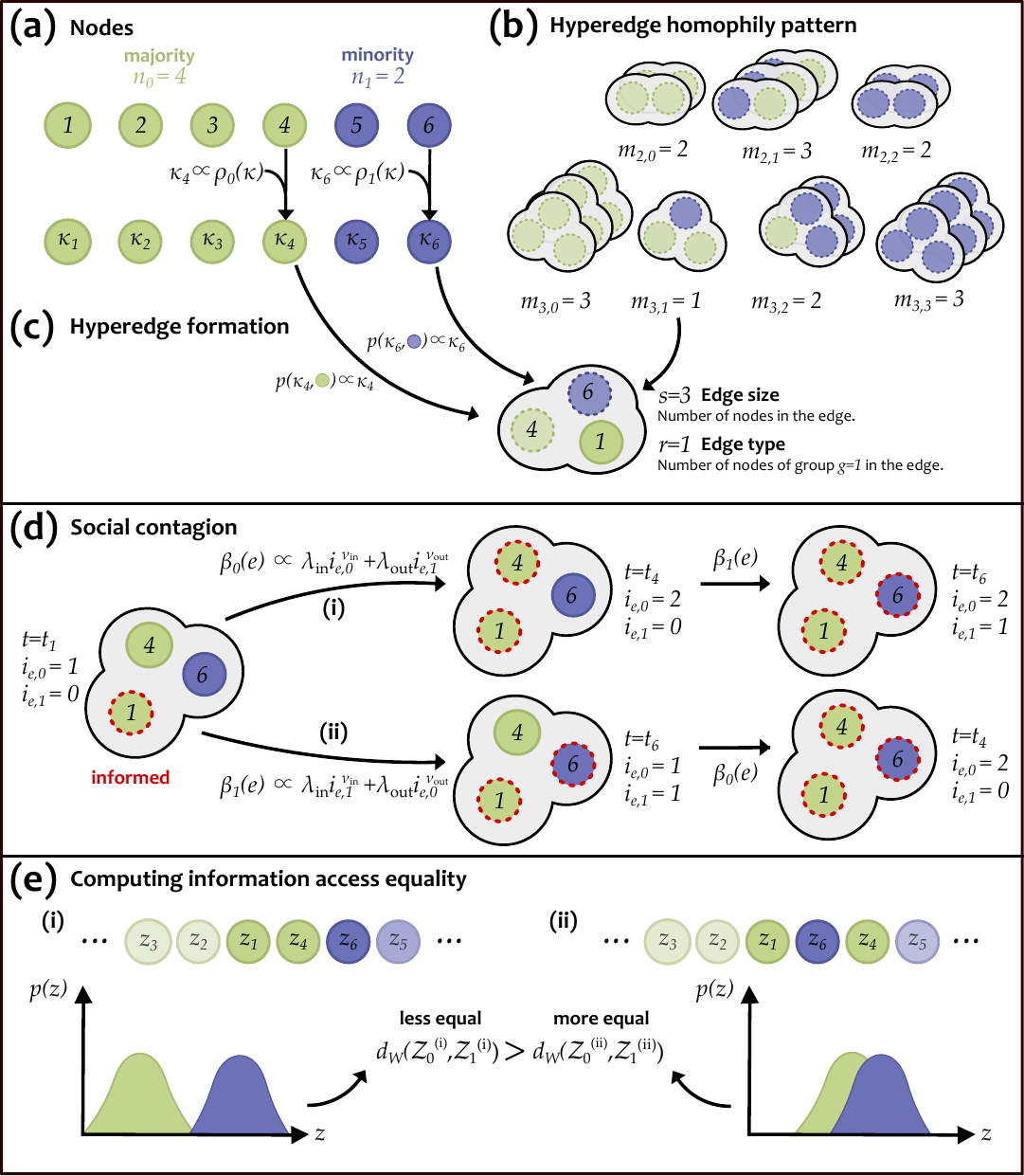}
    \caption{\textbf{Schematic overview of hypergraph formation with the \hypergraph~model, the social contagion dynamics of the \contagion~model, and inequality measurement with $d_W$.} \textbf{(a)} Each node $v$ is assigned a group $g_v \in \{0,1\}$ (green and purple) and a hidden variable $\kappa_v$ from a group-specific distribution $\rho_{g_v}(\kappa)$, encoding its propensity to participate in hyperedges and, therefore, its degree. \textbf{(b)} We fix the hyperedge homophily pattern by setting the number $m_{s,r}$ of hyperedges of size $s$ and type $r$. \textbf{(c)} We randomly place nodes $v$ into hyperedges $e$ with probability $p(\kappa_v,g_v)$ determined by their groups $g_v$ and hidden variables $\kappa_v$. \textbf{(d)} An example contagion step, where node $v=1$ transmits information to either (i) node $v'=4$ or (ii) node $v'=6$ through a shared hyperedge with different rates $\beta_{g_{v'}}(e)$ depending on the group membership of $v'$. Dotted red borders indicate informed nodes. Even though all nodes are informed, \textbf{(e)} shows that, when nodes are ranked by the time they are informed, differences emerge in group-wise rank distributions $\mathcal{Z}_g$, which we compare using the Wasserstein distance $d_W$.}
    \label{fig:schematics}
\end{figure*}
%
%


\section{Results}\label{sec:results}
\subsection{Hypergraph Model}\label{sec:results:hypergraph_model}
%
To investigate how hyperedge homophily and degree heterogeneity shape information access inequality in hypergraphs, we introduce a new generative model, the Hypergraphs with Hyperedge Homophily (\hypergraph) model, designed to encode group membership, tunable homophily that depends on hyperedge size, and heterogeneous degree distributions. Unlike existing models, the \hypergraph~model allows explicit control over homophily composition for all hyperedge sizes, while maintaining computational efficiency. We describe the details of this model in Section~\ref{sec:methods:hypergraph_model} and schematically depict it in Fig.~\ref{fig:schematics}.

In the \hypergraph~model, a node $v$ is assigned a binary value $g_v$ that encodes group membership and a hidden variable $\kappa_v$, sampled from a group-dependent distribution $\rho_{g_v}$, which governs its propensity to appear in hyperedges (Fig.~\ref{fig:schematics}(a)). Thus, the hidden variables $\kappa_v$ allow us to control the expected degree of node $v$, where the degree $k_v=|\{e \in \mathcal{E} : v\in e\}|$ is the number of hyperedges to which $v$ belongs. We emphasize that in hypergraphs, unlike in pairwise networks, the number of edges to which a node belongs and the number of adjacent nodes are generally not the same. The group variable $g$ can be used to encode any binary concept (e.g.,~high/low socio-economic status or male/female gender). While the term group has occasionally been used to refer to a given hyperedge, we reserve the term for the attribute $g \in \{0, 1\}$. We are interested in the case where one group, $g=0$, is larger than the other group, $g=1$. We therefore refer to $g=1$ as the minority group and $g=0$ as the majority throughout the manuscript.

We adopt the notion of homophily introduced by Veldt et al.~\cite{veldt2023_CombinatorialCharacterizationsImpossibilities}(see Section~\ref{sec:methods:hypergraph_homophily}). In hypergraphs, homophily depends on hyperedge size and cannot be summarized by a single number. To emphasize this dependence on hyperedge size, we refer to this measure as hyperedge homophily. The hyperedge homophily $h_{s,r}^{(g)}$ of group $g$ depends on the number of hyperedges $m_{s,r}$ of hyperedge size $s\in\{2,\dots,s_\mathrm{max}\}$ of type $r \in \{0, \dots s\}$. The type of a hyperedge, $r$, counts the number of nodes from group $g=1$ in a hyperedge of size $s$. For example, $m_{3,2}=10$ means that we observe $10$ hyperedges of size $s=3$ that consist of $r=2$ nodes from group $g=1$ and $s-r=1$ node from group $g=0$. As discussed in Section~\ref{sec:methods:hypergraph_homophily}, hyperedge homophily is a normalized version of these counts; however, we use these counts directly as inputs to the \hypergraph~model (Fig.~\ref{fig:schematics}(b)), resulting in a fixed value of hyperedge homophily for given inputs.

For each hyperedge of size $s$ and type $r$, we select $r$ nodes from group $g=1$ and $s-r$ nodes from group $g=0$ with probability proportional to their hidden variable $\kappa_v$ (Fig.~\ref{fig:schematics}(c)). These selected nodes are then assigned to a new hyperedge, and this process is repeated independently for each hyperedge count specified by $m_{s,r}$. This hyperedge formation process ensures that the hypergraphs generated by the model follow the specified hyperedge homophily pattern exactly and maintain the degree distribution in expectation. We report the exact hyperedge counts $m_{s,r}$ used to generate our hypergraphs and show example degree distributions of generated hypergraphs in Appendix~\ref{appendix:parameters_synthetic}.

We design three hyperedge homophily patterns---neutral, homophilous, and heterophilous---for hyperedges of size $s\in\{2,3,4\}$. In the neutral pattern, the hyperedge counts $m_{s,r}$ are as one would expect if nodes were placed into hyperedges randomly and irrespective of group membership. Such hypergraphs exhibit neither homophily nor heterophily. The homophilous pattern captures a node's preference for in-group connections, implying overrepresentation of hyperedges of a single group (i.e., $r=0$ for $g=0$, and $r=s$ for $g=1$) and underrepresentation of hyperedges with nodes from both groups. In the heterophilous pattern, hyperedges containing nodes from only one group are underrepresented, while hyperedges containing nodes from different groups are overrepresented. While we use this model to study information access inequality, the characteristics of the \hypergraph~model make it broadly applicable to other phenomena shaped by group structure, such as polarization, misinformation spread, and innovation diffusion.


\subsection{Social Contagion Model}\label{sec:results:social_contagion_model}
To model how information spreads through group-structured, higher-order networks, we introduce the nonlinear, asymmetric SI (\contagion) model. The \contagion~model extends the nonlinear contagion framework introduced by St-Onge et al.~\cite{stonge2022_InfluentialGroupsSeeding}  to incorporate group-dependent asymmetries in both transmission rates and reinforcement dynamics. This extension allows us to capture realistic spreading behaviors such as in-group preference, echo chamber effects, and social inhibition between groups. As in classical compartmental models, we treat information transmission as a continuous-time stochastic process~\cite{kiss2017_MathematicsEpidemicsNetworks}. Details of the \contagion~model are provided in Section~\ref{sec:methods:contagion}, with a schematic illustration in Fig.~\ref{fig:schematics}(d).

To model information spread, we track whether each node has received a piece of information and define $\beta_g(e)$ as the rate at which previously uninformed nodes from group $g$ acquire information from a given hyperedge $e$ of size $s_e$. The rate is a nonlinear function of the number of informed nodes of the same group, $i_{e,g}$, and the other group, $i_{e,g'}$, in $e$. It is given by
\begin{equation}
    \beta_{g}(e) = (s_e-i_{e,g})(\lambda_\mathrm{in} i_{e,g}(t)^{\nu_\mathrm{in}} + \lambda_\mathrm{out} i_{e,g'}(t)^{\nu_\mathrm{out}}),
\end{equation}
where $\lambda_\mathrm{in}$ and $\lambda_\mathrm{out}$ are the transmission rates to in- and out-group nodes and the exponents $\nu_\mathrm{in}$ and $\nu_\mathrm{out}$ are the nonlinearity parameters. 

Differentiating between $\lambda_\mathrm{in}$ and $\lambda_\mathrm{out}$ allows the \contagion~model to capture asymmetric contagion, i.e., when transmission rates differ based on group membership. We focus on the case $\lambda_\mathrm{in} > \lambda_\mathrm{out}$, where nodes prefer to pass information to members of their own group.

The exponents $\nu_\mathrm{in}$ and $\nu_\mathrm{out}$ govern super- or sublinear increases in the transmission rate based on the numbers of informed individuals in each group. Superlinear spread ($\nu > 1$) captures social reinforcement, while sublinear spread ($\nu < 1$) models social inhibition. As with $\lambda$, group-dependent values of $\nu$ allow the model to represent nuanced spreading behavior---such as in-group reinforcement and out-group inhibition---by setting $\nu_\mathrm{in} > 1 > \nu_\mathrm{out}$.

\begin{table*}[tb]
\centering
\resizebox{0.85\textwidth}{!}{%
\begin{tabular}{c|c|c|c|c|c|c}
\textbf{Measure}                  & \multicolumn{1}{c|}{\textbf{Temporal}} & \multicolumn{1}{c|}{\textbf{Acquisition}} & \multicolumn{1}{c|}{\textbf{Diffusion}} & \multicolumn{1}{c|}{\textbf{Bounded}} & \multicolumn{1}{c|}{\textbf{Directional}} & \multicolumn{1}{c}{\textbf{Single run}} \\ \hline \hline
$d_W(\mathcal{Z}_0, \mathcal{Z}_1)$ & \checkmark & \checkmark & & \checkmark & & \checkmark \\
$\alpha(f)$ & \checkmark & \checkmark & & & \checkmark & \\
$\delta(f)$ & \checkmark & & \checkmark & & \checkmark &  
\end{tabular}
}
\caption{\textbf{Properties of inequality measures.} The properties of our optimal transport-based inequality measure $d_W(\mathcal{Z}_0, \mathcal{Z}_1)$, acquisition fairness $\alpha(f)$, and diffusion fairness $\delta(f)$.  All three measures are temporal. The measures $d_W(\mathcal{Z}_0, \mathcal{Z}_1)$ and $\alpha(f)$ govern information access and $\delta(f)$ captures information diffusion. The measure $d_W(\mathcal{Z}_0, \mathcal{Z}_1)$ is bounded in $[0, n/2]$, but is not directional, whereas $\alpha(f)$ and $\delta(f)$ capture which group is advantaged. Finally, $d_W(\mathcal{Z}_0, \mathcal{Z}_1)$ can be calculated for a single run, while $\alpha(f)$ and $\delta(f)$ must be averaged over runs. These measures differ in scope, directionality, and interpretability, offering complementary perspectives that together support a more comprehensive understanding of inequality in temporal information spread.}
\label{tab:measure_comparison}
\end{table*}

In our simulations, we study three scenarios with symmetric transmission rates ($\lambda_\mathrm{in} = \lambda_\mathrm{out}$): linear contagion ($\nu_\mathrm{in}, \nu_\mathrm{out} = 1$), sublinear contagion ($\nu_\mathrm{in},\nu_\mathrm{out}< 1$), and superlinear contagion ($\nu_\mathrm{in}, \nu_\mathrm{out}> 1$). We also consider a fourth, asymmetric case with in-group transmission bias and in-group social reinforcement ($\lambda_\mathrm{in} > \lambda_\mathrm{out}$ and $\nu_\mathrm{in} > 1 > \nu_\mathrm{out}$). This configuration captures a setting in which nodes preferentially transmit information to their own group and require greater exposure to adopt information from out-group nodes, reflecting common social dynamics such as trust asymmetries, echo chambers, or inter-group friction. Although other asymmetric combinations of $\lambda_\mathrm{in}, \lambda_\mathrm{out}$ and $\nu_\mathrm{in}, \nu_\mathrm{out}$ are possible, we focus on this asymmetric case because it best reflects the diffusion of information in the real world under structural bias. Other parameter combinations may be appropriate for other processes. A full list of the parameter values used in the simulations can be found in Appendix~\ref{appendix:parameters_synthetic}.


\subsection{Inequality Measures}\label{sec:results:inequality_measures}

While many inequality and fairness measures have been proposed in the contagion literature, we focus on three that are particularly well-suited to our setting. Specifically, we seek measures that apply to stochastic, time-sensitive dynamics on hypergraphs and reflect both access and dissemination processes. Because timing shapes opportunity, we prioritize measures that capture not just who is informed, but when. We study three such measures of group-level inequality: an optimal transport–based distance $d_W(\mathcal{Z}_0, \mathcal{Z}_1)$ between group-level rank distributions $\mathcal{Z}_g$, and two established measures from prior work---acquisition fairness $\alpha(f)$ and diffusion fairness $\delta(f)$~\cite{zappala2025_GenderDisparitiesDissemination}, where $f$ is the fraction of informed nodes. Each measure captures temporal inequality from a distinct perspective. Two of the measures, $d_W(\mathcal{Z}_0, \mathcal{Z}_1)$ and $\alpha(f)$, govern information access and $\delta(f)$ captures information diffusion. Additionally, $d_W(\mathcal{Z}_0, \mathcal{Z}_1)$ is bounded, but lacks directionality, whereas $\alpha(f)$ and $\delta(f)$ capture which group is advantaged. Taken together, they offer complementary insights into how structural and dynamical factors shape group-level differences (see Table~\ref{tab:measure_comparison}).

To quantify inequality using optimal transport, we simulate the spread of information on each hypergraph with a seed set $\mathcal{S}$. As information propagates, each node $v$ receives the information at time $\tau_v$, which we convert to a rank $z_v$ indicating the order in which it was reached---e.g., $z_9 = 8$ means node $v=9$ was the eighth to be informed. Fig.\ref{fig:schematics}(d) illustrates how different realizations of the process can produce variation in $\beta_g(e)$ and $i_{e,g}(t)$, resulting in different values of $\tau_v$ and $z_v$. To assess whether one group tends to be reached earlier, we extract the rank distributions for each group and compute the Wasserstein distance $d_W(\mathcal{Z}_0, \mathcal{Z}_1)$. This comparison, shown in Fig.\ref{fig:schematics}(e), quantifies group-level differences in access timing. Unlike threshold-based fairness metrics, the optimal transport distance reflects distributional shifts across the entire contagion process and can be applied directly to individual simulation runs.

We also employ two other measures: Acquisition fairness $\alpha(f)$ compares the fraction of informed nodes in the minority group $g = 1$ to the overall fraction $f$ of informed nodes. Diffusion fairness $\delta(f)$ quantifies how much longer it takes to inform a fraction $f$ of the population when the seeds $\mathcal{S}$ are in the minority ($g = 1$) versus in the majority ($g = 0$). In both cases, values $\alpha(f), \delta(f) > 1$ indicate an advantage for the minority group, while values below $1$ indicate a disadvantage. We define both measures in detail in Section~\ref{sec:methods:measures}, where we also discuss their assumptions and limitations. Notably, prior work evaluated these measures only at a fixed threshold $f = \tilde{f}$, yielding scalar summaries of between-group differences. In contrast, we treat them as functions of $f$, allowing us to track how inequality evolves over the course of the contagion process and uncover dynamic patterns that scalar values may miss.


\begin{figure*}[t!]
    \centering
    \includegraphics[width=0.8\textwidth]{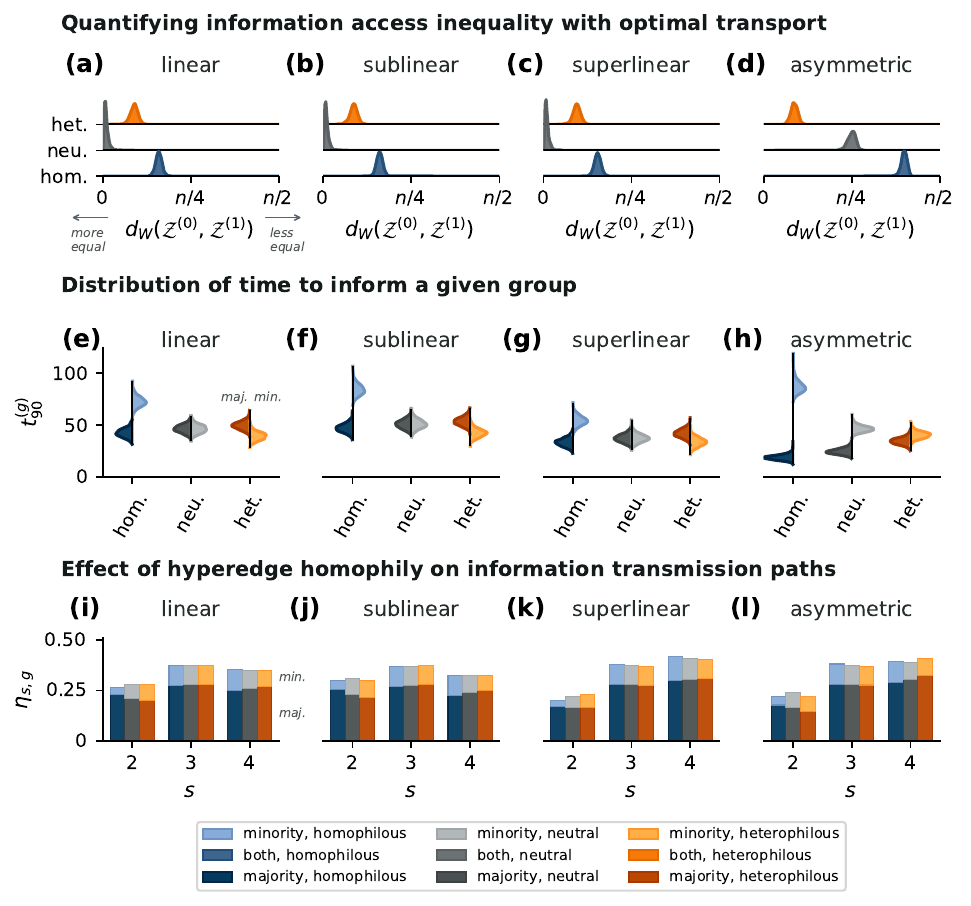}
    \caption{\textbf{Quantifying information access inequality in random hypergraphs.} By simulating different contagion processes on hypergraphs randomly sampled from our model, we measure the effect of hyperedge homophily patterns on information access inequality. Top row: distribution of Wasserstein distances $d_W(\mathcal{Z}_0, \mathcal{Z}_1)$ between group-wise empirical rank distributions under \textbf{(a)} linear, \textbf{(b)} sublinear, \textbf{(c)} superlinear, and \textbf{(d)} asymmetric transmission. Middle row: violin plot distributions of time $t^{(g)}_{90}$ required to inform $90$\% of the majority $g=0$ (darker shade, left side of violin) or minority $g=1$ (lighter shade, right side of violin) under \textbf{(e)} linear, \textbf{(f)} sublinear, \textbf{(g)} superlinear, and \textbf{(h)} asymmetric transmission. Bottom row: average fraction of transmission events involving hyperedges of size $s$, stratified by group, under \textbf{(i)} linear, \textbf{(j)} sublinear, \textbf{(k)} superlinear, and \textbf{(l)} asymmetric transmission. Darker bars indicate transmission among majority nodes; lighter bars indicate transmission among minority nodes. In each subplot, results from homophilous, neutral, and heterophilous hypergraphs are shown in shades of blue, gray, and orange, respectively. All results are averaged over $n_\mathrm{hg}=10^3$ independent simulations of the \contagion~model on hypergraphs generated from the \hypergraph~model with the same structural characteristics. Inequality in information access emerges across hyperedge homophily patterns and social contagion types, with interactions between structure and dynamics driving stark group-level differences, particularly under asymmetric and superlinear processes.}
    \label{fig:emd_synthetic}
\end{figure*}

\subsection{Information Access on Hypergraphs with Basic Hyperedge Homophily Patterns}\label{sec:results:synthetic_emd}
To understand how structural bias and transmission dynamics jointly shape inequality, we begin by analyzing information access on hypergraphs generated from the \hypergraph~model with simple, uniform hyperedge homophily patterns---specifically, hypergraphs that are entirely neutral, homophilous, or heterophilous across all hyperedge sizes $s \in \{2, 3, 4\}$. In homophilous hypergraphs, nodes have more in-group connections than expected, while heterophilous hypergraphs contain an overabundance of out-group connections. Connections in neutral hypergraphs are formed irrespective of group membership. Within and across these hyperedge homophily regimes, we compare four social contagion processes with our \contagion~model: symmetric linear, sublinear, and superlinear dynamics (all with $\lambda_\mathrm{in} = \lambda_\mathrm{out}$), as well as a nonlinear asymmetric case ($\lambda_\mathrm{in} > \lambda_\mathrm{out}$, $\nu_\mathrm{in} > 1 > \nu_\mathrm{out}$). These experiments allow us to quantify when each group gains access to time-sensitive information under different, basic patterns of hyperedge homophily and various nonlinearity settings. 

In the main text, we focus on hypergraphs with unequal group sizes, with $n_0=7500$ nodes in group $g=0$ (majority) and $n_1=2500$ nodes in group $g=1$ (minority). We also restrict our attention to sparse hypergraphs (average degree $\langle k\rangle=11$) with heterogeneous degree distributions, i.e., hidden variables $\kappa$ drawn from a Pareto distribution $\rho_g(\kappa)=\mathrm{Pareto[\bar{\kappa}_g, \gamma]}$ with exponent $\gamma=2.9$ of the probability density function and group-dependent mean $\bar{\kappa}_g$ (see also Section~\ref{sec:methods:hypergraph_model} and Appendix~\ref{appendix:parameters_synthetic}). Inequality in information access can stem from differences in both the order and timing with which individuals receive information. We capture these effects by comparing group-wise rank distributions using the Wasserstein distance $d_W(\mathcal{Z}_0, \mathcal{Z}_1)$, examining the time $t^{(g)}_{90}$ to reach $90\%$ of a given group $g$, and analyzing how hyperedge sizes shape the information spread (Fig.~\ref{fig:emd_synthetic}). The results for equal-sized groups and homogeneous degree distributions are provided in Appendix~\ref{appendix:additional_results_synthetic}.

On both homophilous and heterophilous hypergraphs, we observe unequal outcomes under all simulated contagion dynamics (see Fig.~\ref{fig:emd_synthetic}(a)-(d)). For homophilous hypergraphs, inequality is highest under the asymmetric contagion process, with an average distance of $\langle d_W(\mathcal{Z}_0,\mathcal{Z}_1)\rangle\approx 3979$, which approaches the theoretical maximum of $d_W=\frac{n}{2}=5000$. For heterophilous hypergraphs, inequality peaks under superlinear dynamics, with an average value of $\langle d_W(\mathcal{Z}_0,\mathcal{Z}_1)\rangle\approx 940$. In contrast, neutral hypergraphs yield nearly equal outcomes across all symmetric dynamics, regardless of the level of nonlinearity. However, inequality emerges in the asymmetric case with an average distance of $\langle d_W(\mathcal{Z}_0,\mathcal{Z}_1)\rangle\approx 2489$.

Somewhat surprisingly, the distributions of Wasserstein distances remain highly consistent across all symmetric contagion dynamics (Fig.~\ref{fig:emd_synthetic}(a)-(c)). In these cases, homophilous hypergraphs consistently exhibit the most unequal outcomes, followed by heterophilous, and then neutral. Under asymmetric dynamics this order shifts: inequality is again highest on homophilous hypergraphs, however neutral hypergraphs become more unequal than heterophilous ones.

To further examine the effect of different hyperedge homophily patterns, we analyze the time $t^{(g)}_{90}$ required to inform $90$\% of the nodes in a given group $g$. Across all hyperedge homophily patterns, transmission is fastest under superlinear dynamics and slowest under sublinear dynamics (Fig.~\ref{fig:emd_synthetic}(e)-(g)). This alternative view also reveals an important distinction between the inequality observed in homophilous versus heterophilous hypergraphs. In the homophilous case, the observed inequality consistently favors the majority, regardless of the contagion dynamics. On average, it takes less time to inform $90$\% of the majority than it does to inform $90$\% of the minority. For heterophilous hypergraphs, the timing-based advantage favors the minority under symmetric dynamics---minority nodes reach $90$\% coverage more quickly. This distinction could signal the existence of an elite minority group instead of a disadvantaged minority. However, this pattern reverses under asymmetric dynamics, where the majority becomes favored (Fig.~\ref{fig:emd_synthetic}(h)). A similar reversal occurs in neutral hypergraphs, which show no timing difference under symmetric dynamics, but exhibit a clear majority advantage under asymmetric transmission.

To understand how different dynamics shape information flow, we analyze which hyperedge sizes are most responsible for transmission events. We find that the nature of the contagion process shifts the focus of spread across hyperedge sizes compared to the linear case. Superlinear and asymmetric dynamics amplify the role of larger hyperedges, while sublinear dynamics deemphasize transmission in higher-order hyperedges (Fig.~\ref{fig:emd_synthetic}(i)-(l)). Although group-level differences in these distributions are subtle, such shifts in hyperedge-size activity could interact with structural patterns. Knowing the differences between transmission in hyperedge size can help experimentally determine the presence of higher-order effects and also inform intervention strategies.


\begin{figure*}[t]
    \centering
    \includegraphics[width=0.8\textwidth]{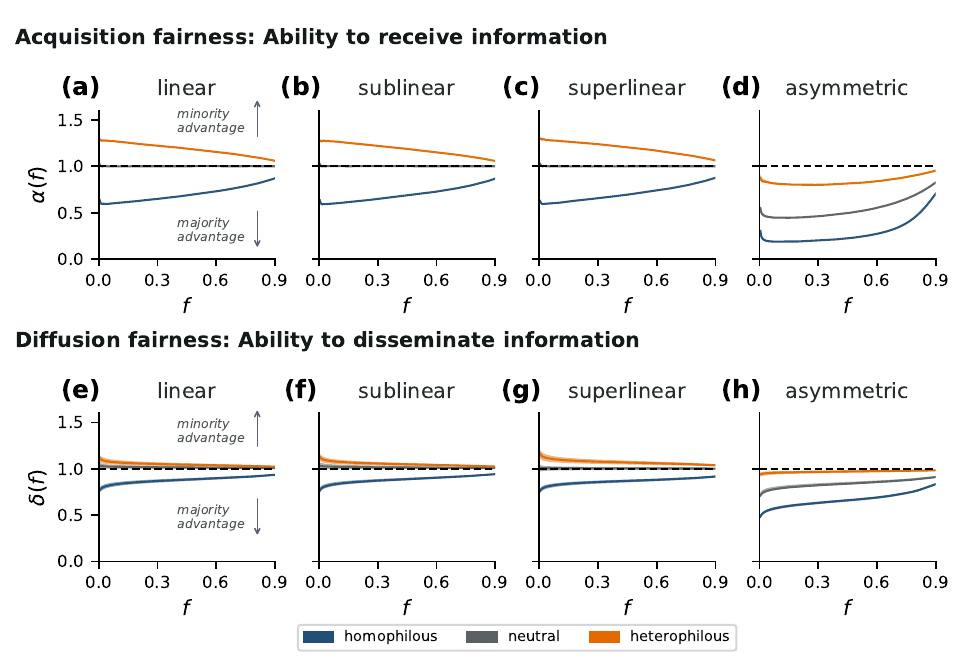}
\caption{\textbf{Measuring group-level differences in acquiring and spreading information.} We assess information access inequality using two fairness measures applied to simulated contagion processes on hypergraphs with varying hyperedge homophily patterns. Top row: acquisition fairness, $\alpha(f)$, which captures a group's ability to receive information, under \textbf{(a)} linear, \textbf{(b)} sublinear, \textbf{(c)} superlinear, and \textbf{(d)} asymmetric contagion dynamics. Bottom row: diffusion fairness, $\delta(f)$, which captures a group's ability to spread information, under \textbf{(e)} linear, \textbf{(f)} sublinear, \textbf{(g)} superlinear, and \textbf{(h)} asymmetric contagion dynamics. Results are averaged over $n_\mathrm{hg}=10^3$ simulations of the \contagion~model on homophilous (blue), heterophilous (orange), and neutral (gray) hypergraphs generated from the \hypergraph~model. The dashed black line indicates equality, while $\alpha(f),\delta(f)>1$ denote a minority advantage and $\alpha(f),\delta(f)<1$ indicate a majority advantage. We estimate $99$\% confidence intervals using $100$ bootstrap samples. The minority group is disadvantaged in both access and spread under homophilous conditions and under asymmetric transmission, but gains an advantage under symmetric, heterophilous dynamics.}
    \label{fig:acquisition_and_diffusion}
\end{figure*}

\subsection{Acquiring Versus Disseminating Information}\label{sec:results:acquisition_and_diffusion}

We use acquisition fairness $\alpha(f)$ and diffusion fairness $\delta(f)$ to investigate how groups differ in their ability to access and spread information throughout the contagion process. Acquisition fairness $\alpha(f)$ captures a group's ability to receive information. It compares the informed proportion of minority to majority nodes for an overall fraction of informed nodes $f$~\cite{zappala2025_GenderDisparitiesDissemination}. Diffusion fairness $\delta(f)$, by contrast, quantifies a group's ability to spread information by comparing the time it takes for a fraction $f$ of the population to become informed when seeding occurs in the minority group versus the majority~\cite{zappala2025_GenderDisparitiesDissemination}. These measures are defined  in Section~\ref{sec:methods:measures}. To understand how access and influence evolve over time, we analyze $\alpha(f)$ and $\delta(f)$ as functions of the informed fraction $f \in (0.0, 0.9)$ across contagion simulations on homophilous, heterophilous, and neutral hypergraphs. We consider the same contagion dynamics as in Section~\ref{sec:results:synthetic_emd}: linear, sublinear, superlinear, and asymmetric transmission.

Focusing on acquisition fairness $\alpha(f)$, we observe results that align closely with our previous observations based on the Wasserstein distance $d_W$: inequality is similar across all symmetric (i.e., linear, sublinear, and superlinear) contagion dynamics (Fig.~\ref{fig:acquisition_and_diffusion}(a)-(c)). For homophilous hypergraphs, we observe $\alpha(f)<1$ across all values of $f$, indicating a persistent majority advantage. Conversely, in heterophilous hypergraphs, $\alpha(f)>1$, signaling a consistent minority advantage. In both cases, these inequalities diminish as $f$ increases---that is, as more nodes receive the information. This decline is expected since $\alpha(f)$ compares the fraction of informed minority nodes to the overall informed fraction and both approach $1$ over time in connected hypergraphs. Neutral hypergraphs show approximately equal access throughout the process with $\alpha(f)\approx 1$ for all $f$.

In contrast, the asymmetric contagion dynamics produce a qualitatively different pattern: all homophilous regimes result in a majority advantage, with $\alpha(f)<1$ throughout most of the spreading process (Fig.~\ref{fig:acquisition_and_diffusion}(d)). Notably, the observed inequality is not only greater---i.e., $\alpha(f)$ is smaller---but also more persistent. Fairness is only restored late in the contagion process, with $\alpha(f)$ approaching $1$ only when $f\gtrsim 0.9$.

Unlike our inequality measure $d_W$ and acquisition fairness which capture how groups receive information, diffusion fairness $\delta(f)$ instead reflects a group's ability to spread information throughout the hypergraph (Fig.~\ref{fig:acquisition_and_diffusion}(e)-(h)). We observe $\delta(f) < 1$ when hypergraphs are homophilous under all contagion dynamics, indicating that information seeded in the minority group takes longer to reach a given fraction of $f$ nodes compared to information seeded in the majority. These differences are largest for small values of $f$ and gradually diminish as $f$ increases, meaning that we approach equality as more nodes are informed. For heterophilous hypergraphs under symmetric dynamics, we observe a slight minority advantage, $\delta(f)>1$, while we still see a majority advantage under asymmetric dynamics (Fig.~\ref{eq:acquisition_fairness}(h)). However, these differences are relatively small compared to those observed under the homophilous pattern. In neutral hypergraphs, diffusion is essentially equal under symmetric dynamics $\delta(f)\approx 1$, but asymmetric transmission still hinders dissemination from the minority group.

Together, these results highlight the importance of jointly considering hyperedge homophily patterns and contagion dynamics when evaluating group-level fairness. Homophilous connectivity consistently disadvantages the minority group, particularly under asymmetric dynamics, while heterophily can provide a modest minority advantage, especially in terms of information access. Importantly, the magnitude and direction of these effects evolve throughout the contagion process, underscoring the need for dynamic, time-aware notions of fairness in information spread.

\subsection{The Effects of Mixed Hyperedge Homophily on Random Hypergraphs}
\label{sec:results:mixed}

\begin{figure*}[p!]
    \centering
    \includegraphics[width=0.9\textwidth]{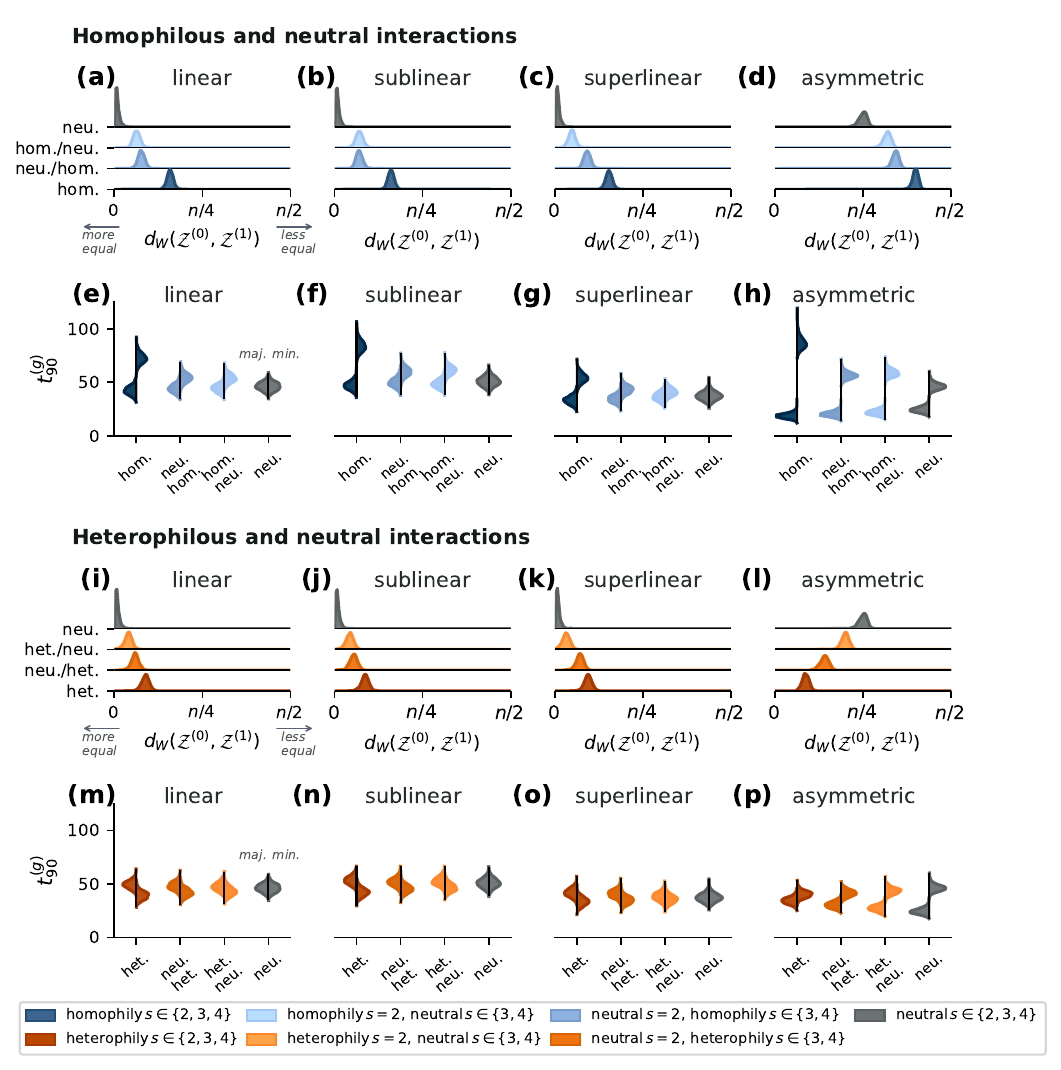}
    \caption{\textbf{Quantifying information access inequality on hypergraphs with mixed hyperedge homophily patterns.} We simulate four contagion processes on synthetic hypergraphs that combine neutral connectivity with either homophilous or heterophilous interactions at specific hyperedge sizes, and assess group-level differences in time-critical information access. Rows $1$ and $3$: distributions of Wasserstein distances $d_W(\mathcal{Z}_0,\mathcal{Z}_1)$ between majority and minority rank distributions. We display results for mixed homophilous-neutral hypergraphs under \textbf{(a)} linear, \textbf{(b)} sublinear, \textbf{(c)} superlinear, and \textbf{(d)} asymmetric contagion and results for heterophilous-neutral hypergraphs under \textbf{(i)} linear, \textbf{(j)} sublinear, \textbf{(k)} superlinear, and \textbf{(l)} asymmetric contagion. Rows $2$ and $4$: violin plots of $t^{(g)}_{90}$, the time to reach $90$\% of nodes in the majority $g=0$ (darker, left violin) and minority $g=1$ (lighter, right violin) groups. We display results for mixed homophilous-neutral hypergraphs under \textbf{(e)} linear, \textbf{(f)} sublinear, \textbf{(g)} superlinear, and \textbf{(h)} asymmetric contagion and results for heterophilous-neutral hypergraphs under \textbf{(m)} linear, \textbf{(n)} sublinear, \textbf{(o)} superlinear, and \textbf{(p)} asymmetric contagion. Mixed homophilous patterns are shown in shades of blue, while mixed heterophilous patterns are shown in orange. All results are averaged over $n_\mathrm{hg}=10^3$ independent simulations of the \contagion~model on hypergraphs generated from the \hypergraph~model with the same structural characteristics. Inequality is amplified when homophilous or heterophilous patterns are localized in hyperedges that reinforces the dominant social contagion pathway, meaning that hyperedge-level structure, as well as dynamical factors jointly shape information access.}
    \label{fig:emd_mixed}
\end{figure*}

Prior studies of real-world higher-order networks have shown that homophily can vary with the scale of interaction~\cite{veldt2023_CombinatorialCharacterizationsImpossibilities, sarker2024_HigherorderHomophilySimplicial, rizi2024_homophilywithinacross, gallo2024_HigherOrderModelingFacetoFace}. We refer to this phenomenon as mixed hyperedge homophily, where the homophily pattern differs across hyperedge sizes. Although such patterns have been documented empirically, their consequences for information diffusion remain less clear. The \hypergraph~model allows us to reproduce heterogeneous structures in a controlled setting, making it possible to disentangle their role in shaping time-critical information access.

To examine how mixed hyperedge homophily influences inequality in information access, we build on the basic homophily patterns considered in Sections~\ref{sec:results:synthetic_emd} and~\ref{sec:results:acquisition_and_diffusion} and use the \hypergraph~model to generate hypergraphs where the type of connectivity varies with hyperedge size. Concretely, we define four mixed patterns that combine neutral, homophilous, and heterophilous connectivity at different hyperedge sizes $s \in \{2,3,4\}$. Each pattern is constructed by varying the hyperedge count inputs $m_{s,r}$ across hyperedge sizes while maintaining the same degree distribution (Pareto, $\gamma=2.9$). Two of the patterns combine homophilous and neutral hyperedges: homophily-neutral features homophilous interactions at $s=2$ and neutral interactions at $s=3,4$, while neutral-homophily has neutral interactions at $s=2$ and homophilous hyperedges of size $s=3,4$. Similarly, heterophily-neutral has heterophilous pairwise edges and neutral higher-order hyperedges, while neutral-heterophily exhibits the opposite pattern. As with the basic patterns, the homophilous case is implemented by overrepresenting single-group hyperedges, and the heterophilous case by underrepresenting them. We report the exact hyperedge counts $m_{s,r}$ for each pattern in Appendix~\ref{appendix:parameters_synthetic}. These mixed patterns allow us to probe how different combinations of hyperedge homophily at different scales affect inequality in time-critical information access.

We then investigate how these mixed structures interact with contagion dynamics, examining both the timing and order of information access. Consider again the example of job opportunities spreading through online social platforms: individuals may primarily share such information with close contacts through direct messages, leading to homophilous pairwise connections (e.g., along socio-economic lines). In contrast, larger interactions, such as feeds or interest-based forums, may be more heterophilous, as participation is less constrained by background. As in our analysis of basic hyperedge homophily patterns (Section~\ref{sec:results:synthetic_emd}), we quantify inequality using the Wasserstein distance $d_W(\mathcal{Z}_0,\mathcal{Z}_1)$ between group-wise empirical rank distributions and the time $t_{90}^{(g)}$ required for $90$\% of nodes in group $g$ to receive information (see Fig.~\ref{fig:emd_mixed}).

Across all homophilous patterns, we observe a consistent majority advantage in information access (Fig.~\ref{fig:emd_mixed}(a)-(h)). Inequality is highest when all hyperedges are homophilous, regardless of contagion dynamics. In contrast, neutral connectivity leads to roughly equal outcomes, except in the asymmetric case. Importantly, mixed hyperedge homophily patterns yield intermediate levels of inequality, with effects depending on which hyperedge sizes carry homophilous interactions. Under superlinear contagion, where information spreads more readily through larger groups, homophilous higher-order edges result in more inequality than homophilous pairwise edges (Fig.~\ref{fig:emd_mixed}(c)). Conversely, under sublinear contagion, homophilous higher-order edges lack this effect because contagion is distributed more evenly across hyperedge sizes (Fig.~\ref{fig:emd_mixed}(b)).

In hypergraphs with heterophilous connections, symmetric contagion dynamics consistently favor the minority group (Fig.~\ref{fig:emd_mixed}(i)-(k),(m)-(o)), while asymmetric dynamics reverse this pattern and favor the majority (Fig.~\ref{fig:emd_mixed}(l),(p)). Fully heterophilous and neutral hypergraphs define the extremes of this behavior, but mixed patterns, with heterophilous and neutral hyperedges, yield intermediate levels of inequality. As with homophilous patterns, the hyperedge size in which heterophilous interactions appear plays a critical role. Under symmetric transmission, larger heterophilous hyperedges produce a stronger minority advantage than heterophilous pairwise edges. This effect is particularly pronounced under superlinear contagion, where large group interactions dominate spread (Fig.~\ref{fig:emd_mixed}(j),(k)). In contrast, under asymmetric dynamics, heterophilous pairwise connections are more effective at offsetting inequality than heterophilous higher-order edges (Fig.~\ref{fig:emd_mixed}(l)).

Taken together, our results show that the effects of mixed hyperedge homophily on inequality depend on which hyperedge sizes carry homophilous or heterophilous interactions, and how those sizes interact with the dynamics of spreading. These findings carry implications for structural interventions: reducing homophily in large-group interactions is most effective under superlinear contagion, where much of the spread occurs through larger groups, whereas pairwise-focused changes may be simpler to implement and equally effective under sublinear dynamics. While the success of such interventions depends on accurately characterizing spreading dynamics in real-world hypergraphs, our analysis underscores the importance of accounting for both higher-order structure and diffusion mechanisms when reasoning about inequalities in information access.


\subsection{Real-world Hypergraphs}
\label{sec:results:real_mixed_homophily}

\begin{figure*}[tb]
\centering
\includegraphics[width=0.9\textwidth]{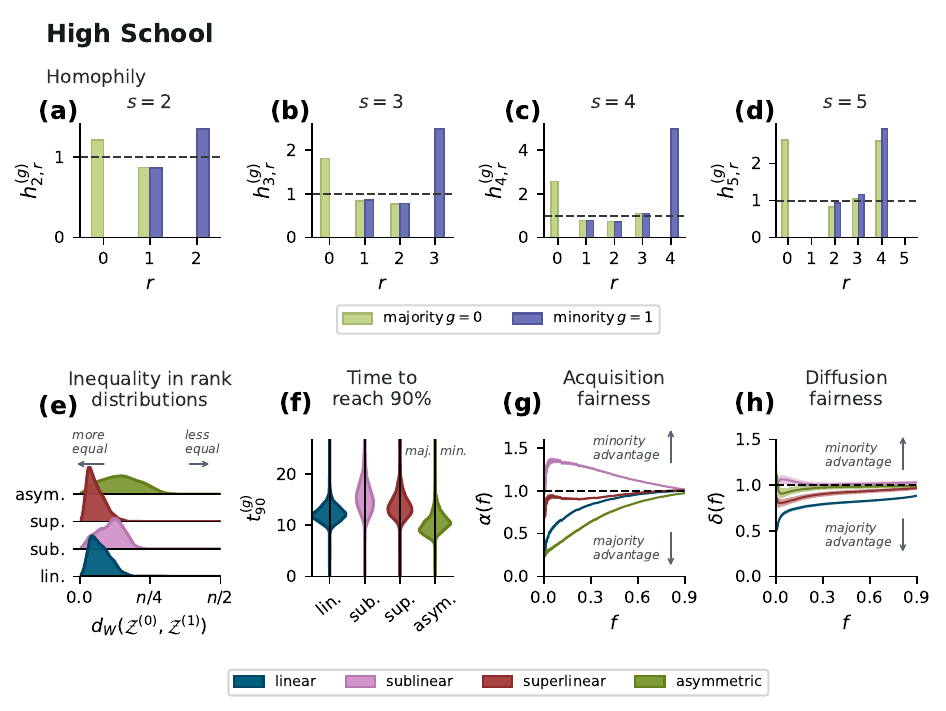}
\caption{\textbf{Homophily patterns and information diffusion inequality in the High School hypergraph.} \textbf{(a)-(d)} Hyperedge homophily $h_{s,r}^{(g)}$ is shown for hyperedge sizes $s\in\{2,3,4,5\}$ for majority ($g=0$, green) and minority ($g=1$, purple) groups. The dashed line shows the expected value under random mixing; values above indicate over-representation. Inequality is captured by \textbf{(e)} distributions of Wasserstein distances $d_W(\mathcal{Z}_0,\mathcal{Z}_1)$, \textbf{(f)} violin plots of the time $t^{(g)}_{90}$ to inform 90\% of majority (left) and minority (right) nodes, \textbf{(g)} acquisition fairness $\alpha(f)$, and \textbf{(h)} diffusion fairness $\delta(f)$. Panels (e)–(h) average results over $n_\mathrm{hg}=10^3$ simulations for linear (blue), sublinear (pink), superlinear (red), and asymmetric (green) contagion. Confidence intervals in (g),(h) are estimated from $100$ bootstrap samples. The unique homophily pattern yields a majority advantage under asymmetric contagion, but a minority advantage under sublinear contagion.}
\label{fig:highschool}
\end{figure*}

\begin{figure*}[tb]
\centering
\includegraphics[width=0.9\textwidth]{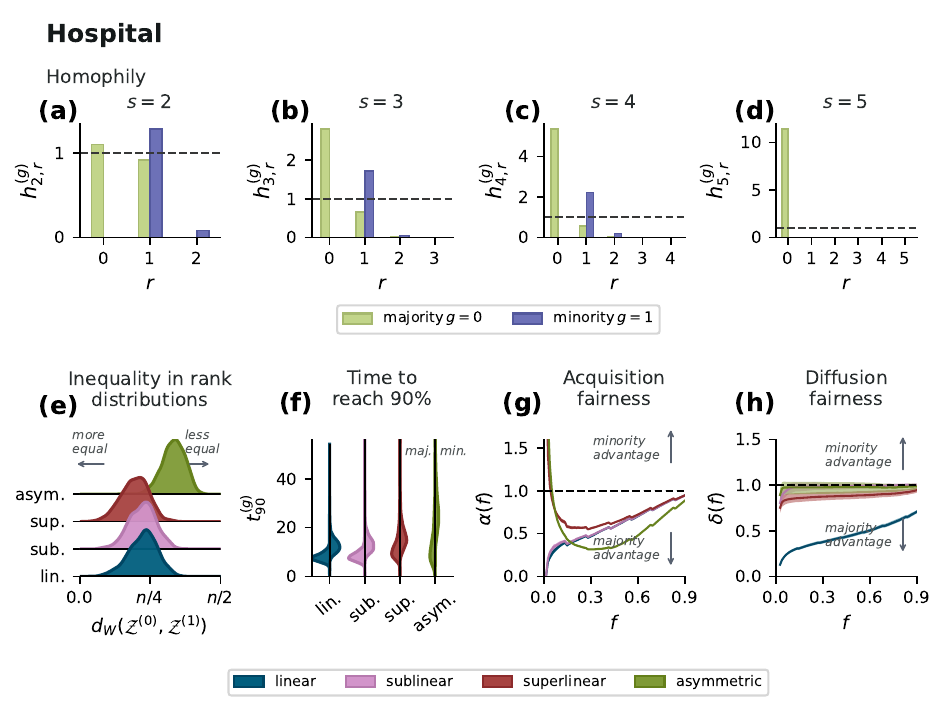}
\caption{\textbf{Homophily patterns and information diffusion inequality in the Hospital hypergraph.} \textbf{(a)-(d)} Hyperedge homophily $h_{s,r}^{(g)}$ is shown for hyperedge sizes $s\in\{2,3,4,5\}$ for majority ($g=0$, green) and minority ($g=1$, purple) groups. The dashed line shows the expected value under random mixing; values above indicate over-representation. Inequality is captured by \textbf{(e)} distributions of Wasserstein distances $d_W(\mathcal{Z}_0,\mathcal{Z}_1)$, \textbf{(f)} violin plots of the time $t^{(g)}_{90}$ to inform 90\% of majority (left) and minority (right) nodes, \textbf{(g)} acquisition fairness $\alpha(f)$, and \textbf{(h)} diffusion fairness $\delta(f)$. Panels (e)–(h) average results over $n_\mathrm{hg}=10^3$ simulations for linear (blue), sublinear (pink), superlinear (red), and asymmetric (green) contagion. Confidence intervals in (g),(h) are estimated from $100$ bootstrap samples. Homophily concentrated in the majority group produces extreme inequality under asymmetric contagion, approaching the theoretical maximum of $n/2$.}
\label{fig:hospital}
\end{figure*}

We now turn from synthetic experiments to case studies on real-world hypergraphs, providing complementary evidence of how structural patterns translate into inequalities in information access. We focus on two face-to-face interaction datasets: a high school contact network~\cite{mastrandrea2015_ContactHighSchool} and a hospital contact network~\cite{vanhems2013_Estimating}. These examples show how mechanisms identified in synthetic settings manifest in empirical systems. Additional datasets from domains such as scientific collaboration and legislative activity are analyzed in Appendices~\ref{appendix:real-world_hypergraph_construction}--\ref{appendix:additional_results_real_world}.

The High School dataset records contacts among $327$ students over five days, with hyperedges formed when groups of students are within $1.5$ meters for at least $20$ seconds~\cite{benson2018_SimplicialClosurePrediction, mastrandrea2015_ContactHighSchool}. Hyperedges include up to five students ($s \in \{2,3,4,5\}$), and the gender distribution is moderately imbalanced ($55\%$ male). The Hospital dataset captures interactions among $75$ individuals, including $46$ healthcare workers and $29$ patients~\cite{vanhems2013_Estimating}. Like the High School network, hyperedges are defined by co-presence, but group imbalance is more pronounced ($61\%$ staff, $39\%$ patients). Further construction details appear in Appendix~\ref{appendix:real-world_hypergraph_construction}, and the experimental setup for the real-world simulations is detailed in Appendix~\ref{appendix:parameters_real_world}.

\paragraph{High School. } The High School hypergraph’s homophily pattern resembles, but is not identical to, the synthetic homophilous case (Fig.~\ref{fig:highschool}(a)-(d)). For $s \in \{2,3,4\}$, both groups show homophily: same-gender edges are overrepresented and mixed-gender edges underrepresented. At $s=5$, however, all-minority hyperedges are absent, and mixed groups with four minority and one majority student are overrepresented alongside all-majority groups. Notably, the minority is more homophilous than the majority at smaller sizes. Such irregularities highlight a key feature of real-world systems: interaction patterns rarely conform perfectly to idealized models, instead reflecting contextual contingencies.

Simulations on the High School hypergraph confirm that these structures shape inequalities in line with synthetic predictions. Asymmetric contagion produces the largest inequalities, consistently favoring the majority group (Fig.~\ref{fig:highschool}(e),(g)), as in our synthetic homophilous cases (Fig.~\ref{fig:emd_synthetic}(d)). Sublinear contagion also creates inequality, but here the minority gains the advantage. Our synthetic results suggest that this reversal arises because sublinear dynamics downweight large hyperedges, balancing transmission events across sizes (Fig.~\ref{fig:emd_synthetic}(j)). In synthetic settings, this mechanism reduces but does not reverse the majority advantage (Fig.~\ref{fig:emd_mixed}(b)). In the High School hypergraph, however, minority homophily is stronger at smaller sizes, which may help explain why the reduction is strong enough to yield a minority advantage. Other measures of inequality, including diffusion fairness and the time to inform $90\%$ of nodes, support this interpretation without revealing additional effects (Fig.~\ref{fig:highschool}(f),(h)).

\paragraph{Hospital. } In the Hospital hypergraph, the homophily pattern is stark. Hospital staff (the majority) form homophilous groups across all hyperedge sizes, while patients rarely interact with one another, instead appearing almost exclusively in mixed groups with staff (Fig.~\ref{fig:hospital}(a)-(d)). This reflects the functional organization of the hospital confirmed by the analysis in Vanhems et al.~\cite{vanhems2013_Estimating}: patients interact primarily with staff, while staff interact both with patients and with one another.

The structural asymmetries in the Hospital hypergraph generate extreme inequalities under contagion. In the asymmetric case, inequality measured in terms of $d_W(\mathcal{Z}_0, \mathcal{Z}_1)$ approaches the theoretical upper bound ($n/2$), implying that nearly all staff receive information before any patient (Fig.~\ref{fig:hospital}(e)). Interestingly, acquisition fairness initially shows a minority advantage in superlinear and asymmetric contagion (Fig.~\ref{fig:hospital}(g)). Larger homophilous hyperedges drive the spread under these dynamics (Fig.~\ref{fig:emd_synthetic}(k),(l)), yet in the Hospital hypergraph only staff form such large homophilous groups. One plausible mechanism is that information saturates subsets of staff before spreading further, forcing transmission through patients en route to other staff groups (e.g., across shifts). Ultimately, however, strong staff–staff homophily dominates, producing a sharp majority advantage. Time-to-inform distributions corroborate this interpretation (Fig.~\ref{fig:hospital}(f)). With respect to diffusion fairness, outcomes are approximately balanced except under linear symmetric contagion (Fig.~\ref{fig:hospital}(h)). This likely reflects the unique homophily structure: in superlinear and asymmetric cases, staff saturation makes patients instrumental conduits, while in sublinear contagion, reduced reliance on large hyperedges similarly routes spread through patients. By contrast, the linear case neither prioritizes nor suppresses large staff-dominated hyperedges, allowing staff homophily to drive outcomes unchecked.

Together, these case studies show how the patterns identified in synthetic settings appear in real-world contexts, though with added complexities. In the High School hypergraph, inequality depends on the interplay between small-group minority and large-group majority homophily, with sublinear contagion sometimes reversing the expected majority advantage. In the Hospital hypergraph, structural asymmetry between patients and staff drives extreme inequalities under asymmetric contagion. These results highlight the value of synthetic analysis as a framework for interpreting real-world outcomes, while also underscoring that the magnitude and direction of inequalities is determined by the precise combination of group sizes, interaction structures, and contagion dynamics.

\section{Discussion}\label{sec:discussion}

Timely access to information---whether about opportunities, news, or knowledge---can shape both individual success and group-level differences. Understanding how access to information varies between groups is therefore a crucial step toward explaining disparities in individual and collective outcomes. Because information transmission often involves complex dynamics, including social reinforcement, social inhibition, and higher-order interactions, it is essential to move beyond the classical models of social contagion on networks and develop generative higher-order network models that encode characteristics such as group structure, hyperedge homophily, and degree heterogeneity.

In this work, we introduce the Hypergraphs with Hyperedge Homophily (\hypergraph) model, along with a nonlinear asymmetric SI (\contagion) social contagion model that extends the model of St-Onge et al.~\cite{stonge2022_InfluentialGroupsSeeding} to incorporate asymmetry in the transmission of information between in-group and out-group nodes in hypergraphs. Combining the \hypergraph~model and \contagion~model allows us to systematically probe how time-sensitive information spreads on hypergraphs. Our results indicate that information access inequality is shaped by the interplay of mesoscopic hypergraph structure and the social contagion process.

While we focus here on inequalities in time-critical information access, the models we introduce have broader applications across a variety of settings. In innovation adoption, nonlinear contagion dynamics and hyperedge homophily often shape the spread of new behaviors, technologies, and ideas. In public health campaigns, such as vaccine uptake, social reinforcement within higher-order structures like families or workplaces can critically influence behavior change. In misinformation and rumor propagation, asymmetric transmission and group polarization can amplify biases in information spread, highlighting opportunities for targeted interventions.

Beyond applied settings, the \contagion~model offers a methodological tool for benchmarking fairness criteria for the spread of information. In addition, the \hypergraph~model enables evaluating structural interventions such as rewiring hypergraphs to reduce homophilous interactions and stress-testing classical diffusion models under more realistic higher-order and group-structured conditions. By enabling fine-grained control over both structure and dynamics, the \hypergraph~and \contagion~models open new avenues for studying how network inequalities manifest and how they might be mitigated.

In this paper, we examined when and how group-level  differences can arise under a wide range of conditions that stem from a combination of structure and dynamics. Consistent with prior work on pairwise networks, we find that combining homophilous interactions with asymmetric transmission yields the starkest inequality, disadvantaging certain groups in both acquiring and disseminating information~\cite{wang2022_InformationAccessEquality, zappala2025_GenderDisparitiesDissemination}. However, inequality also emerges in more subtle ways. For instance, heterophilous hypergraphs can produce minority advantages that might reflect a group of elites rather than a disadvantaged population. Even in symmetric dynamics, the interaction between hyperedge sizes and nonlinear contagion can create group-level differences, depending on which types of interactions dominate the spread. Homophily and group imbalance also act together: in group-balanced hypergraphs, biased transmission seems to be a prerequisite for unequal outcomes, whereas imbalance amplifies disparities even under symmetric contagion. These variations highlight the importance of understanding not just the existence of inequality, but its underlying mechanisms.

Our synthetic analysis demonstrates that the consequences of homophily depend on how it is distributed across hyperedge sizes and how those sizes interact with contagion dynamics. Under superlinear contagion, much of the spread is channeled through large-group interactions, so inequality is amplified when those higher-order interactions are homophilous. By contrast, under sublinear contagion, large hyperedges play a smaller role because transmission is distributed more evenly across sizes. As a result, homophily in higher-order interactions does not generate the same degree of inequality. Taken together, these patterns suggest that interventions could operate differently across regimes: reducing homophily in large-group edges is most effective in superlinear settings, while even modest adjustments to pairwise connections can be effective in sublinear settings, where neutral edges at any size reduce inequality.

Our real-world case studies of the High School and Hospital hypergraphs demonstrate how synthetic analysis can guide interpretation, while also revealing the challenges of working with empirical data. In the High School hypergraph, sublinear contagion reduces the majority’s advantage by limiting the influence of their larger homophilous groups to the point that the minority gains earlier access. In the Hospital hypergraph, by contrast, stark structural asymmetries between staff and patients drive extreme inequalities under asymmetric contagion. More broadly, these cases underscore that real-world systems rarely display clean patterns: homophily may be stronger for the minority than the majority, certain hyperedge types may be absent altogether, and higher-order patterns become increasingly difficult to characterize as hyperedge size grows. In addition, extreme group imbalances can induce high variance in contagion outcomes, exacerbating the difficulty of reliably quantifying inequality in information access. Real-world hypergraphs also include many structural properties outside the scope of the \hypergraph~model, such as clustering and nestedness. These complexities highlight the need for caution when interpreting real-world outcomes, while also showing the value of synthetic analysis for generating intuitions that help disentangle which structural features are most consequential.\\

While our work highlights potential sources of inequality in information access, it has a few limitations. Communication is a complex process shaped by psychological, interpersonal, and platform-level factors. Although we incorporate some empirically motivated features, such as in-group preference and transmission nonlinearities, the \contagion~model cannot capture the full richness of real-world information flows. Likewise, while the \hypergraph~model encodes hyperedge homophily and degree heterogeneity, it omits other structural features that may be important, such as hyperedge overlap, clustering, temporal dynamics, or feedback loops. These omissions become especially salient when analyzing real-world data, where interaction patterns rarely align neatly with idealized structures and where higher-order homophily can be difficult to measure due to sparsity in large hyperedges. Finally, all measures of information access inequality that we use (i.e., $d_W$ as well as acquisition and diffusion fairness) quantify group-level differences but do not capture fairness at the individual level or make normative claims about which outcomes are desirable or just.\\

Our work opens many avenues for future research. First, a more complete theoretical characterization of the \hypergraph~model could help clarify the range of structures it can produce and support analytical study of its limiting behavior. Second, our real-world results suggest that extreme group imbalance and large hyperedges can have meaningful effects, with imbalance tending to exacerbate inequality and large hyperedges amplifying nonlinear contagion dynamics. Further study is needed to precisely understand when and how these structural changes alter information access in real-world systems. Third, enriching the model with additional structural features observed in real-world systems, such as clustering or nestedness, could improve its ability to approximate real-world contexts and disentangle which properties most strongly drive inequality. Finally, the broader applicability of the model invites exploration of settings beyond information access, including innovation diffusion, political polarization, and changes in health behavior, where higher-order structure and group dynamics play a central role. These extensions, among others, would strengthen the utility of our modeling approach as both a practical tool and a foundation for new theoretical insights.


\section{Conclusion}\label{sec:conclusion}
Taken together, our findings reveal how inequalities in information access emerge from the interplay between higher-order network structure and the dynamics of information transmission. By connecting structural features to time-sensitive outcomes, our work contributes to a growing effort in network science and algorithmic fairness to understand---and eventually mitigate---the mechanisms behind unequal information access in socio-technical systems. Our modeling approach complements this effort to design more equitable platforms by enabling systematic exploration of how group structure, hyperedge homophily and its dependence on hyperedge size, and nonlinear contagion jointly shape who gains access, when, and why disparities persist.


\section{Methods}\label{sec:methods}

In this section, we introduce the tools which allow us to define hypergraphs with fixed hyperedge homophily and tunable degree distributions. First, we explain the notion of hypergraph homophily introduced by Veldt et al.~\cite{veldt2023_CombinatorialCharacterizationsImpossibilities} and define our generative model of hypergraphs, the \hypergraph~model. We then introduce the nonlinear asymmetric SI model, the \contagion~model, and describe how we conduct stochastic simulations of the social contagion process. Finally, we define the three measures of information access inequality we use throughout this work. We provide an overview of the mathematical notation used in this section in Table~\ref{tab:notation} in Appendix~\ref{appendix:notation}.


\subsection{Hyperedge Homophily}\label{sec:methods:hypergraph_homophily}

We use the measure of higher-order homophily $h$ introduced by Veldt et al.~\cite{veldt2023_CombinatorialCharacterizationsImpossibilities} to quantify the level of homophily in a hypergraph $H=(\mathcal{V}, \mathcal{E})$, where $\mathcal{V}$ is the set of nodes and $\mathcal{E}$ is the set of hyperedges. To emphasize that homophily in hypergraphs can vary with hyperedge size, we refer to this measure as hyperedge homophily. Let $g_v \in \{0, 1\}$ denote the group membership of node $v \in \mathcal{V}$; this binary label can represent attributes such as gender, employment status, or other binary groupings. The size of hyperedge $e$ is denoted by $s_e = |e|$. We define the type of a given hyperedge $e$ as the number of group $g=1$ nodes it contains, $r_e=|\{v\in e : g_v = 1\}|$. When we are not referring to a specific node or hyperedge, we drop the subscripts and write $g$, $s$, and $r$ for group, hyperedge size and hyperedge type, respectively. We assume $s_e \geq 2$ throughout the paper.

Following Veldt et al.~\cite{veldt2023_CombinatorialCharacterizationsImpossibilities}, we define the affinity score $a^{(0)}_{s,r}$ for nodes of group $g=0$ in hyperedges of size $s$ and type $r\in\{0,\dots,s-1\}$, as
\begin{equation}
    a^{(0)}_{s,r} = \frac{m_{s,r}}{\sum_{r^\prime = 0}^{s-1}m_{s,r^\prime}},
\end{equation}
where $m_{s,r}$ is the number of hyperedges of size $s$ with $r$ nodes from group $g=1$.

The affinity score $a^{(0)}_{s,r}$ of group $g=0$ captures the number of hyperedges of size $s$ that contain $s-r>0$ nodes from group $g=0$, normalized by the total number of hyperedges that contain at least one node from group $g=0$. Similarly, we define the affinity score $a^{(1)}_{s,r}$ for nodes of group $g=1$ in hyperedges of size $s$ and type $r\in\{1,\dots,s\}$, as
\begin{equation}
    a^{(1)}_{s,r} = \frac{m_{s,r}}{\sum_{r^\prime = 1}^{s} m_{s,r^\prime}}.
\end{equation}

The affinity score $a^{(1)}_{s,r}$ of group $g=1$ captures the number of hyperedges of size $s$ that contain $r>0$ nodes from group $g=1$, normalized by the total number of edges that contain at least one node of group $g=1$. The two affinity scores measure the tendency of nodes of a given group to appear in edges of a certain type.

To calculate hyperedge homophily $h_{s,r}^{(g)}$, affinity scores are normalized using appropriate baseline scores $b_{s,r}^{(g)}$. These baselines are derived from a null model in which the number of hyperedges of size $s$, denoted $m_s$, is fixed, but nodes are assigned to hyperedges at random, independent of group membership. For nodes in group $g=0$, and for hyperedges of size $s$ with type $r\in\{0,\dots,s-1\}$, the baseline score is
\begin{equation}\label{eq:baseline_group_0}
    b^{(0)}_{s,r} = \frac{1}{1 - \left(1-\frac{n_0}{n}\right)^s}\binom{s}{s-r}\left(\frac{n_0}{n}\right)^{s - r} \left(1 - \frac{n_0}{n}\right)^r~.
\end{equation}

For nodes in group $g=1$, and for hyperedges of size $s$ with type $r\in\{0,\dots,s-1\}$, the baseline score is
\begin{equation}\label{eq:baseline_group_1}
    b^{(1)}_{s,r} = \frac{1}{1 - \left(1-\frac{n_1}{n}\right)^s}\binom{s}{r}\left(\frac{n_1}{n}\right)^r \left(1 - \frac{n_1}{n}\right)^{s-r}~.
\end{equation}

Equation~\eqref{eq:baseline_group_1} states that, under the null model, the distribution over type $r$ hyperedges among all size $s$ hyperedges that include at least one node from group $g=1$ follows a renormalized binomial distribution. Intuitively, this corresponds to performing $s$ Bernoulli trials with success probability $n_1/n$, i.e., the proportion of nodes in group $g=1$. Since we condition on the hyperedge containing at least one node from group $g=1$, we must renormalize the distribution. This renormalization is captured by the factor in front of the binomial coefficient in Equation~\eqref{eq:baseline_group_1}. The intuition for Equation~\eqref{eq:baseline_group_0} is analogous, but we condition on the presence of at least one node from group $g=0$.

The homophily scores $h_{s,r}^{(g)}$ that constitute the hyperedge homophily $h$ are the ratios of the affinity and baseline scores:
\begin{equation}\label{eq:hypergraph_homophily}
    h^{(g)}_{s,r} = \frac{a^{(g)}_{s,r}}{b^{(g)}_{s,r}}~.
\end{equation}

For each hyperedge size $s\in\{2,\dots,s_\mathrm{max}\}$, the type $r$ ranges over $r\in\{0, \dots, s-1\}$ when $g=0$, and $r\in\{1, \dots, s\}$ when $g=1$.
For brevity, we refer to these quantities as affinity, baseline, and homophily scores, although they are referred to as alternative affinity, alternative baseline, and alternative homophily scores by Veldt et al.~\cite{veldt2023_CombinatorialCharacterizationsImpossibilities}.

Equation~\eqref{eq:hypergraph_homophily} describes the hyperedge homophily pattern of a single hypergraph of maximum hyperedge size $s_\mathrm{max}$ using $\sum_{s=2}^{s_\mathrm{max}}2s=s_\mathrm{max}^2 + s_\mathrm{max} -2$ values. This added complexity enables the expression of richer homophily structures than are possible in pairwise networks---for example, homophily that varies with hyperedge size. However, the high dimensionality of this representation poses challenges for both interpretation and visualization. We also note that when $s_\mathrm{max}=2$, the hyperedge homophily measure $h^{(g)}_{s,r}$ reduces to classical notions of pairwise homophily and dyadicity~\cite{coleman1958_RelationalAnalysisStudy,park2007_DistributionNodeCharacteristics}. The hyperedge homophily measure bears similarity to a generalized homophily measure on pairwise networks which defines homophily for cliques of arbitrary size~\cite{rizi2024_homophilywithinacross} instead of hyperedges.


\subsection{Hypergraph Model}\label{sec:methods:hypergraph_model}

To study the influence of homophilous connections and degree heterogeneity in systems with higher-order interactions, we need to generate hypergraphs with fixed levels of hyperedge homophily and tunable degree distributions. However, this presents several challenges. First, a hypergraph's hyperedge size sequence $(s_e)_{e\in\mathcal{E}}$ and degree sequence $(k_v)_{v\in\mathcal{V}}$ must satisfy a joint graphicality constraint~\cite{ruggeri2024_FrameworkGenerateHypergraphs}, i.e., they need to be such that a hypergraph with this exact degree and edge size sequence exists. This poses a challenge for the stub matching schemes that are commonly employed in micro-canonical configuration models for graphs~\cite{fosdick2018_ConfiguringRandomGraph}. Second, Veldt et al.~\cite{veldt2023_CombinatorialCharacterizationsImpossibilities} present combinatorial impossibility results for some values $\tilde{h}$ of their proposed homophily measure $h$. This means that there may not exist a hypergraph that realizes a given hyperedge homophily pattern $\tilde{h}$. Finally, the number of possible hyperedges of size $s$ increases as $\binom{n}{s}$. This makes it infeasible to generate hypergraphs by sampling an independent Bernoulli random variable at each possible hyperedge with success probability dependent on the properties of the nodes in that hyperedge~\cite{ruggeri2024_FrameworkGenerateHypergraphs}.

To overcome these challenges, we propose a generative model, the Hypergraphs with Hyperedge Homophily (\hypergraph) model, that creates hypergraphs with fixed levels of hyperedge homophily and degree heterogeneity in a computationally efficient way.
The model takes as input the number of nodes $n_g$ in each group $g$, the number of hyperedges $m_{s,r}$ of a given size $s \in \{2,\dots,s_\mathrm{max}\}$ and type $r \in \{0,\dots,s\}$, and the group-dependent distributions $\rho_g$ on $[0,\infty)$ of hidden variables that control the expected degree of nodes. We refer to the means of these distributions as $\bar{\kappa}_g$. This information is sufficient to exactly fix the hyperedge homophily to a value $\tilde{h}$ as per Equation~\eqref{eq:hypergraph_homophily} and to softly constrain the degree distribution.

Given these input values, we construct a hypergraph as follows: 

First, we order the nodes arbitrarily and assign the first $n_0$ nodes to group $g = 0$ and the remaining $n_1 = n - n_0$ nodes to group $g = 1$. This ensures that we have exactly $|\mathcal{V}_g| = n_g$ nodes in each group, where $\mathcal{V}_g$ is the subset of nodes in group $g$.

We then sample a hidden variable $\kappa_v$ at random for each node $v$ from the group-dependent distribution $\rho_{g_v}(\kappa)$. This value $\kappa_v$ governs the node's propensity to join hyperedges. As such, the distributions $\rho_g(\kappa)$ influence the degree heterogeneity of the hypergraphs.

Then, we create $m_{s,r}$ hyperedges for each size $s$ and type $r$ by randomly sampling $r$ nodes from $\mathcal{V}_1$ with probability proportional to $\kappa_v$ and sampling the remaining $s-r$ nodes from $\mathcal{V}_0$, also with probability proportional to $\kappa_v$. The probability of choosing a node $v\in\mathcal{V}_g$ then becomes 
\begin{equation}
    \pi_{v} = \frac{\kappa_v}{\sum_{v'\in\mathcal{V}_{g_v}}\kappa_{v'}}\approx\frac{\kappa_v}{n_{g_v}\bar{\kappa}_g}~,
\end{equation}
where the sample mean of $\{\kappa_v\}_{v\in\mathcal{V}_g}$ is replaced with the mean $\bar{\kappa}_g$ of the distribution $\rho_g(\kappa)$ from which the nodes are sampled. This serves as a good approximation as $n_g\rightarrow \infty$. The number of chances for a node of group $g$ to join a hyperedge is then 
\begin{equation}
\begin{split}
    \sigma_1 = \sum_{s=2}^{s_\mathrm{max}}\sum_{r=1}^{s} r m_{s,r} ~~~\text{if}~g=1~,\\
    \sigma_0 = \sum_{s=2}^{s_\mathrm{max}}\sum_{r=0}^{s-1} (s-r) m_{s,r} ~~~\text{if}~g=0~.\\
\end{split}
\end{equation}

Therefore, the expected degree $\bar{k}_v$ of a node is
\begin{equation}
\begin{split}
    \bar{k}_v = \pi_v \sigma_1= \frac{\kappa_v}{n_{1}\bar{\kappa}_1} \sum_{s=2}^{s_\mathrm{max}}\sum_{r=1}^{s} r m_{s,r}~~~\text{if}~g=1~,\\
    \bar{k}_v = \pi_v \sigma_0 = \frac{\kappa_v}{n_{0}\bar{\kappa}_0} \sum_{s=2}^{s_\mathrm{max}}\sum_{r=0}^{s-1} (s-r) m_{s,r}~~~\text{if}~g=0~.\\
\end{split}
\end{equation}

This means that we can ensure a node's expected degree under the model $\bar{k}_v$ corresponds to its hidden variable $\kappa_v$ by adjusting the mean of the distribution $\rho_g(\kappa)$ to $\bar{\kappa}_g=\frac{\sigma_g}{n_g}$ for $g\in\{0,1\}$. We ensure that this consistency condition is always met in our simulations.

To ensure that hypergraphs do not have multi-edges and that nodes appear at most once within a given hyperedge, we take the following steps: If a node appears more than once in a given hyperedge, we disregard that hyperedge and resample until the hyperedge contains unique nodes. Once all hyperedges are sampled, we erase multi-edges. As the probability of multi-edges decays with an increasing number of nodes $n$, this results in a minor perturbation to the hypergraph~\cite{chodrow2020_ConfigurationModelsRandom}. We report the parameter values used to generate the hypergraphs we use in our simulations in Appendix~\ref{appendix:parameters_synthetic}.


\subsection{Contagion Model}\label{sec:methods:contagion}

We introduce a nonlinear social contagion model that accounts for asymmetry in information transmission and, therefore, allows us to capture social reinforcement and social inhibition. As this model adds nonlinearity and asymmetry to the classical Susceptible-Infected (SI) model, we refer to it as the nonlinear, asymmetric SI (\contagion) model.

We represent whether a node $v \in \mathcal{V}$ is aware of a piece of information using a time-dependent state variable $x_v(t) \in \{0, 1\}$, where $x_v(t)=1$ indicates that node $v$ is informed at time $t$, and $x_v(t)=0$ indicates that it is still unaware of the information. The number of informed nodes in the hypergraph is $i(t)=\sum_{v \in \mathcal{V}}x_v(t)$. To track group-specific dynamics, we quantify the number of informed nodes in group $g$ as $i_g(t)=\sum_{v\in \mathcal{V}_g}x_v(t)$. Finally, we refer to the number of informed nodes from group $g$ in a specific hyperedge $e$ as $i_{e,g}(t)=\sum_{v\in e\cap \mathcal{V}_g}x_v(t)$. The state of each node at time $t=0$ is given as an initial condition $x_v(t=0)=\tilde{x}_v$, and we are primarily concerned with the time-evolution of the node states for times $t>0$.

We assume that once a node becomes informed, it does not forget the information over the time scale of the contagion process. Informed nodes can spread the information to uninformed nodes within the same hyperedge, leading to a single type of transition: from uninformed to informed. We parameterize the transmission dynamics using eight parameters: four transmission rates $\lambda_{gg'}$, representing the rate at which a node from group $g$ informs a node from group $g'$, and four nonlinearity exponents $\nu_{gg'}$ that modulate the influence of the number of informed nodes involved in such transmissions. The total rate at which nodes from group $g$ acquire information within a hyperedge $e$ is
\begin{equation}\label{eq:transmission_rate_full}
    \beta_{g}(e) = (s_e-i_{e,g})(\lambda_{gg} i_{e,g}(t)^{\nu_{gg}} + \lambda_{g'g} i_{e,g'}(t)^{\nu_{g'g}}).
\end{equation}

To simplify the \contagion~model, we assume symmetry in both the base transmission rates and the nonlinearity parameters. Specifically, we define the in-group transmission rates as $\lambda_\mathrm{{in}} = \lambda_{00}=\lambda_{11}$ and the out-group transmission rate as $\lambda_\mathrm{{out}}=\lambda_{01}=\lambda_{10}$. Similarly, we set $\nu_\mathrm{{in}}=\nu_{00}=\nu_{11}$ and $\nu_\mathrm{{out}}=\nu_{01}=\nu_{10}$ and refer to them as the in-group and out-group nonlinearity parameters, respectively. When $\lambda_\mathrm{{in}} > \lambda_\mathrm{{out}}$, the model captures a preference for spreading information within groups. The nonlinearity parameters $\nu_\mathrm{{in}}$ and $\nu_\mathrm{{out}}$ account for phenomena such as social reinforcement or inhibition by allowing the transmission rate to exhibit superlinear or sublinear dependence on the number of informed in-group or out-group members. Under this parameterization, the transmission rate in Equation~\eqref{eq:transmission_rate_full} becomes
\begin{equation}\label{eq:transmission_rate_inout}
    \beta_{g}(e) = (s_e-i_{e,g})(\lambda_\mathrm{in} i_{e,g}(t)^{\nu_\mathrm{in}} + \lambda_\mathrm{out} i_{e,g'}(t)^{\nu_\mathrm{out}}).
\end{equation}

Equation~\eqref{eq:transmission_rate_inout} is closely related to the transmission rate function used by St-Onge et al.~\cite{stonge2022_InfluentialGroupsSeeding} to study contagion on a hypergraph with only a single group of nodes. In the special case where $n_0=n$ and $n_1=0$, the \contagion~model reduces exactly to theirs: $\beta_0(e) = (s_e-i_{e,0})\lambda_\mathrm{{in}}i_{e,0}^{\nu_\mathrm{{in}}}$. However, note that the \contagion~model does not generally reduce to this form in the case of two indistinguishable groups of equal size, i.e., when $n_0=n_1=n/2$, $\lambda_\mathrm{{out}}=\lambda_\mathrm{{in}}=\lambda$, and $\nu_\mathrm{{out}}=\nu_\mathrm{{in}}=\nu$. This is because
\begin{equation}
    \lambda (i_{e,0}(t)+i_{e,1}(t))^\nu \neq \lambda (i_{e,0}^\nu(t)+i_{e,1}^{\nu}(t))~~~\text{for}~\nu\neq 1.
\end{equation}
Only under the additional condition of linear spread (i.e., $\nu=1$) do the models coincide exactly.

The case of linear contagion admits an important structural equivalence: when the hypergraph is locally tree-like, i.e., the hyperedges overlap at most at a single node, the \contagion~model with $\nu_\mathrm{{in}}=\nu_\mathrm{{out}}=1$ is equivalent to contagion on the clique projection of the hypergraph, where each hyperedge of size $s$ is replaced by an $s$-clique in the corresponding pairwise network~\cite{stonge2022_InfluentialGroupsSeeding}.


\subsection{Stochastic Simulation}\label{sec:methods:simulation}
We describe the procedure used to simulate the contagion process under the \contagion~model on both synthetic and real-world hypergraphs. The initialization details vary slightly depending on whether the hypergraph is synthetic, has ground-truth labels, or includes probabilistically inferred labels. Full parameter settings and implementation details are provided in Appendix~\ref{appendix:parameters_synthetic} for synthetic hypergraphs and Appendix~\ref{appendix:parameters_real_world} for real-world hypergraphs.

For synthetic hypergraphs, each simulation run begins with an independently sampled hypergraph generated from a fixed set of parameters. In real-world datasets, by contrast, the hypergraph structure is fixed. If node labels are inferred rather than observed, we resample the labels before each simulation using the procedure described in Appendix~\ref{appendix:parameters_real_world}. For datasets with ground-truth node labels, both the structure and the labeling remain fixed across runs. Once node labels are assigned, we select a seed set $\mathcal{S}\subset \mathcal{V}$ either by sampling uniformly from all nodes or by sampling uniformly within each group-level composition of the seed set. We discuss these seeding strategies in more detail in Appendix~\ref{appendix:parameters_synthetic} and Appendix~\ref{appendix:parameters_real_world}.

To simulate the continuous-time stochastic contagion process defined by the transmission rate function in Equation~\eqref{eq:transmission_rate_inout}, we use Gillespie's algorithm as described below~\cite{kiss2017_MathematicsEpidemicsNetworks, masuda2023_GillespieMultiagentDynamics}. At time $t=0$, we initialize all seed nodes $v \in \mathcal{S}$ as informed, i.e., set $x_v(0)=1$ for all $v \in \mathcal{S}$. To determine the time $t$ of the next transmission event, we first compute the total transmission rate over all hyperedges and groups,
\begin{equation}
    \beta = \sum_{e \in \mathcal{E}}\sum_{g\in \{0, 1\}}\beta_g(e).
\end{equation}

We then draw a time increment $\Delta t \sim \mathrm{Exp}(\beta)$ from the exponential distribution $\mathrm{Exp}$ and advance the simulation time via $t \leftarrow t + \Delta t$.

To determine which node becomes informed, we proceed in two steps. First, we select a hyperedge $e$ and target group $g$ with probability proportional to their contribution $\beta_g(e)$ to the total rate $\beta$. Then, we select a node $v$ uniformly at random from the set of uninformed nodes in group $g$ within hyperedge $e$, i.e., $v\sim \mathrm{U}[\{v^\prime\in e:g_{v^\prime}=g \wedge x_{v^\prime}(t)=0\}]$, where $\mathrm{U}[\mathcal{X}]$ denotes the uniform distribution on a set. The selected node $v$ updates its state from $x_v(t)=0$ to $x_v(t)=1$, and remains in this state at all future times $t'>t$. After each event, we update the relevant transmission rates $\beta_g(e)$ based on the new node states. This process is repeated until no further events can occur, i.e., until $\beta = 0$.


\subsection{Measuring Information Access Inequality with Optimal Transport}\label{sec:methods:measures}

To quantify between-group differences in the timing of information access, we introduce a measure based on the Wasserstein distance between empirical distributions of the times at which nodes are informed.

Simulating the contagion process described in Sections~\ref{sec:methods:contagion} and~\ref{sec:methods:simulation} yields a time $\tau_v$ at which each node $v \in \mathcal{V}$ becomes informed. The set of these times $\mathcal{T}=\{\tau_v :v\in \mathcal{V}\}$ gives rise to a corresponding set of ranks $\mathcal{Z}=\{z_v : v\in \mathcal{V\}}$, where $z_v=1$ indicates that node $v$ was the first to receive the information. To study access patterns by group $g$, we define group-specific rank distributions $\mathcal{Z}_g=\{z_v : v\in \mathcal{V}_g\}$.

We use the Wasserstein distance between rank distributions $d_W(\mathcal{Z}_0, \mathcal{Z}_1)$ to quantify inequality in information access, that is, the extent to which one group tends to receive information earlier than the other. Larger values of $d_W$ indicate a greater separation between the two distributions, meaning that one group is systematically informed before the other. However, because the Wasserstein distance is a symmetric measure, it does not indicate which group is advantaged. On a hypergraph with $n$ nodes, the distance is bounded above by $n/2$. This corresponds to the maximal possible separation in ranks between the two groups. 

Because both the contagion dynamics (Section~\ref{sec:methods:contagion}) and the hypergraph generation process (Section~\ref{sec:methods:hypergraph_model}) are stochastic, we simulate $n_{hg}$ independent realizations and compute the Wasserstein distance $d_W(\mathcal{Z}_0, \mathcal{Z}_1)$ for each run. This yields a distribution of inequality values across different instances of the spreading process and hypergraph structure, which we can analyze to assess typical outcomes, variability, and sensitivity to modeling choices.


\subsection{Measuring Information Access Inequality with Acquisition \& Diffusion Fairness}

An alternative way to quantify a group's ability to access information is through acquisition fairness, a measure introduced by Zappalà et al.~\cite{zappala2025_GenderDisparitiesDissemination}. We assume that group $g=1$ denotes the group of interest---typically a minority or historically disadvantaged group. Let $t_f$ be the time at which a fraction $f$ of all nodes is informed, i.e., $i(t_f)/n=f$. Acquisition fairness $\alpha(f)$ is defined as the expected value of the ratio between the proportion of informed group $g=1$ nodes and the overall proportion of informed nodes at time $t_f$, under uniformly random seeding:
\begin{equation}\label{eq:acquisition_fairness}
    \alpha(f) = \left\langle \frac{i_1(t_f)/n_1}{i(t_f)/n}\right\rangle_{\mathcal{S}\sim\mathrm{U}[\mathcal{V}]}.
\end{equation}

Here, $\mathcal{S}\sim\mathrm{U}[\mathcal{V}]$ denotes (with slight abuse of notation) that all nodes $v$ in the seed set $\mathcal{S}$ are chosen uniformly at random from $\mathcal{V}$.
Values $\alpha(f) > 1$ indicate an advantage for the group of interest, while $\alpha(f) < 1$ reflects a disadvantage. A value of $\alpha(f) = 1$ corresponds to equal access between groups at fraction $f$.

Beyond a group's ability to receive information, we may also ask about its ability to disseminate information throughout a pairwise or higher-order network. To capture this phenomenon, Zappalà et al.~\cite{zappala2025_GenderDisparitiesDissemination} introduce diffusion fairness $\delta(f)$. This measure compares how quickly information spreads when seeded within different groups. Specifically, it is defined as the ratio between the expected time to reach a fraction $f$ of all nodes when the seed set is chosen uniformly at random from group $g=0$, versus from group $g=1$:
\begin{equation}
    \delta(f)=\frac{\langle t_f\rangle_{\mathcal{S}\sim U[{\mathcal{V}_0}]}}{\langle t_f\rangle_{\mathcal{S}\sim U[{\mathcal{V}_1}]}}~.
\end{equation}

A value $\delta(f)>1$ indicates that information spreads more quickly when seeded in group $g=1$, suggesting a dissemination advantage for that group. Conversely, $\delta(f) < 1$ implies a disadvantage and $\delta(f)=1$ corresponds to equal spreading efficiency.

While Zappalà et al.~\cite{zappala2025_GenderDisparitiesDissemination} focus on acquisition fairness $\alpha$ and diffusion fairness $\delta$ measured at the end of the contagion process, our analysis considers  how $\alpha(f)$ and $\delta(f)$ evolve as functions of the informed fraction $f$. This dynamic perspective allows us to track how fairness emerges and shifts throughout the spreading process.

\FloatBarrier


\section*{Acknowledgments}

This work was motivated by Wang et al.'s paper~\cite{wang2022_InformationAccessEquality} on information access inequality on generative models of complex networks. We thank Giovanni Petri, Iacopo Iacopini, Guillaume St-Onge, and Dakota Murray for their feedback and stimulating discussions.
This work was supported in part by a Volkswagen Foundation grant on ``Reclaiming Individual Autonomy and Democratic Discourse Online: How to Rebalance Human and Algorithmic Decision Making'' and by funds from the Joseph E.~Aoun endowed chair.


\section*{Competing interests}

The authors declare no competing interests.


\section*{Data availability}

The co-sponsorship data for House and Senate~\cite{chodrow2021_GenerativeHypergraphClustering, fowler2006_ConnectingCongress, fowler2006_CosponsorshipHouseSenate} are available at~\url{https://www.cs.cornell.edu/~arb/data/}. Hypergraphs for Primary School~\cite{chodrow2021_GenerativeHypergraphClustering, stehl2011_ContactPrimarySchool}, High School~\cite{benson2018_SimplicialClosurePrediction, mastrandrea2015_ContactHighSchool}, and Hospital~\cite{vanhems2013_Estimating} are constructed from publicly available contact data from~\url{http://www.sociopatterns.org/}. The raw data for the DBLP collaborations~\cite{dblp} is available at~\url{https://dblp.org/}. The APS co-authorship and citation data~\cite{aps_datasets} are available upon request from APS~\url{https://journals.aps.org/datasets}. We describe the hypergraph construction process in Appendix~\ref{appendix:datasets}.


\section*{Code availability}

The code used to construct real-world and synthetic hypergraphs, run information contagion simulations, and evaluate the equality of the simulations is publicly available at~\url{https://doi.org/10.5281/zenodo.17546299} and in the repository~\url{https://github.com/moritz-laber/hypergraph-information-access-equality}  The authors used generative AI to help style portions of the figures, data cleaning, and automate bulk simulation scripts. All code generated or modified by AI was checked for correctness.


\section*{Author contributions}

M.L., S.D., and J.E. began this work as a class project in T.E-R.’s course at Northeastern University on machine learning with graphs (\url{https://eliassi.org/fa23nets.html}). When the decision was made to expand the work for publication, B.K. joined the project. B.K. and T.E-R. supervised the work. M.L., S.D., and J.E. contributed equally to this work, including study conception and design, implementation of computational experiments, and analysis of results. M.L., S.D., and J.E. wrote the first draft of the manuscript. All authors contributed to the revision and review of the manuscript and approved the final version.

\bibliography{bib.bib}

\clearpage


\FloatBarrier
\begin{appendices}
\onecolumn

\setcounter{figure}{0}
\setcounter{table}{0}
\renewcommand{\thefigure}{A\arabic{figure}}
\renewcommand{\thetable}{A\arabic{table}}


\section{Notation}\label{appendix:notation}

We summarize the mathematical notation used throughout the manuscript in Table~\ref{tab:notation}.
\vspace{1cm}
\noindent

\begin{table}[H]
    \centering
    \resizebox{0.99\textwidth}{!}{%
    \begin{tabular}{r|l}
        Symbol & Description \\ \hline
        $H = (\mathcal{V}, \mathcal{E})$                                   & a hypergraph  \\
        $\mathcal{V}$                                                      & the set of nodes \\
        $\mathcal{V}_g = \{v\in \mathcal{V}:g_v=g\}$                       & the set of nodes in group $g$ \\
        $n = |\mathcal{V}|$                                                & the number of nodes \\
        $n_g = |\mathcal{V}_g|$                                            & the number of nodes in group $g$ \\
        $v \in \mathcal{V} $                                               & a node               \\
        $g,g_v \in\{0,1\}$                                                 & group membership (of a node $v$) \\
        $k_v=|\{e \in \mathcal{E} : v\in e\}|$                             & the degree of node $v$, i.e., the number of hyperedges to which node $v$ belongs \\
        $\bar{k}=\frac{1}{n}\sum_{v\in \mathcal{V}}k_v$, $\bar{k}_g=\frac{1}{n_g}\sum_{v\in\mathcal{V}_g}k_v$                                             & the average degree (conditioned on a group), i.e., the average number of hyperedges to which a node belongs\\
        $k'_v=\left |\cup_{e \in \mathcal{E} : v \in e} (e \setminus \{v\} )\right|$                             & the number of unique nodes to which node $v$ is connected \\
        $\bar{k}'=\frac{1}{n}\sum_{v\in \mathcal{V}}k'_v$, $\bar{k}'_g=\frac{1}{n_g}\sum_{v\in\mathcal{V}_g}k'_v$                                             & the average number of nodes to which a node is connected (conditioned on a group) \\
        $\overline{k^2}$, $\overline{k^2}_g$                               & the second moment of the degree distribution (conditioned on a group)\\
        $p_g(k)$                                                           & the distribution of degrees $k$ within group $g$\\
        $\kappa_v$                                                         & the hidden variable of node $v$\\
        $\bar{\kappa}$, $\bar{\kappa}_g$                                   & the average of the hidden variables (conditioned on a group)\\
        $\bar{k}_v$ & the expected degree of node $v$ with hidden variable $\kappa_v$ under the \hypergraph~model \\
        $\rho_g(\kappa)$                                                   & the distribution of hidden variables within group $g$\\
        $\mathcal{E}$                                                      & the set of hyperedges \\
        $e \in \mathcal{E}$                                                & a hyperedge \\
        $s, s_e = |e|$                                                     & the size (of a hyperedge $e$)\\
        $s_\mathrm{max} = \max \{s_e:e\in\mathcal{E}\}$                    & the size of the largest hyperedge \\
        $r, r_e = |\{v\in e: g_v=1\}|$                                     & type (of hyperedge $e$, i.e., the number of group $g=1$ nodes in it)\\
        $m = |\mathcal{E}|$                                                & the number of hyperedges \\
        $m_s$                                                              & the number of hyperedges of size $s$ \\
        $m_{s,r}$                                                          & the number of hyperedges of size $s$ and type $r$ \\
        $a_{s,r}^{(g)}$                                                    & the affinity score of nodes of type $g$ in hyperedges of size $s$ and type $r$ \\
        $b_{s,r}^{(g)}$                                                    & the baseline score of nodes of type $g$ in hyperedges of size $s$ and type $r$ \\
        $h, h_{s,r}^{(g)}$                                                 & the hyperedge homophily score (of nodes of type $g$ in hyperedges of size $s$ and type $r$)\\
        $x_v(t)\in\{0,1\}$                                                  & the state of node $v$ at time $t$\\
        $i(t)$                                                              & the total number of informed nodes at time $t$\\
        $i_g(t)$                                                            & the total number of informed nodes in group $g$ at time $t$\\
        $i_{e,g}(t)$                                                        & the number of informed nodes in group $g$ in hyperedge $e$ at time $t$\\
        $t\in[0,\infty)$                                                    & a point in time \\
        $t^{(g)}_{90}$                                                      & the time at which $90\%$ of nodes are informed\\
        $t_f$                                                               & the time at which a fraction $f$ of nodes are informed\\
        $\Delta t$                                                          & a time increment \\
        $\tau_v$                                                            & the time at which node $v$ receives information \\
        $\mathcal{T}=\{\tau_v:v\in\mathcal{V}\}$                            & the set of times at which nodes are informed  \\
        $z_v$                                                               & the rank of node $v$ in terms of time to be informed \\
        $\mathcal{Z}=\{z_v: v\in \mathcal{V}\}$                             & the set of node ranks in terms of time to be informed \\
        $\mathcal{Z}_g=\{z_v : v \in \mathcal{V}_g\}$                       & the set of node ranks in terms of time to be informed for nodes in group $g$ \\
        $\lambda_{gg'}$                                                     & the spreading rate parameter for transmission from group $g$ to $g'$  \\
        $\lambda_\mathrm{in}$                                               & the spreading rate parameter for transmission to in-group nodes \\
        $\lambda_\mathrm{out}$                                              & the spreading rate parameter for transmission to out-group nodes \\
        $\nu_{gg'}$                                                         & the nonlinearity parameter for transmission from group $g$ to $g'$  \\
        $\nu_\mathrm{in}$                                                   & the nonlinearity parameter for transmission to in-group nodes \\
        $\nu_\mathrm{out}$                                              & the nonlinearity parameter for transmission to out-group nodes \\
        $\beta_g(e)$                                                        & the rate at which nodes of type $g$ receive information in hyperedge $e$ \\
        $\beta=\sum_{g\in\{0,1\}}\sum_{e\in\mathcal{E}}\beta_g(e)$          & the rate at which any transmission event occurs \\
        $\mathcal{S}\subset\mathcal{V}$                                     & the set of seed nodes \\
        $n_\mathrm{seed}=|\mathcal{S}|$                                     & the number of seed nodes\\
        $n_\mathrm{seed}^{(g)}=|\mathcal{S}\cap\mathcal{V}_g|$              & the number of seed nodes in group $g$ \\
        $\mathrm{U}[\mathcal{X}]$                                           & the uniform distribution on a set $\mathcal{X}$\\
        $\mathrm{Poisson}[\bar{x}]$                                         & the Poisson distribution with mean $\bar{x}$ \\
        $\mathrm{Exp}[\mu]$                                                 & the Exponential distribution with rate $\mu$ \\
        $\mathrm{Pareto}[\bar{x},\gamma]$                                   & the Pareto distribution with mean $\bar{x}$ and exponent $\gamma$ of the probability density function\\
        $d_W(\mathcal{X},\mathcal{Y})$                                      & Wasserstein distance between sets $\mathcal{X},\mathcal{Y}$ interpreted as empirical distributions\\
        $\alpha(f)$                                                         & acquisition fairness as a function of the fraction of informed nodes $f$\\
        $\delta(f)$                                                         & diffusion fairness as a function of the fraction of informed nodes $f$\\
        $n_\mathrm{hg}$                                                     & the number of independent simulations\\
        $c$                                                                 & confidence score for the predicted gender produced by a gender-labeling API\\
        $p_v(g)$                                                            & probability that a given node is male $g=0$ or female $g=1$ based on the scores $c$ predicted by a gender-labeling API
    \end{tabular}%
    }
    \caption{Mathematical notation used throughout the manuscript.}
    \label{tab:notation}
\end{table}

\newpage
\FloatBarrier
\section{Synthetic Hypergraph Properties and Parameters}\label{appendix:parameters_synthetic}

Here we describe the parameters used in the experiments on synthetic hypergraphs, including the parameters of the hypergraph model, those of the social contagion model, and those of the seeding strategies. We also provide additional details on the simulations.

\paragraph{Hypergraph Model Parameters}
The inputs to the hypergraph model are the number of nodes $n_g$ in each group, the group-wise degree distributions $\rho_g(\kappa)$, and the number of hyperedges $m_{s,r}$ of size $s\in\{2,\dots,s_\mathrm{max}\}$ and type $r\in\{0,\dots,s\}$, where $s_\mathrm{max}$ is the largest hyperedge size. The type $r$ of a hyperedge is the number of group $g=1$ nodes contained in it.

In the main text, we provide results for group imbalanced hypergraphs with $n_0=7500$ nodes in the majority group ($g=0$) and $n_1=2500$ nodes in the minority group ($g=1$). We focus on degree heterogeneous hypergraphs with degrees sampled from a Pareto distribution $\mathrm{Pareto}[\bar{\kappa}_g,\gamma]$ with exponent $\gamma=2.9$ of the probability density function and means $\bar{\kappa}_g$ set equal to

\begin{equation}\label{eq:appendix:degree_consistency}
    \begin{split}
    \bar{\kappa}_0 &= \frac{1}{n_0}\sum_{s\in\{2,\dots,s_\mathrm{max}\}}\sum_{r\in\{0,\dots,s-1\}}(s-r)m_{s,r},\\
    \bar{\kappa}_1 &= \frac{1}{n_1}\sum_{s\in\{2,\dots,s_\mathrm{max}\}}\sum_{r\in\{1,\dots,s\}}r m_{s,r},\\
    \end{split}
\end{equation}
where the hyperedge counts $m_{s,r}$ are chosen according to a given hyperedge homophily pattern. 

In Appendix~\ref{appendix:additional_results_synthetic}, we present results for group balanced hypergraphs with $n_0=n_1=5000$ and degree homogeneous hypergraphs with $\kappa$ sampled from a Poisson distribution, $\rho_g(\kappa)=\mathrm{Poisson}[\bar{\kappa}_g]$ with mean $\bar{\kappa}_g$ determined by Equation~\eqref{eq:appendix:degree_consistency}.

We also create mixed hyperedge homophily patterns by selecting different combinations of the basic hyperedge homophily patterns for each hyperedge size $s$. We restrict ourselves to fixing one pattern for pairwise interactions ($s=2$), and another pattern for all higher-order interactions ($s>2$). For example, the homophily-neutral mixed pattern uses the homophilous basic pattern for the pairwise interactions and the neutral pattern for the higher-order edges. The resulting hyperedge counts are listed in Table~\ref{tab:homophily_patterns}.

\begin{figure*}[tb]
\centering
\includegraphics[width=0.9\textwidth]{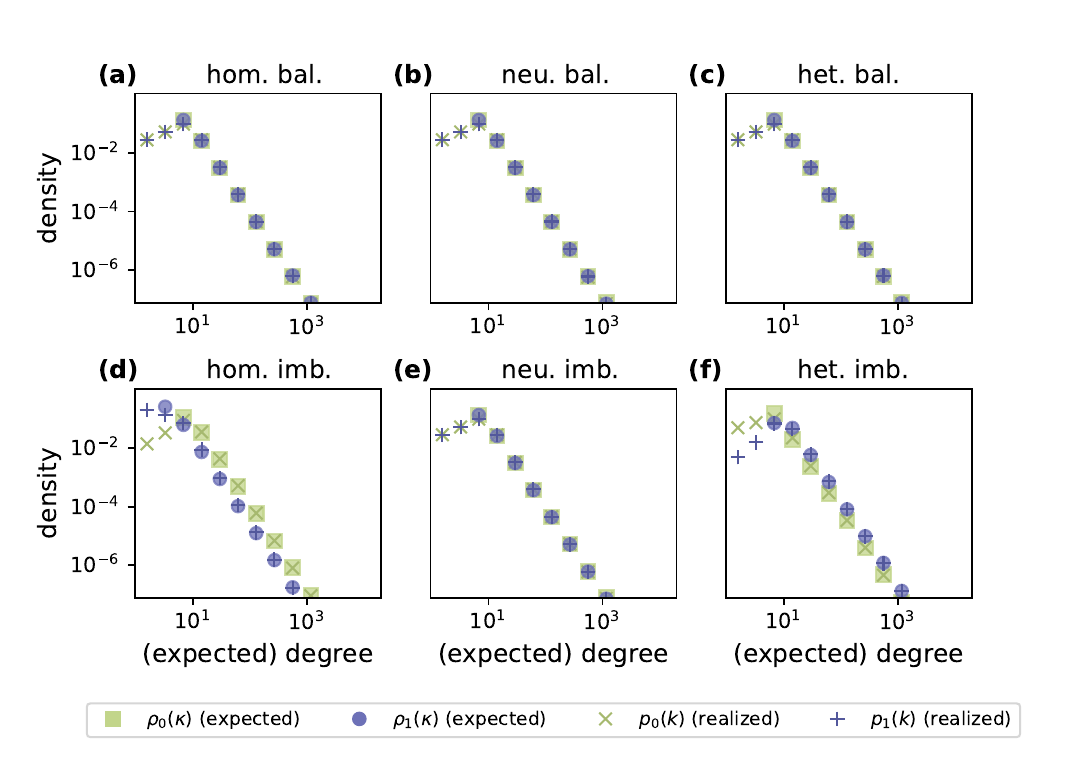}
\caption{\textbf{Realized and expected degree distributions in synthetic hypergraphs.} The distribution $\rho_g(\kappa)$ of expected degrees $\kappa$ of the majority group $g=0$ (green squares) and minority group $g=1$ (purple circles), as well as the distributions $p_{g}(k)$ of realized degrees $k$ of the majority group $g=0$ (green crosses) and minority group $g=1$ (purple pluses). Top row: group balanced hypergraphs ($n_0=n_1=5000$) with \textbf{(a)} homophilous, \textbf{(b)} neutral, and \textbf{(c)} heterophilous hyperedge homophily patterns in all hyperedge sizes $s\in\{2,3,4\}$. Bottom row: group imbalanced hypergraphs ($n_0=7500$ and $n_1=2500$) with \textbf{(d)} homophilous, \textbf{(e)} neutral, and \textbf{(f)} heterophilous hyperedge homophily patterns in all hyperedge sizes $s\in\{2,3,4\}$. All distributions are based on $n_\mathrm{hg}=10^3$ hypergraphs. We observe strong agreement between expected and realized degree distributions.}
\label{fig:degree_distribution}
\end{figure*}

To verify that our hypergraph sampling procedure faithfully generates the heavy‑tailed degree sequence that we imposed as an input, we compare the expected degrees $\kappa$ drawn from a group dependent Pareto distribution $\rho_g(\kappa)$ for $g\in\{0,1\}$ with the group-dependent distribution $p_g(k)$ of degrees $k$ that are realized in generated hypergraphs. As in the main text, we set the exponent of the probability density function of the Pareto distribution to $\gamma=2.9$ and choose the group-dependent mean $\bar{\kappa}_g$ according to the prescribed hyperedge homophily pattern using Equation~\eqref{eq:appendix:degree_consistency}.  We consider both group-balanced ($n_0=n_1=5000$) and group-imbalanced ($n_0=7500$, $n_1=2500$) hypergraphs with homophilous, neutral, or heterophilous connectivity patterns for hyperedges of all sizes $s\in\{2,3,4\}$.
The former isolates the effects of homophily patterns by removing the influence of group imbalance, providing a clean baseline. The latter captures the interaction between homophily patterns and group imbalance, yielding a more realistic scenario that highlights the qualitative differences relative to the balanced case. 
Figure~\ref{fig:degree_distribution} shows excellent agreement between the distribution of expected degrees and realized degrees on group-balanced homophilous (Figure~\ref{fig:degree_distribution} (a)), neutral (Figure~\ref{fig:degree_distribution} (b)), and heterophilous (Figure~\ref{fig:degree_distribution} (c)) hypergraphs, as well as group-imbalanced homophilous (Figure~\ref{fig:degree_distribution} (d)),
neutral (Figure~\ref{fig:degree_distribution} (e)), and heterophilous (Figure~\ref{fig:degree_distribution} (f)) hypergraphs.

\begin{table*}[tb]
    \centering
    \resizebox{\textwidth}{!}{%
    \begin{tabular}{c|c|c|c|c|c|c|c|c|c|c|c|c|c|c}
    \textbf{Name}      & $n_0$ & $n_1$ & $m_{2,0}$ & $m_{2,1}$ & $m_{2,2}$ & $m_{3,0}$ & $m_{3,1}$ & $m_{3,2}$ & $m_{3,3}$ & $m_{4,0}$ & $m_{4,1}$ & $m_{4,2}$ & $m_{4,3}$ & $m_{4,4}$   \\ \hline \hline
    neu.      & 5000  & 5000  & 6250     & 12500    & 6250     & 1500     & 4500     & 4500     & 1500     & 375      & 1500     & 2250     & 1500     & 375 \\
    hom.      & 5000  & 5000  & 8334     & 8332     & 8334     & 2122     & 3878     & 3878     & 2122     & 546      & 1403     & 2102     & 1403     & 546 \\
    het.      & 5000  & 5000  & 3575     & 17850    & 3575     & 800      & 5200     & 5200     & 800      & 195      & 1604     & 2404     & 1604     & 195 \\
    hom.-neu. & 5000  & 5000  & 8334     & 8332     & 8334     & 1500     & 4500     & 4500     & 1500     & 375      & 1500     & 2250     & 1500     & 375 \\
    neu.-hom. & 5000  & 5000  & 6250     & 12500    & 6250     & 2122     & 3878     & 3878     & 2122     & 546      & 1403     & 2102     & 1403     & 546 \\
    neu.-het  & 5000  & 5000  & 6250     & 12500    & 6250     & 800      & 5200     & 5200     & 800      & 195      & 1604     & 2404     & 1604     & 195 \\
    het.-neu. & 5000  & 5000  & 3575     & 17850    & 3575     & 1500     & 4500     & 4500     & 1500     & 375      & 1500     & 2250     & 1500     & 375 \\
    neu.      & 7500  & 2500  & 14062    & 9375     & 1563     & 5062     & 5063     & 1687     & 188      & 1898     & 2533     & 1266     & 280      & 23   \\
    hom.      & 7500  & 2500  & 21919    & 2421     & 660      & 7626     & 3144     & 1052     & 178      & 2850     & 1938     & 970      & 215      & 27   \\
    het.      & 7500  & 2500  & 7200     & 16520    & 1280     & 2570     & 6976     & 2326     & 128      & 951      & 3126     & 1561     & 348      & 14 \\
    hom.-neu. & 7500  & 2500  & 21919    & 2421     & 660      & 5062     & 5063     & 1687     & 188      & 1898     & 2533     & 1266     & 280      & 23   \\
    neu.-hom. & 7500  & 2500  & 14062    & 9375     & 1563     & 7626     & 3144     & 1052     & 178      & 2850     & 1938     & 970      & 215      & 27   \\
    neu.-het  & 7500  & 2500  & 14062    & 9375     & 1563     & 2570     & 6976     & 2326     & 128      & 951      & 3126     & 1561     & 348      & 14 \\
    het.-neu  & 7500  & 2500  & 7200     & 16520    & 1280     & 5062     & 5063     & 1687     & 188      & 1898     & 2533     & 1266     & 280      & 23 
    \end{tabular}
    }
    \caption{\textbf{The number of hyperedges $m_{s,r}$ of size $s$ and type $r$ used to define hyperedge homophily patterns.} We report hyperedge counts for balanced ($n_0 = n_1 = 5000$) and imbalanced ($n_0 = 7500,~n_1 = 2500$) hypergraphs. Basic hyperedge homophily patterns include neutral (neu.), homophilous (hom.), and heterophilous (het.). Mixed hyperedge homophily patterns are indicated by a dash, e.g., neu.-het. denotes a neutral pattern for $s=2$ hyperedges and a heterophilous pattern for hyperedges with $s>2$.}
    \label{tab:homophily_patterns}
\end{table*}

\paragraph{Social Contagion Model Parameters}
We consider four settings for the contagion dynamics: symmetric linear, symmetric superlinear, symmetric sublinear, and asymmetric nonlinear. These are defined using the parsimonious parameterization introduced in Section~\ref{sec:methods:contagion}, which involves in-group and out-group transmission rates $\lambda_\mathrm{in}, \lambda_\mathrm{out}$ and nonlinearity parameters $\nu_\mathrm{in}, \nu_\mathrm{out}$. A contagion process is considered symmetric when $\lambda_\mathrm{in}=\lambda_\mathrm{out}$ and $\nu_\mathrm{in}=\nu_\mathrm{out}$. We classify the symmetric dynamics as linear, superlinear, or sublinear depending on whether $\nu_\mathrm{in}$ and $\nu_\mathrm{out}$ are equal to, greater than, or less than $1$, respectively. In contrast, we define the asymmetric case by $\lambda_\mathrm{out}<\lambda_\mathrm{in}$ along with $\nu_\mathrm{out}<1$ and $\nu_\mathrm{in}>1$. We refer to this case as asymmetric nonlinear. Exact parameter values are listed in Table~\ref{tab:dynamics_parameters_synthetic}.

\begin{table*}[tb]
    \centering
    \begin{tabular}{l|c|c|c|c}
        \textbf{Dynamics}     & $\lambda_\mathrm{in}$ & $\lambda_\mathrm{out}$ & $\nu_\mathrm{in}$ & $\nu_\mathrm{out}$ \\ \hline \hline
        symmetric linear       &  0.01 & 0.01  & 1.0 & 1.0 \\
        symmetric sublinear    &  0.01 & 0.01  & 0.5 & 0.5 \\
        symmetric superlinear  &  0.01 & 0.01  & 2.0 & 2.0 \\ 
        asymmetric nonlinear   &  0.02 & 0.005 & 2.0 & 0.5 \\
    \end{tabular}
    \caption{\textbf{Social contagion parameters for synthetic hypergraphs.} In- and out-group transmission rates are denoted by $\lambda_\mathrm{in}$ and $\lambda_\mathrm{out}$, respectively, while nonlinearity parameters are denoted by $\nu_\mathrm{in}$ and $\nu_\mathrm{out}$.}
    \label{tab:dynamics_parameters_synthetic}
\end{table*}

\paragraph{Seeding Strategy}
We specify initial conditions for the social contagion process by selecting a set of nodes $\mathcal{S}$. We refer to these nodes as seed nodes or seeds. We denote the size of this set as $n_\mathrm{seed} = |\mathcal{S}|$, and the number of seed nodes belonging to group $g$ as $n_\mathrm{seed}^{(g)}$. We use $n_\mathrm{seed}=4$ and apply a random proportional seeding strategy, where seeds are chosen uniformly at random from each group in proportion to group size: $n^{(g)}_\mathrm{seed}/n_\mathrm{seed}=n_g/n$. In the group-imbalanced case, this yields $n^{(0)}_\mathrm{seed}=3$ and $n^{(1)}_\mathrm{seed}=1$, while both groups receive two seeds in the balanced case.

The only exception arises in the computation of diffusion fairness $\delta(f)$, which compares the time $t_f$ required to inform a fraction $f$ of nodes when seeding occurs exclusively in one group. For this analysis, we maintain the total seed count as $n_\mathrm{seed}=4$ and select all seeds from one group, either $n_\mathrm{seed}^{(0)}=4, n_\mathrm{seed}^{(1)}=0$ or $n_\mathrm{seed}^{(0)}=0, n_\mathrm{seed}^{(1)}=4$.

\paragraph{Simulation}
Our results are based on $n_\mathrm{hg}=10^3$ independent realizations of the stochastic information spreading process for each hyperedge homophily pattern and set of contagion parameters. Each simulation involves three steps: sampling a hypergraph from the generative model with a specified group size and hyperedge homophily pattern, selecting seed nodes according to the seeding strategy, and simulating the contagion process on the hypergraph.


\section{Supplementary Results on Synthetic Hypergraphs}\label{appendix:additional_results_synthetic}

We quantify the extent of information access inequality in imbalanced, degree-heterogeneous hypergraphs in Sections~\ref{sec:results:synthetic_emd},~\ref{sec:results:acquisition_and_diffusion}, and~\ref{sec:results:real_mixed_homophily} of the main text. Here, we provide supplementary results for hypergraphs that are either group-balanced or degree-homogeneous. We consistently find that the effect of group balance on inequality is much stronger than that of degree distribution.

\begin{figure*}[p]
\centering
 \includegraphics[width=0.9\textwidth]{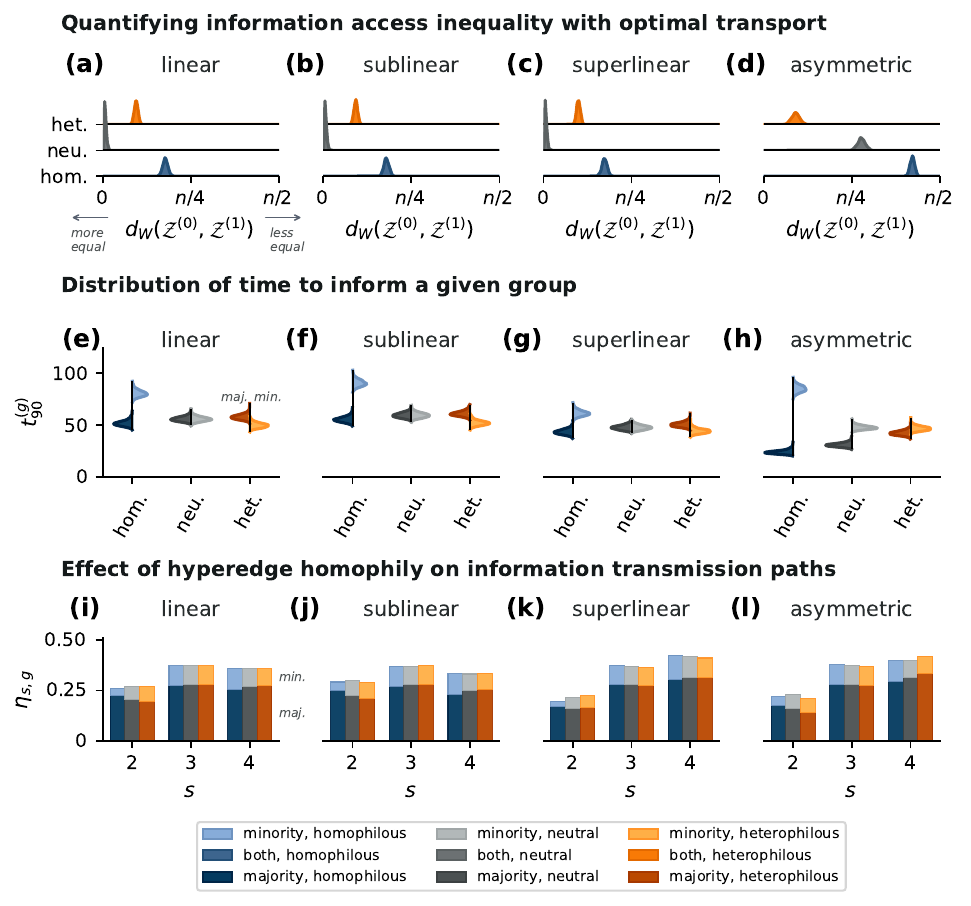}
\caption{\textbf{Quantifying information access inequality in imbalanced, degree-homogeneous hypergraphs.} Top row: $d_W(\mathcal{Z}_0, \mathcal{Z}_1)$ under (a) linear, (b) sublinear, (c) superlinear, and (d) asymmetric contagion. Middle row: $t^{(g)}_{90}$ time to inform $90$\% of a group $g$ under (e) linear, (f) sublinear, (g) superlinear, and (h) asymmetric contagion. Bottom row: fraction of transmission events by hyperedge size and group under (i) linear, (j) sublinear, (k) superlinear, and (l) asymmetric contagion. Colors indicate hyperedge homophily: orange (heterophilous), gray (neutral), and blue (homophilous). All results are averaged over $n_\mathrm{hg}=10^3$ simulations per setting.}
\label{fig:emd_synthetic_imbalanced_poisson}
\end{figure*}

\begin{figure*}[p]
\centering
 \includegraphics[width=0.9\textwidth]{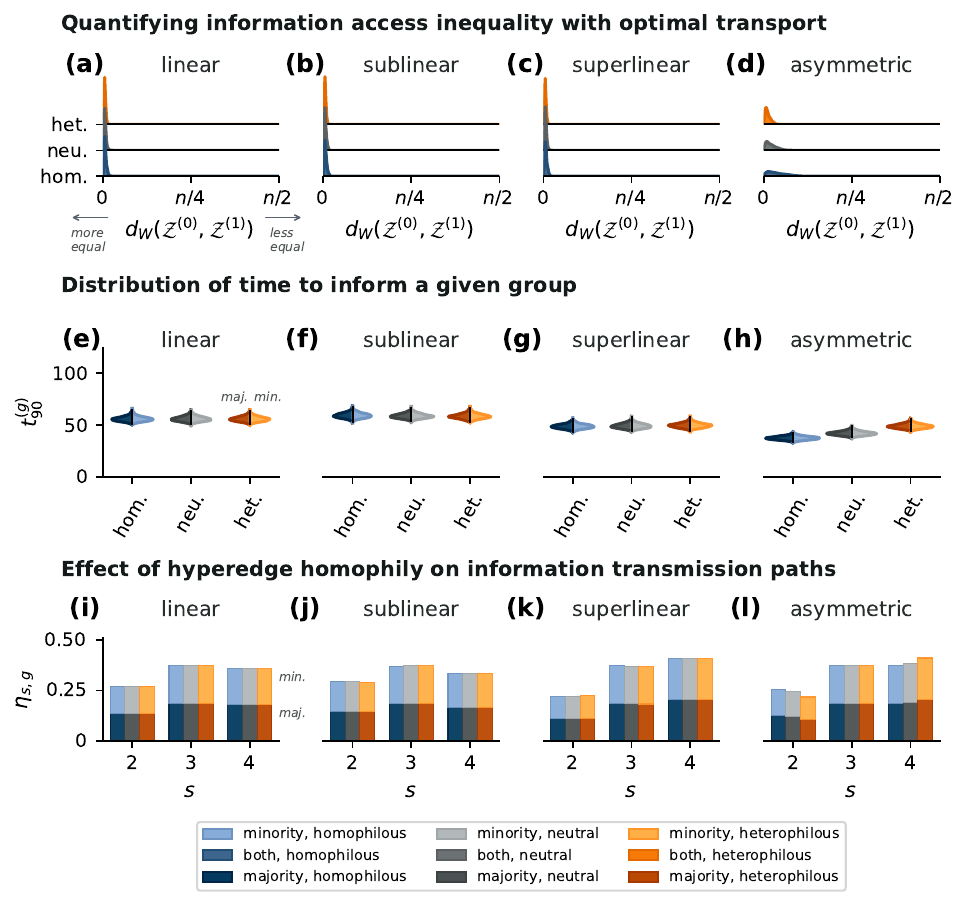}
\caption{\textbf{Quantifying information access inequality in balanced, degree-homogeneous hypergraphs.} Top row: $d_W(\mathcal{Z}_0, \mathcal{Z}_1)$ under (a) linear, (b) sublinear, (c) superlinear, and (d) asymmetric contagion. Middle row: $t^{(g)}_{90}$ time to inform $90$\% of a given group $g$ under (e) linear, (f) sublinear, (g) superlinear, and (h) asymmetric contagion. Bottom row: fraction of transmission events by hyperedge size and group under (i) linear, (j) sublinear, (k) superlinear, and (l) asymmetric contagion. Colors indicate hyperedge homophily: orange (heterophilous), gray (neutral), and blue (homophilous). All results are averaged over $n_\mathrm{hg}=10^3$ simulations per setting.}
\label{fig:emd_synthetic_balanced_poisson}
\end{figure*}

\begin{figure*}[p]
\centering
 \includegraphics[width=0.9\textwidth]{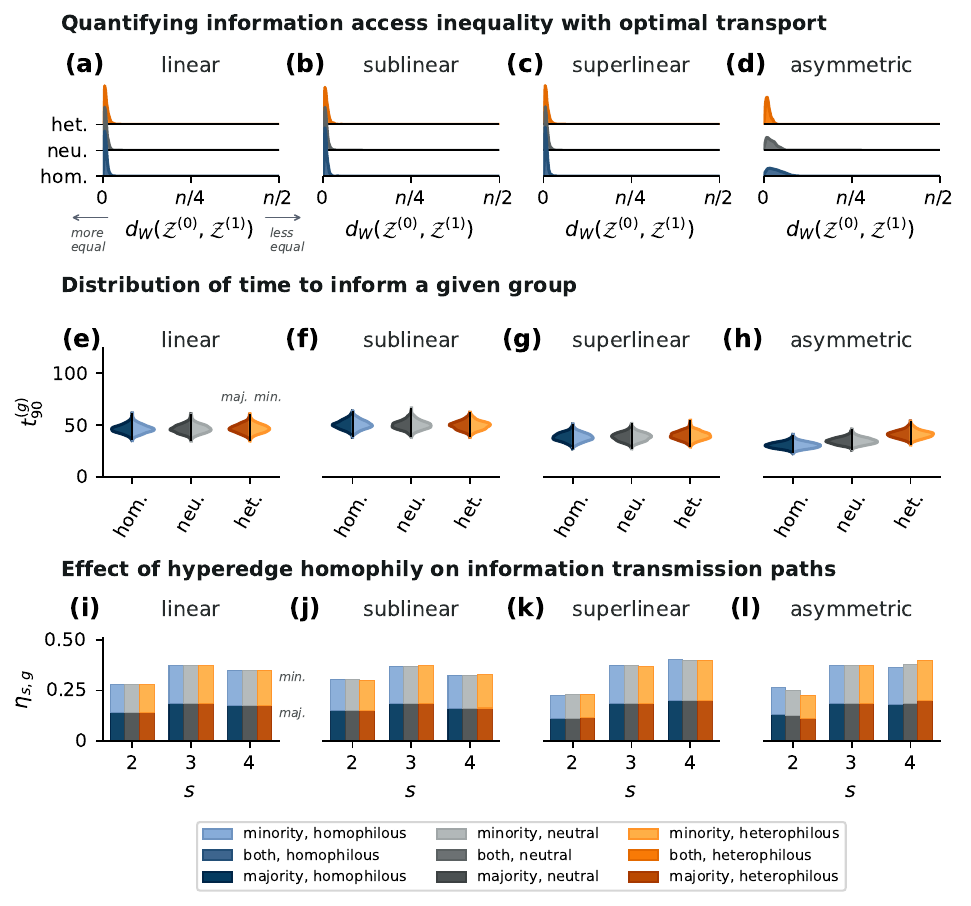}
\caption{\textbf{Quantifying information access inequality in balanced, degree-heterogeneous hypergraphs.} Top row: $d_W(\mathcal{Z}_0, \mathcal{Z}_1)$ under (a) linear, (b) sublinear, (c) superlinear, and (d) asymmetric contagion. Middle row: $t^{(g)}_{90}$ time to inform $90$\% of a group $g$ under (e) linear, (f) sublinear, (g) superlinear, and (h) asymmetric contagion. Bottom row: fraction of transmission events by hyperedge size and group under (i) linear, (j) sublinear, (k) superlinear, and (l) asymmetric contagion. Colors indicate hyperedge homophily: orange (heterophilous), gray (neutral), and blue (homophilous). All results are averaged over $n_\mathrm{hg}=10^3$ simulations per setting.}
\label{fig:emd_synthetic_balanced_powerlaw}
\end{figure*}

\begin{figure*}[p]
\centering
\includegraphics[width=0.9\textwidth]{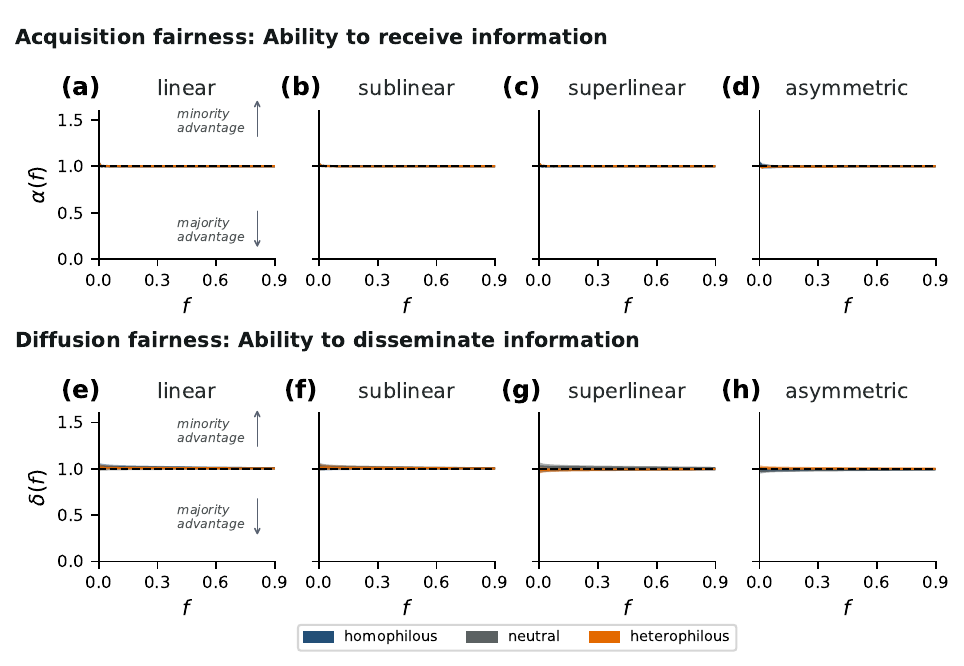}
\caption{\textbf{Acquisition and diffusion fairness in balanced, degree-heterogeneous hypergraphs.} Top row: acquisition fairness $\alpha(f)$ under (a) linear, (b) sublinear, (c) superlinear, and (d) asymmetric contagion. Bottom row: diffusion fairness $\delta(f)$ under (e) linear, (f) sublinear, (g) superlinear, and (h) asymmetric contagion. Results are averaged over $n_\mathrm{hg}=10^3$ simulations. Curves for homophilous, heterophilous, and neutral hypergraphs are shown in blue, orange, and gray, respectively. The dashed black line indicates equality, while $\alpha(f),\delta(f)>1$ indicate a minority advantage. We estimate $99\%$ confidence intervals from $100$ bootstrap samples.}\label{fig:acquisition_and_diffusion_balanced_powerlaw}
\end{figure*}

For hypergraphs with basic hyperedge homophily patterns, imbalanced but degree-homogeneous hypergraphs exhibit inequality patterns qualitatively similar to those in the main text (Fig.~\ref{fig:emd_synthetic_imbalanced_poisson}). In contrast, group-balanced hypergraphs show minimal inequality regardless of whether degree distributions are homogeneous (Fig.~\ref{fig:emd_synthetic_balanced_poisson}) or heterogeneous (Fig.~\ref{fig:emd_synthetic_balanced_powerlaw}). In both cases, the group-wise rank distributions are nearly indistinguishable (Fig.~\ref{fig:emd_synthetic_balanced_poisson}(a)-(d), Fig.~\ref{fig:emd_synthetic_balanced_powerlaw}(a)-(d)), and the time $t^{(g)}_{90}$ to reach $90\%$ of the nodes in each group is nearly identical (Fig.~\ref{fig:emd_synthetic_balanced_poisson}(e)-(h), Fig.~\ref{fig:emd_synthetic_balanced_powerlaw}(e)-(h)). The main exception is under asymmetric contagion. Here, we see greater variability in the Wasserstein distance $d_W(\mathcal{Z}_0, \mathcal{Z}_1)$ and inequality for homophilous hypergraphs, likely due to in-group transmission bias (Fig.~\ref{fig:emd_synthetic_balanced_poisson}(d), Fig.~\ref{fig:acquisition_and_diffusion_balanced_powerlaw}(d)).

%

%

\begin{figure*}[p]
\centering
\includegraphics[width=0.9\textwidth]{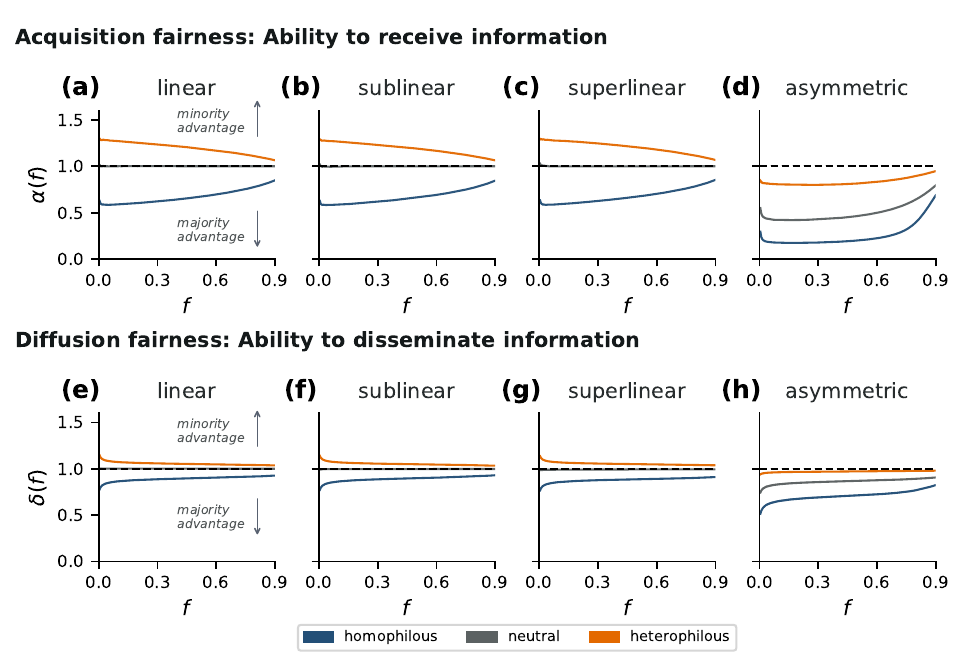}
\caption{\textbf{Acquisition and diffusion fairness in imbalanced, degree-homogeneous hypergraphs.} Top row: acquisition fairness $\alpha(f)$ under (a) linear, (b) sublinear, (c) superlinear, and (d) asymmetric contagion. Bottom row: diffusion fairness $\delta(f)$ under (e) linear, (f) sublinear, (g) superlinear, and (h) asymmetric contagion. Results are averaged over $n_\mathrm{hg}=10^3$ simulations. Curves for homophilous, heterophilous, and neutral hypergraphs are shown in blue, orange, and gray, respectively. The dashed black line indicates equality, while $\alpha(f),\delta(f)>1$ indicate a minority advantage. We estimate $99\%$ confidence intervals from $100$ bootstrap samples.}\label{fig:acquisition_and_diffusion_imbalanced_poisson}
\end{figure*}

\begin{figure*}[p]
\centering
\includegraphics[width=0.9\textwidth]{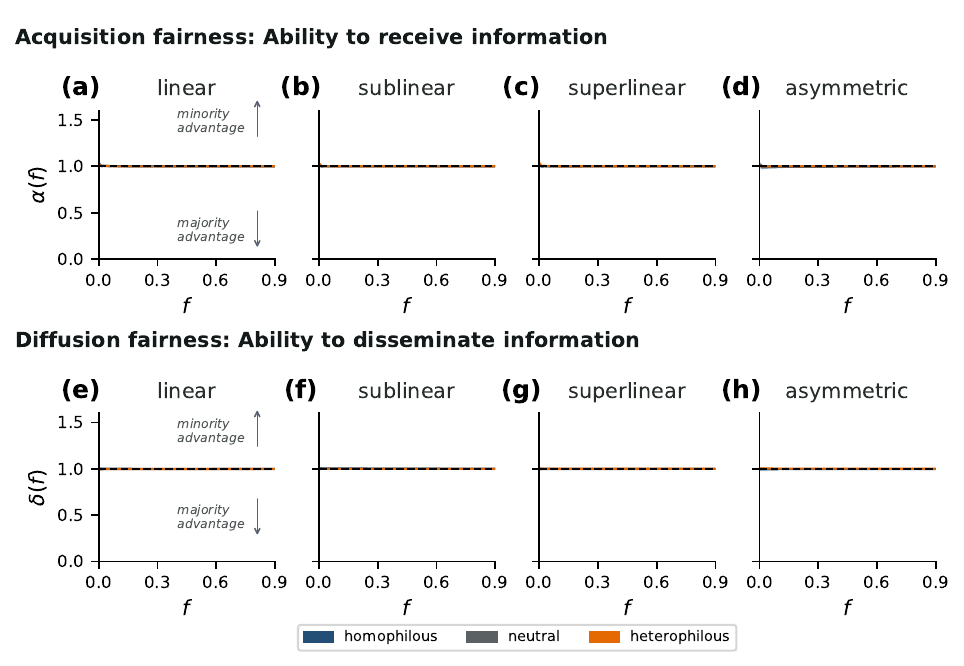}
\caption{\textbf{Acquisition and diffusion fairness in balanced, degree-homogeneous hypergraphs.} Top row: acquisition fairness $\alpha(f)$ under (a) linear, (b) sublinear, (c) superlinear, and (d) asymmetric contagion. Bottom row: diffusion fairness $\delta(f)$ under (e) linear, (f) sublinear, (g) superlinear, and (h) asymmetric contagion. Results are averaged over $n_\mathrm{hg}=10^3$ simulations. Curves for homophilous, heterophilous, and neutral hypergraphs are shown in blue, orange, and gray, respectively. The dashed black line indicates equality, while $\alpha(f),\delta(f)>1$ indicate a minority advantage. We estimate $99\%$ confidence intervals from $100$ bootstrap samples.}\label{fig:acquisition_and_diffusion_balanced_poisson}
\end{figure*}

These trends extend to our fairness metrics. For imbalanced, degree-homogeneous hypergraphs, both acquisition fairness and diffusion fairness closely mirror the patterns observed in imbalanced, degree-heterogeneous hypergraphs (Fig.~\ref{fig:acquisition_and_diffusion_imbalanced_poisson}). In contrast, group-balanced hypergraphs show little-to-no inequality across either measure, regardless of degree distribution. In both the degree-homogeneous case (Fig.~\ref{fig:acquisition_and_diffusion_balanced_poisson}) and the degree-heterogeneous case (Fig.~\ref{fig:acquisition_and_diffusion_balanced_powerlaw}),  both $\alpha(f)$ and $\delta(f)$ remain close to one throughout the entire contagion process.

%

%

\begin{figure*}[p]
         \centering
         \includegraphics[width=0.9\textwidth]{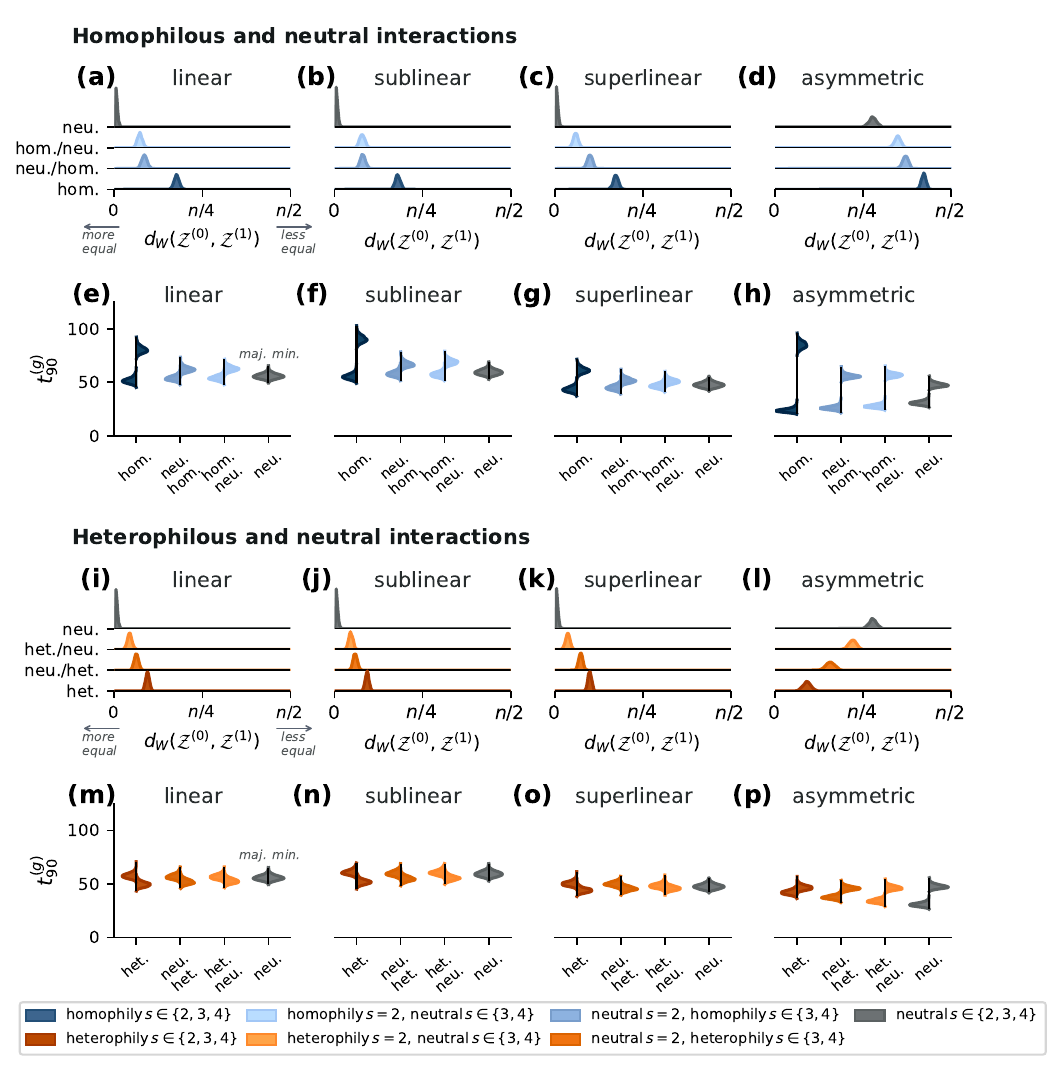}
\caption{\textbf{Information access inequality on imbalanced, degree-homogeneous hypergraphs with mixed hyperedge homophily.} Rows 1 and 3: distributions of Wasserstein distances $d_W(\mathcal{Z}_0,\mathcal{Z}_1)$ for mixed homophilous-neutral hypergraphs under (a) linear, (b) sublinear, (c) superlinear, and (d) asymmetric contagion, and for mixed heterophilous-neutral hypergraphs under (i) linear, (j) sublinear, (k) superlinear, and (l) asymmetric contagion. Rows 2 and 4: violin plots of $t^{(g)}_{90}$, the time to inform $90$\% of nodes in the majority $g=0$ (darker, left) and minority $g=1$ (lighter, right) groups, under (e)-(h) and (m)-(p) for the same settings. Mixed homophilous patterns are shown in shades of blue, while mixed heterophilous patterns are orange. All results are averaged over $n_\mathrm{hg}=10^3$ simulations on hypergraphs with consistent structure.}\label{fig:emd_mixed_imbalanced_poisson}
\end{figure*}

\begin{figure*}[p]
         \centering
         \includegraphics[width=0.9\textwidth]{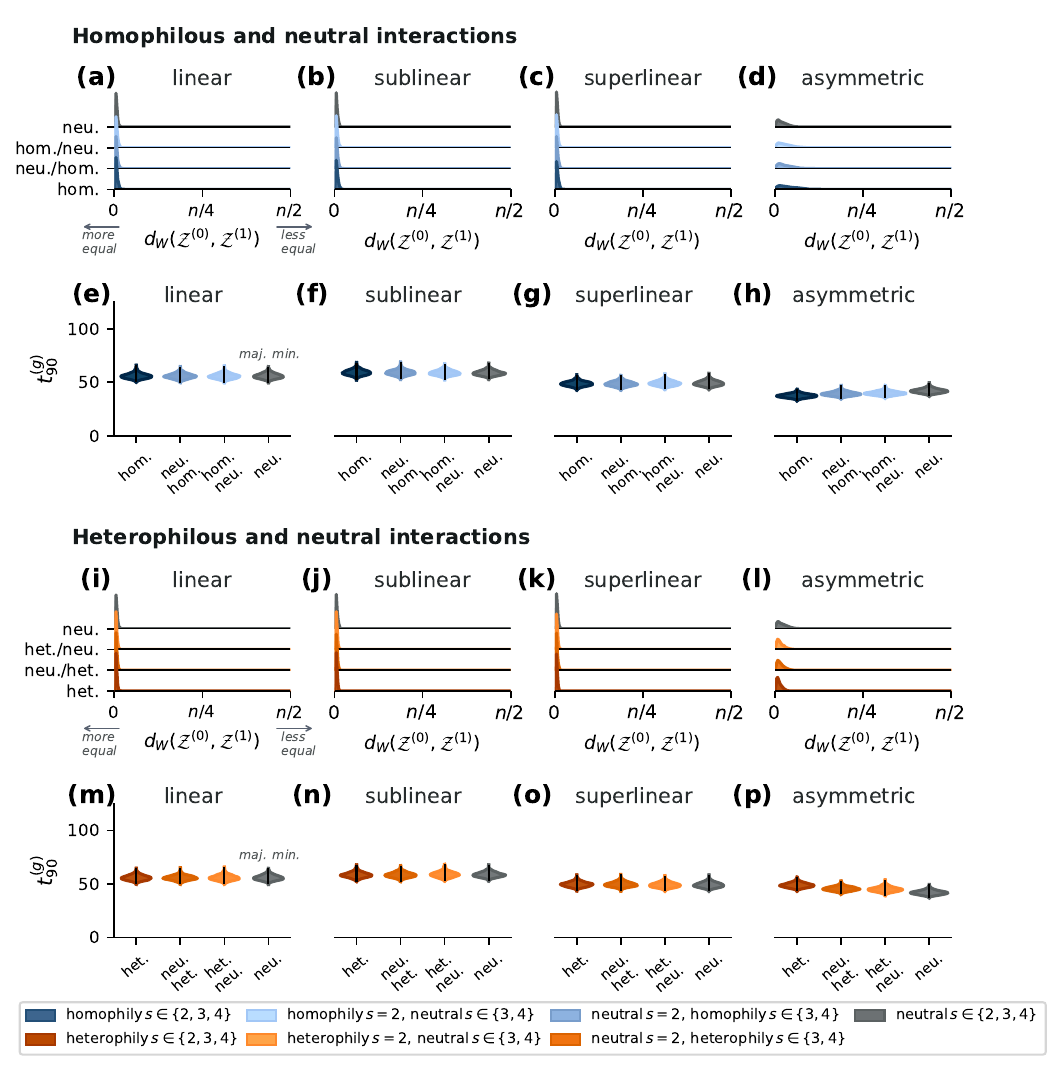}
\caption{\textbf{Information access inequality on balanced, degree-homogeneous hypergraphs with mixed hyperedge homophily.} Rows 1 and 3: distributions of Wasserstein distances $d_W(\mathcal{Z}_0,\mathcal{Z}_1)$ for mixed homophilous-neutral hypergraphs under (a) linear, (b) sublinear, (c) superlinear, and (d) asymmetric contagion, and for mixed heterophilous-neutral hypergraphs under (i) linear, (j) sublinear, (k) superlinear, and (l) asymmetric contagion. Rows 2 and 4: violin plots of $t^{(g)}_{90}$, the time to inform $90$\% of nodes in the majority $g=0$ (darker, left) and minority $g=1$ (lighter, right) groups, under (e)-(h) and (m)-(p) for the same settings. Mixed homophilous patterns are shown in shades of blue, while mixed heterophilous patterns are orange. All results are averaged over $n_\mathrm{hg}=10^3$ simulations on hypergraphs with consistent structure.}\label{fig:emd_mixed_balanced_poisson}
\end{figure*}

\begin{figure*}[p]
         \centering
         \includegraphics[width=0.9\textwidth]{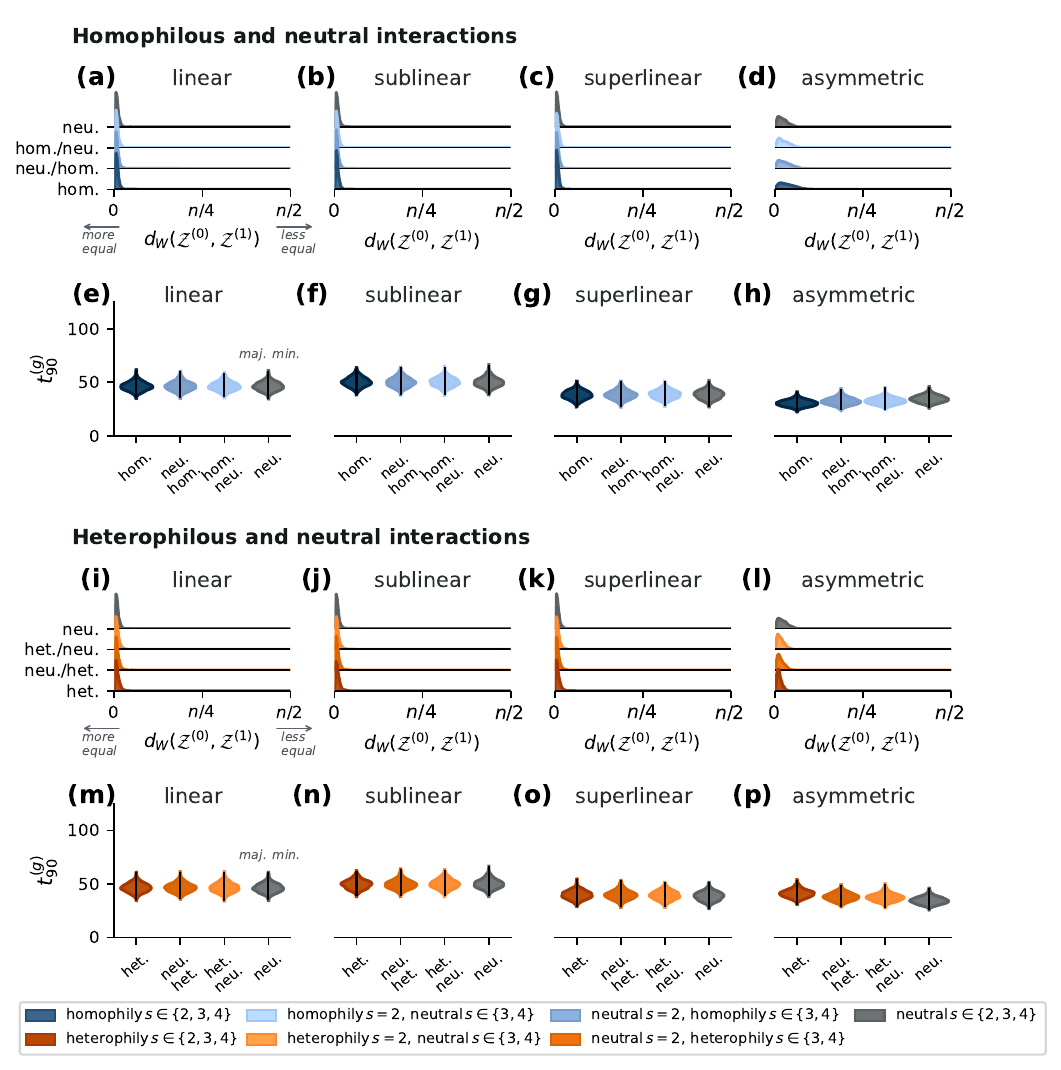}
\caption{\textbf{Information access inequality on balanced, degree-heterogeneous hypergraphs with mixed hyperedge homophily.} Rows 1 and 3: distributions of Wasserstein distances $d_W(\mathcal{Z}_0,\mathcal{Z}_1)$ for mixed homophilous-neutral hypergraphs under (a) linear, (b) sublinear, (c) superlinear, and (d) asymmetric contagion, and for mixed heterophilous-neutral hypergraphs under (i) linear, (j) sublinear, (k) superlinear, and (l) asymmetric contagion. Rows 2 and 4: violin plots of $t^{(g)}_{90}$, the time to inform $90$\% of nodes in the majority $g=0$(darker, left) and minority $g=1$ (lighter, right) groups, under (e)-(h) and (m)-(p) for the same settings. Mixed homophilous patterns are shown in shades of blue, while mixed heterophilous patterns are orange. All results are averaged over $n_\mathrm{hg}=10^3$ simulations on hypergraphs with consistent structure.}\label{fig:emd_mixed_balanced_powerlaw}
\end{figure*}

The same pattern holds for mixed hyperedge homophily hypergraphs. Imbalanced, degree-homogeneous hypergraphs exhibit qualitatively similar outcomes to those in Section~\ref{sec:results:real_mixed_homophily} (Fig.~\ref{fig:emd_mixed_imbalanced_poisson}), while group-balanced hypergraphs again show minimal inequality, both in the degree-homogeneous case (Fig.~\ref{fig:emd_mixed_balanced_poisson}) and the degree-heterogeneous case (Fig.~\ref{fig:emd_mixed_balanced_powerlaw}). The only significant deviations arise under asymmetric, nonlinear contagion, which produces small but detectable differences in the homophilous setting (Fig.~\ref{fig:emd_mixed_balanced_poisson}(d), Fig.~\ref{fig:emd_mixed_balanced_powerlaw}(d)).

%

%

%


\FloatBarrier
\section{Real-World Hypergraph Construction}\label{appendix:real-world_hypergraph_construction}

In this section, we describe how we construct real-world hypergraphs from empirical data and define the group labels used for analysis.

The Primary School~\cite{chodrow2021_GenerativeHypergraphClustering, stehl2011_ContactPrimarySchool}, High School~\cite{benson2018_SimplicialClosurePrediction, mastrandrea2015_ContactHighSchool}, and Hospital~\cite{vanhems2013_Estimating} hypergraphs are based on physical proximity data collected via wearable sensors. These datasets are available at \url{http://www.sociopatterns.org/datasets/}. Nodes represent individuals: students in the Primary and High School datasets, and either healthcare workers or patients in the Hospital dataset. Hyperedges encode sets of individuals who were within $1.5$ meters of each other for at least $20$ seconds during a single day. While the edge data for Primary and High School is also available at \url{https://www.cs.cornell.edu/~arb/data/}, we use the Sociopatterns source to access metadata, including node labels. Following the procedure described at \url{https://www.cs.cornell.edu/~arb/data/}, we construct simplicial complexes from the raw proximity data and retain only the maximal simplices to form hyperedges. We validate our construction by matching the reported max simplex counts to our hyperedge counts. Node labels, gender for school datasets and role (patient or healthcare worker) for the Hospital dataset, are provided in the metadata and original publications.

The Senate and House datasets~\cite{chodrow2021_GenerativeHypergraphClustering, fowler2006_ConnectingCongress, fowler2006_CosponsorshipHouseSenate} are available at \url{https://www.cs.cornell.edu/~arb/data/}. In both datasets, nodes represent members of the United States Senate or House of Representatives, and hyperedges capture sets of legislators who co-sponsored the same bill. We construct two versions of each hypergraph: one labeled by party affiliation (available in the data) and one labeled by gender (inferred using the method described in Appendix~\ref{appendix:gender_labeling}). The datasets are provided as edge lists representing maximal simplices of a simplicial complex. Unlike hypergraphs, simplicial complexes have a downward closure property (e.g., if $\{v_1, v_2, v_3\}$ is listed, all lower-order interactions among these nodes are assumed to exist). Because our analysis is based on hypergraphs, we treat each maximal simplex as a single hyperedge and do not enforce downward closure.

The final two datasets, DBLP~\cite{dblp} and APS~\cite{aps_datasets}, represent scientific co-authorship hypergraphs in computer science and physics, respectively. In both cases, nodes correspond to researchers and hyperedges to sets of co-authors on individual publications. DBLP includes unique author identifiers, whereas APS relies on author names, requiring the disambiguation procedure described in Appendix~\ref{appendix:author_disambiguation}. Following previous work which has identified gender homophily in scientific collaboration~\cite{jaramillo2023structure}, we use gender labels for both datasets. The gender labels are inferred using the approach detailed in Appendix~\ref{appendix:gender_labeling}. The DBLP data is available at \url{https://dblp.org/}, and APS data can be requested from \url{https://journals.aps.org/datasets}.

In all datasets, our analysis and simulations are performed on the largest connected component.


\FloatBarrier
\section{Author Disambiguation}\label{appendix:author_disambiguation}

The APS dataset does not provide unique author identifiers, so we perform name disambiguation to reconstruct individual publication histories. Our procedure builds on a pipeline initially developed by Huang et al.~\cite{huang2020_HistoricalComparisonGender} and later refined by Bachmann et al.~\cite{bachmann2024_CumulativeAdvantageBrokerage}. 

Following Huang et al., we assign the same author identity to names sharing a last name and first initial, provided they also satisfy at least one of three conditions: shared co-authors, mutual citations, or common affiliations. This rule-based method tends to over-merge distinct individuals with similar names, so we adopt the refinement introduced by Bachmann et al., which splits author identities when the first names are not all unique prefixes of one another. While this post-processing step reduces erroneous merges, it may increase the complementary error of failing to link different name variants that correspond to the same person.

In contrast to some earlier studies, we do not filter out publications with large numbers of co-authors as we view them as meaningful and legitimate hyperedges in our framework. Large-scale collaborations can serve as key conduits for spreading ideas, methods, and influence across scientific communities. Excluding them would remove important instances of higher-order interaction, precisely the kinds of structures our model aims to capture. Accordingly, we include all publications, regardless of author count, in the hypergraph.


\FloatBarrier
\section{Gender Labeling}\label{appendix:gender_labeling}

The data sources used to construct the House, Senate, DBLP, and APS hypergraphs do not provide information on node attributes such as gender. In this section, we describe the procedure used to infer gender labels. Specifically, we use two proprietary gender-labeling APIs: \texttt{GenderAPI}\footnote{\url{https://gender-api.com/}} and \texttt{genderize.io}\footnote{\url{https://genderize.io/}}.

We begin by extracting unique names from each dataset and cleaning them to remove non-alphabetic characters. For example, we strip numeric suffixes from DBLP identifiers and remove periods following initials. The resulting name list is submitted to the APIs for gender prediction. Each API returns a predicted gender, \texttt{male}, \texttt{female}, or \texttt{unknown}, along with a confidence score $c \in [0.5, 1.0]$ and the number of samples used to make the prediction. A score $c = 0.5$ corresponds to an unknown label, while $c=1.0$ reflects full confidence. We encode the \texttt{male} group as $g=0$. For each node $v$, we define a probability distribution $p_v(g)$ over group membership based on the predicted label and confidence score. If a node is labeled \texttt{male} with confidence $c$, we set $p_v(g=0)=c$ and $p_v(g=1) = 1-c$. If labeled \texttt{female}, we set $p_v(g=1)=c$ and $p_v(g=0) = 1-c$. For nodes labeled \texttt{unknown}, we set $p_v(g=0)=p_v(g=1)=0.5$.

\begin{figure*}[tb]
\centering
\includegraphics[width=0.9\textwidth]{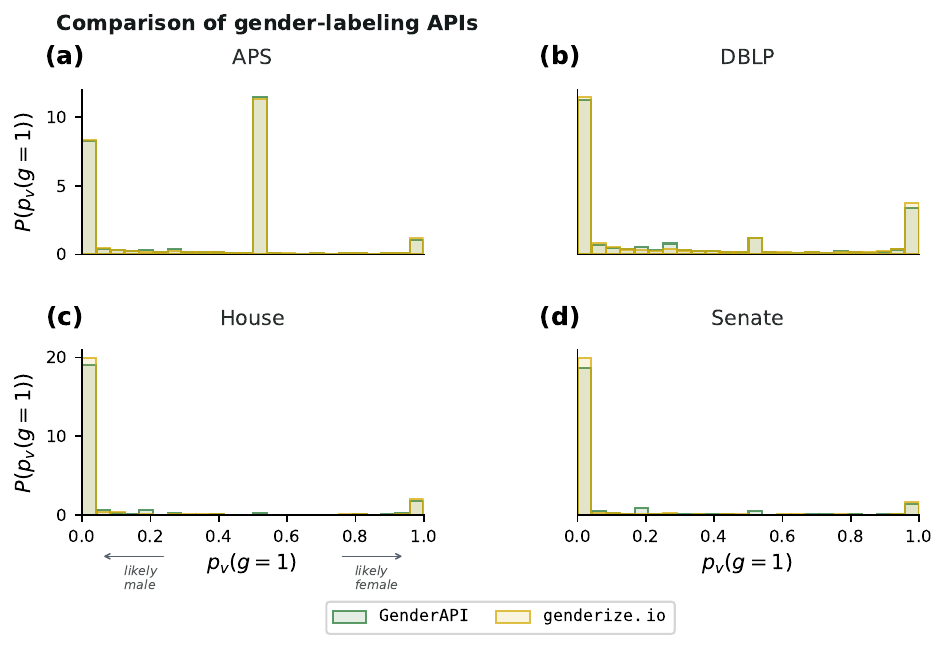}
\caption{\textbf{Distributions of predicted gender probabilities.} The node-level gender probabilities $p_v(g=1)$ for \textbf{(a)} APS, \textbf{(b)} DBLP, \textbf{(c)} House, and \textbf{(d)} Senate. Here, $p_v(g=1)=0$ implies the node is male, while $p_v(g=1)=1$ suggests the node is female. Distributions of probabilities generated by \texttt{GenderAPI} are displayed in green, while those generated by \texttt{genderize.io} are displayed in yellow. We see strong agreement across the two APIs.}\label{fig:gender_labeling}
\end{figure*}

The distributions of probabilities $p_v(g=1)$ according to \texttt{GenderAPI} and \texttt{genderize.io} are qualitatively similar for the DBLP, House, and Senate hypergraphs, with the majority of the predictions being male with high probability, i.e., $p_v(g=1) \approx 0.0$ (Fig.~\ref{fig:gender_labeling}(b)-(d)). While the distributions of probabilities $p_v(g=1)$ are similar according to \texttt{GenderAPI} and \texttt{genderize.io} for the APS dataset, the majority of the predictions are unknown, i.e., $p_v(g=1) \approx 0.5$ (Fig.~\ref{fig:gender_labeling}(a)).

While this probabilistic sampling procedure allows ambiguous nodes to be represented by both male and female attributes over repeated runs, it may also affect group composition and interaction patterns. In particular, sampling can increase the apparent size of the minority group and thereby raise the prevalence of heterophilous edges. For example, in the APS dataset many nodes are labeled as \texttt{unknown}, and probabilistic sampling yields an average composition of roughly $65\%$ men and $35\%$ women. This proportion likely overestimates the number of women in physics, but it avoids discarding ambiguous cases or systematically excluding individuals (e.g., those with names of Asian origin) due to algorithmic bias. Thus, while our procedure helps mitigate the distortions introduced by deterministic labeling or node deletion, it cannot fully eliminate the underlying uncertainties.

Because the APIs likely draw from overlapping sources and methodologies, we treat their predictions as non-independent and do not ensemble them. Instead, we validate the robustness of our results by comparing outputs across both services. We present results based on labels generated from \texttt{GenderAPI} in the main text (Section~\ref{sec:results:real_mixed_homophily}) and list the \texttt{genderize.io}-based results in Appendix~\ref{appendix:additional_results_real_world}.

\FloatBarrier
\section{Dataset Characteristics}\label{appendix:datasets}

\begin{table*}[tb]
\resizebox{\textwidth}{!}{%
\begin{tabular}{l|l|c|c|c|c|c|c|c|c|c|c|c|c|c}
\textbf{Dataset}   & \textbf{Groups} ($g=0$/$g=1$)  & $n$  & $n_{0}$ & $n_1$ & $\bar{k}$ & $\bar{k}_0$ & $\bar{k}_1$ & $s_\mathrm{max}$ & $m$ & $\frac{\bar{k}_1}{\bar{k}_0}$ & $\frac{\overline{k^2}_1}{\overline{k^2}_0}$ & $\bar{k}'$ & $\bar{k}'_0$ & $\bar{k}'_1$ \\ \hline \hline
Hospital      & Healthcare Worker/Patient       & $75$      &$46$       & $29$       & $59$  & $79$  & $28$         & $5$                    & $1825$      & $0.35$ & $0.11$ & $30$ & $37$ & $21$         \\ \hline
High School   & Male/Female                     & $327$     &$182$      & $145$      & $55$  & $53$  & $58$         & $5$                    & $7818$      & $1.10$ & $1.19$ & $36$ & $35$ & $36$         \\ \hline
Primary School& Male/Female                     & $242$     &$130$      & $112$      & $127$ & $132$ & $121$        & $5$                    & $12704$     & $0.92$ & $0.82$ & $69$ & $72$ & $65$         \\ \hline
Senate Bills  & Democrat/Republican             & $294$     &$150$      & $144$      & $732$ & $806$ & $655$        & $99$                   & $21721$     & $0.81$ & $0.67$ & $156$ & $159$ & $153$         \\ \hline
Senate Bills* & Male/Female (GenderAPI)         & $294$     &$259$      & $35$       & $732$ & $743$ & $649$        & $99$                   & $21721$     & $0.87$ & $0.73$  & $156$ & $158$ & $139$       \\ \hline
Senate Bills* & Male/Female (genderize.io)      & $294$     &$264$      & $30$       & $732$ & $739$ & $665$        & $99$                   & $21721$     & $0.90$ & $0.77$ & $156$ & $158$ & $137$        \\ \hline
House Bills   & Democrat/Republican             & $1494$    &$792$      & $702$      & $814$ & $962$ & $648$        & $399$                  & $54933$     & $0.67$ & $0.43$ & $664$ & $686$ & $639$         \\ \hline
House Bills*  & Male/Female (GenderAPI)         & $1494$    &$1303$     & $191$      & $814$ & $797$ & $932$        & $399$                  & $54933$     & $1.17$ & $1.25$ & $664$ & $666$ & $651$        \\ \hline
House Bills*  & Male/Female (genderize.io)      & $1494$    &$1323$     & $171$      & $814$ & $797$ & $951$        & $399$                  & $54933$     & $1.19$ & $1.24$ & $664$ & $666$ & $651$       \\ \hline
DBLP*         & Male/Female (GenderAPI)         & $3386798$ & $2399515$ & $987282$   & $5$   & $6$   & $4$          & $450$                  & $4591505$   & $0.80$  & $0.79$ & $15$ & $15$ & $13$       \\ \hline
DBLP*         & Male/Female (genderize.io)      & $3386798$ & $2367643$ & $1019155$  & $5$   & $6$   & $5$          & $450$                  & $4591505$   & $0.82$  & $0.84$ & $15$ & $15$ & $13$        \\ \hline
APS*          & Male/Female (GenderAPI)         & $482651$  & $320386$  & $162265$   & $37$  & $37$  & $37$         & $1666$                 & $504360$    & $1.00$ & $1.07$ & $1205$ & $1190$ & $1236$        \\ \hline
APS*          & Male/Female (genderize.io)     & $482651$  & $319057$  & $163594$   & $35$  & $39$  & $37$         & $1666$                 & $504360$    & $1.11$ & $1.23$ & $1205$ & $1148$ & $1317$         \\
\end{tabular}
}
\caption{\textbf{Descriptive statistics for the real-world hypergraph datasets.} For each real-world hypergraph, we report the majority group ($g=0$), minority group ($g=1$), total number of nodes ($n$), group-specific node counts ($n_g$), average degree ($\bar{k}$), group-specific average degrees ($\bar{k}_g$), maximum hyperedge size $s_\mathrm{max}$, total number of hyperedges $m$, the power inequality $\bar{k}_1/\bar{k}_0$, moment glass ceiling $\overline{k^2}_1/\overline{k^2}_0$, average number of unique nodes a node is connected to ($\bar{k}'$), and the group-specific average number of unique nodes a node is connected to ($\bar{k}'_g$)}. For datasets with inferred node labels, all values are averaged over $n_\mathrm{hg} = 10^3$ hypergraphs with sampled group labels, as described in Appendix~\ref{appendix:parameters_real_world}.
\label{tab:real_world_data}
\end{table*}

\begin{figure*}[p]
\centering
\includegraphics[width=\textwidth]{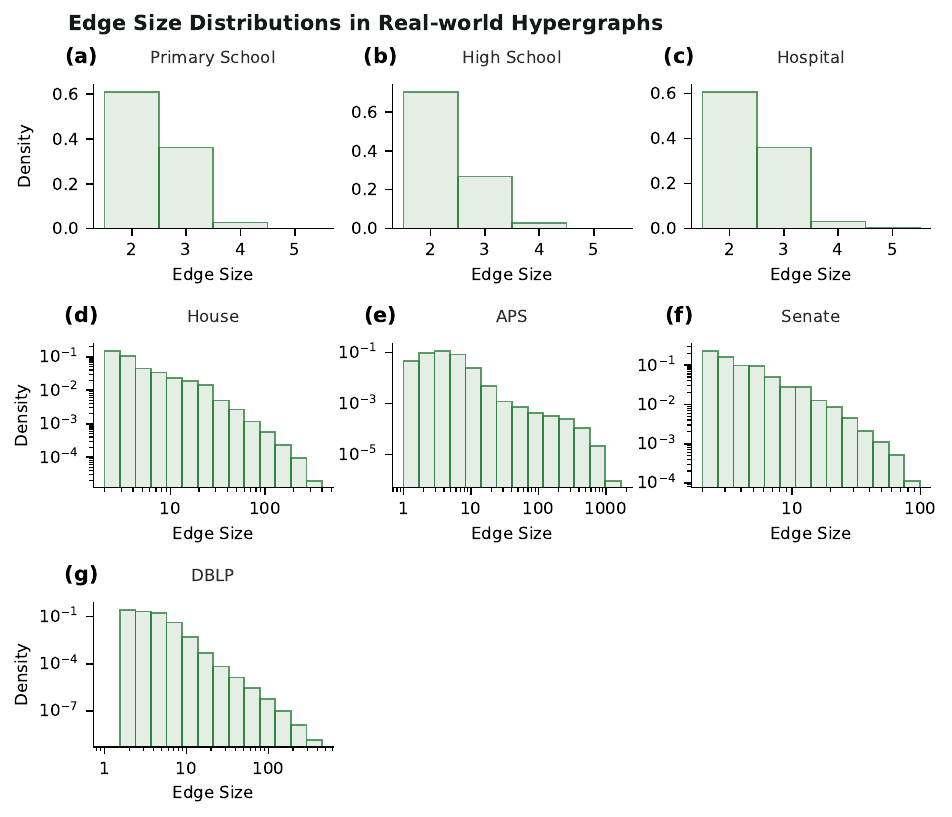}
\caption{\textbf{Distributions of hyperedge sizes in real-world hypergraphs.} The hyperedge size distributions for \textbf{(a)} Primary School, \textbf{(b)} High School, \textbf{(c)} Hospital, \textbf{(d)} House, \textbf{(e)} APS, \textbf{(f)} Senate, and \textbf{(g)} DBLP. We only show one plot each for House, Senate, APS, and DBLP since differences in node labels affect hyperedge type, not hyperedge size. Panels (a)–(c) are binned linearly, while panels (d)–(g) are binned logarithmically in hyperedge size and plotted with log-log axes to capture the broader heterogeneity in hyperedge sizes.}\label{fig:edgesize_distribution}
\end{figure*}

\begin{figure*}[p]
\centering
\includegraphics[width=\textwidth]{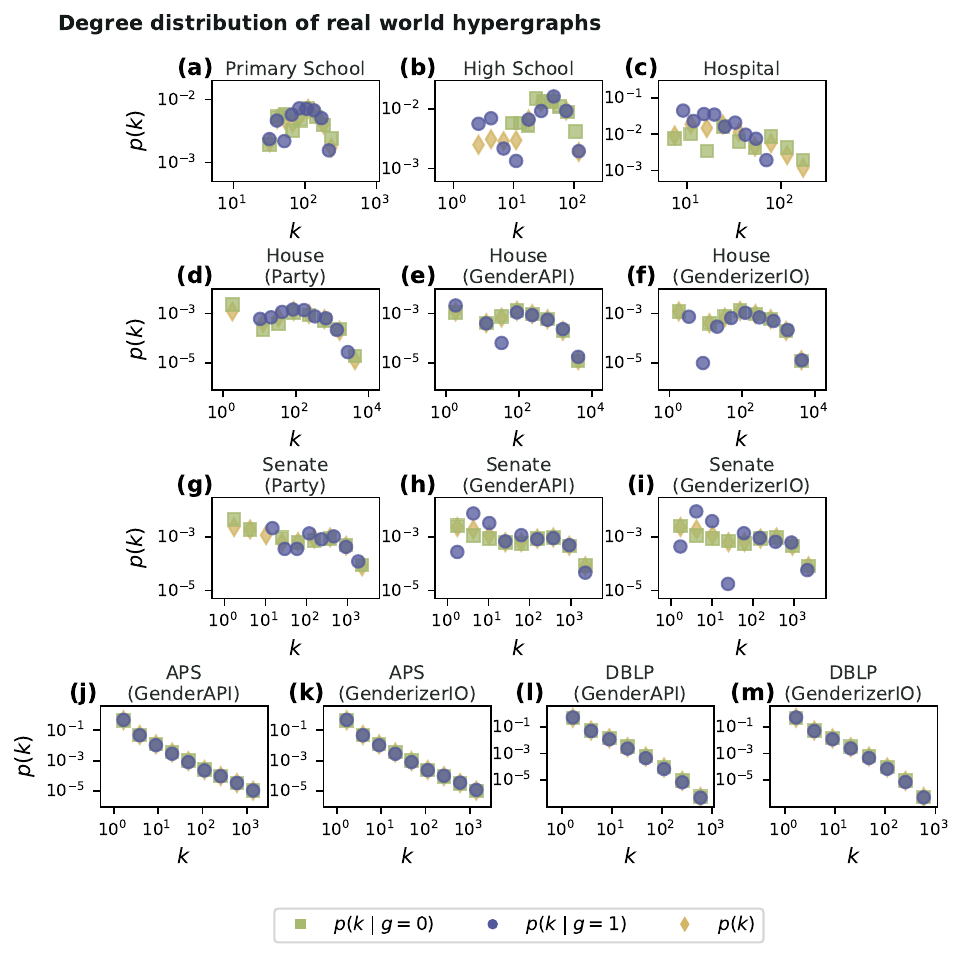}
\caption{\textbf{Degree distributions of nodes in real-world hypergraphs.} The distribution $p(k)$ of degree of the majority group $g=0$ (green squares), the minority group $g=1$ (purple circles), and overall (yellow diamonds) for \textbf{(a)} Primary School, \textbf{(b)} High School, \textbf{(c)} Hospital, \textbf{(d)} House (Party), \textbf{(e)} House (Gender - \texttt{GenderAPI}), \textbf{(f)} House (Gender - \texttt{genderize.io}), \textbf{(g)} Senate (Party), \textbf{(h)} Senate (Gender - \texttt{GenderAPI}), \textbf{(i)} Senate (Gender - \texttt{genderize.io}), \textbf{(j)} APS - \texttt{GenderAPI}, \textbf{(k)} APS - \texttt{genderize.io}, \textbf{(l)} DBLP - \texttt{GenderAPI}, and \textbf{(m)} DBLP - \texttt{genderize.io}. All panels are logarithmically binned in degree and plotted with log-log axes to capture the broad heterogeneity in degree distribution.}
\label{fig:real_degree_distribution}
\end{figure*}

\begin{figure*}[p]
\centering
\includegraphics[width=\textwidth]{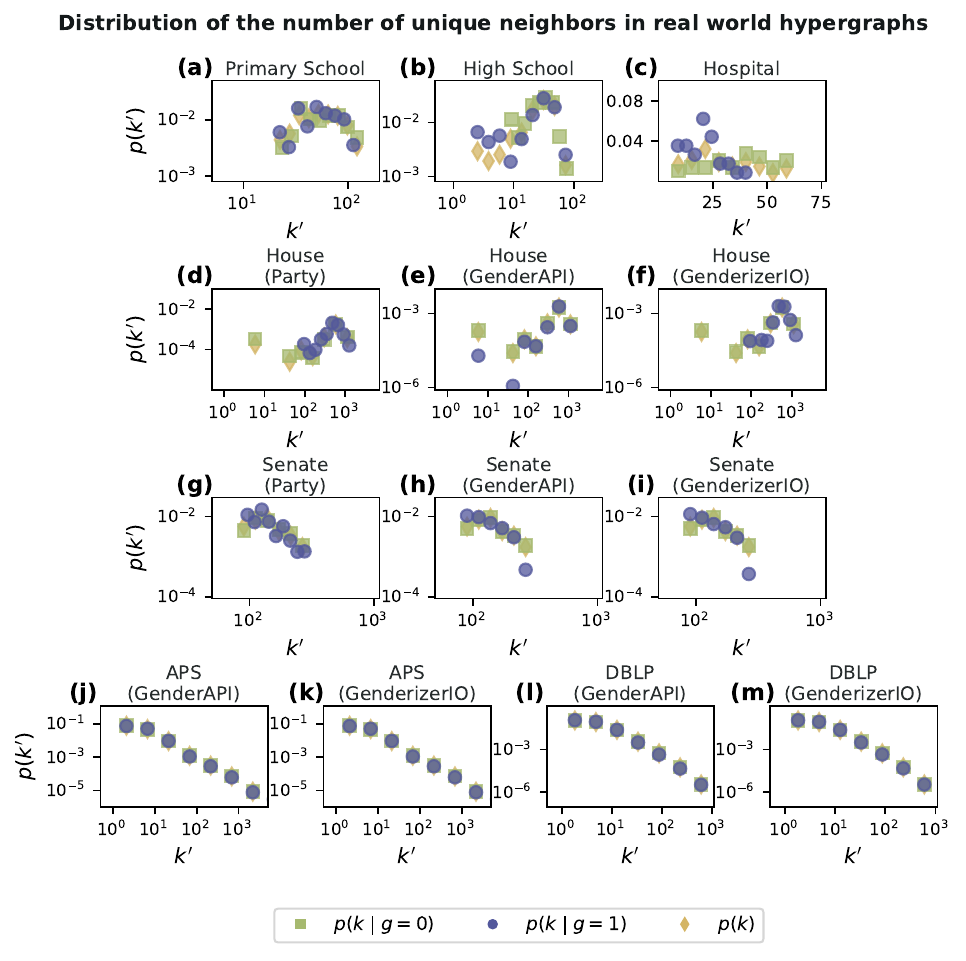}
\caption{\textbf{Distribution of the number of unique neighbors in real-world hypergraphs.} The distribution $p(k')$ of the number of unique nodes to which a given node is connected for \textbf{(a)} Primary School, \textbf{(b)} High School, \textbf{(c)} Hospital, \textbf{(d)} House (Party), \textbf{(e)} House (Gender - \texttt{GenderAPI}), \textbf{(f)} House (Gender - \texttt{genderize.io}), \textbf{(g)} Senate (Party), \textbf{(h)} Senate (Gender - \texttt{GenderAPI}), \textbf{(i)} Senate (Gender - \texttt{genderize.io}), \textbf{(j)} APS - \texttt{GenderAPI}, \textbf{(k)} APS - \texttt{genderize.io}, \textbf{(l)} DBLP - \texttt{GenderAPI}, and \textbf{(m)} DBLP - \texttt{genderize.io}. Distributions for the majority group $g=0$ are displayed in green, those for the minority group are displayed in purple, and the overall distribution is displayed in yellow. Aside from panel (c) which is linearly binned, all panels are logarithmically binned in degree and plotted with log-log axes to capture the broad heterogeneity in degree distribution.}
\label{fig:real_pairwise_degree_distribution}
\end{figure*}

The real-world hypergraphs vary substantially in size, group composition, connectivity, and structural measures of inequality---including differences in node count, group proportions, hyperedge count, average degree, maximum hyperedge sizes, and degree-based differences (Tab.~\ref{tab:real_world_data}). The hypergraphs also exhibit diverse hyperedge size distributions (Fig.~\ref{fig:edgesize_distribution}), degree distributions (Fig.~\ref{fig:real_degree_distribution}), and distributions of numbers of unique neighbors (Fig.~\ref{fig:real_pairwise_degree_distribution}). While $\bar{k}$ captures the average number of hyperedges to which a node belongs, $\bar{k}'$ denotes the average number of unique nodes to which a given node is connected. In addition to displaying summary information on the hypergraph structure, we include two structural inequality measures introduced by Avin et al.~\cite{avin2015_HomophilyGlassCeiling}: the power inequality, defined as the ratio of group-wise average degrees $\bar{k}_1/\bar{k}_0$, and the moment glass ceiling, defined as the ratio of group-wise second moments of the degree distributions $\langle k^2\rangle_1/\langle k^2\rangle_0$. In both cases, values below $1$ indicate a structural advantage for the majority group (assumed to be $g=0$).


\section{Parameters for Simulations on Real-World Hypergraphs}\label{appendix:parameters_real_world}
 
\begin{table*}[tb]
\centering
\begin{tabular}{l|l|c|c|c|c}
\textbf{Datasets}                                                                                                                                 & \textbf{Dynamics}     & $\lambda_\mathrm{in}$ & $\lambda_\mathrm{out}$ & $\nu_\mathrm{in}$ & $\nu_\mathrm{out}$ \\ \hline \hline
\multirow{4}{*}{\begin{tabular}[c]{@{}c@{}} APS, DBLP, House (Gender), \\ Senate (Gender), Hospital, \\ Primary School, High School\end{tabular}} & symmetric linear       & $0.01$         & $0.01$          & $1.00$     & $1.00$      \\
                                                                                                                                         & symmetric sublinear   & $0.01$         & $0.01$          & $0.90$     & $0.90$      \\
                                                                                                                                         & symmetric superlinear & $0.01$         & $0.01$          & $1.11$     & $1.11$      \\
                                                                                                                                         & asymmetric nonlinear   & $0.02$         & $0.005$         & $1.11$     & $0.90$      \\ \hline
\multirow{4}{*}{House (Party), Senate (Party)}                                                                                           & symmetric linear       & $0.01$         & $0.01$          & $1.00$     & $1.00$      \\
                                                                                                                                         & symmetric sublinear   & $0.01$         & $0.01$          & $0.75$     & $0.75$      \\
                                                                                                                                         & symmetric superlinear & $0.01$         & $0.01$          & $1.33$     & $1.33$      \\
                                                                                                                                         & asymmetric nonlinear  & $0.02$         & $0.005$         & $1.33$     & $0.75$     
\end{tabular}
\caption{\textbf{Social contagion parameters for real-world hypergraphs.} In- and out-group transmission rates are denoted by $\lambda_\mathrm{in}$ and $\lambda_\mathrm{out}$, respectively, while nonlinearity parameters are denoted by $\nu_\mathrm{in}$ and $\nu_\mathrm{out}$.} \label{tab:dynamics_parameters_real}
\end{table*}

We describe the parameters used in the experiments on real-world hypergraphs, including the parameters of the social contagion model and those of the seeding strategies. We also provide additional details about the real-world simulations.

\paragraph{Social Contagion Parameters}
We use the parameterization of the social contagion process introduced in Section~\ref{sec:methods:contagion}, which defines separate in-group and out-group transmission rates ($\lambda_\mathrm{in}$, $\lambda_\mathrm{out}$) and nonlinearity parameters ($\nu_\mathrm{in}$, $\nu_\mathrm{out}$). We explore the same qualitative contagion scenarios as in the synthetic experiments (Appendix~\ref{appendix:parameters_synthetic}): a symmetric linear setting with $\lambda_\mathrm{in} = \lambda_\mathrm{out}$ and $\nu_\mathrm{in} = \nu_\mathrm{out} = 1$, a symmetric sublinear setting with $\lambda_\mathrm{in} = \lambda_\mathrm{out}$ and $\nu_\mathrm{in} = \nu_\mathrm{out} < 1$, a symmetric superlinear setting with $\lambda_\mathrm{in} = \lambda_\mathrm{out}$ and $\nu_\mathrm{in} = \nu_\mathrm{out} > 1$, and an asymmetric nonlinear setting with $\lambda_\mathrm{in} > \lambda_\mathrm{out}$, $\nu_\mathrm{in} > 1$, and $\nu_\mathrm{out} < 1$. Because real-world hypergraphs differ in their largest hyperedge size $s_\mathrm{max}$ and the frequency of large hyperedges, we select dataset-specific values for $\lambda_\mathrm{in}$, $\lambda_\mathrm{out}$, $\nu_\mathrm{in}$, and $\nu_\mathrm{out}$, summarized in Table~\ref{tab:dynamics_parameters_real}.

\paragraph{Seeding Strategy}
We use a single seed $n_\mathrm{seed}=1$ chosen uniformly at random from either the minority group $g=1$ ($n_\mathrm{seed}^{(0)}=0$ and $n_\mathrm{seed}^{(1)}=1$) or the majority group $g=0$ ($n_\mathrm{seed}^{(0)}=1$ and $n_\mathrm{seed}^{(1)}=0$). We refer to the former condition as minority seeding and the latter as majority seeding. We present results for majority seeding in the main text and results for minority seeding in Appendix~\ref{appendix:additional_results_real_world}.

\paragraph{Simulation}
We differentiate between real-world hypergraphs with ground truth node labels (Primary School, High School, Hospital, Senate (Party), and House (Party)) and those with inferred gender labels (DBLP, APS, Senate (Gender), and House (Gender)). Details of the gender labeling process are provided in Appendix~\ref{appendix:gender_labeling}.

For hypergraphs with ground-truth labels, we simulate $n_\mathrm{hg} = 10^3$ independent realizations of the contagion process. In each realization, we sample the seed nodes according to the specified seeding strategy and simulate the spread of information using a fixed set of contagion parameters.

The gender labeling procedure yields confidence scores for group membership, which we normalize to obtain node-specific probability distributions $p_v(g)$ over group assignments (see Appendix~\ref{appendix:gender_labeling}). For each of the hypergraphs with inferred labels, we simulate $n_\mathrm{hg} = 10^3$ realizations. We first sample a group label $g_v$ for each node $v \in \mathcal{V}$ from its corresponding distribution $p_v(g)$. We then select seed nodes according to the specified seeding strategy and simulate the contagion process using fixed parameter settings.


\section{Supplementary Results on Real-World Hypergraphs}
\label{appendix:additional_results_real_world}

We present additional results for the real-world hypergraphs introduced in Section~\ref{sec:results:real_mixed_homophily}. The main text focuses on two case studies, High School and Hospital, using majority seeding. Here, we supplement those results by considering minority seeding. We also discuss the homophily patterns and information access inequality of seven additional hypergraphs: Primary School, House (by party), Senate (by party), House (by gender), Senate (by gender), DBLP, and APS. We make three main points: (1) results tend to be robust to minority seeding, (2) outcomes on hypergraphs with predicted gender labels remain qualitatively consistent when node labels are generated using \texttt{genderize.io} instead of \texttt{GenderAPI}, and (3) the analysis of real-world hypergraphs becomes increasingly difficult with the hypergraph size due to interactions between complicated homophily patterns and structural properties.

\begin{figure*}[tb]
\centering
\includegraphics[width=0.9\textwidth]{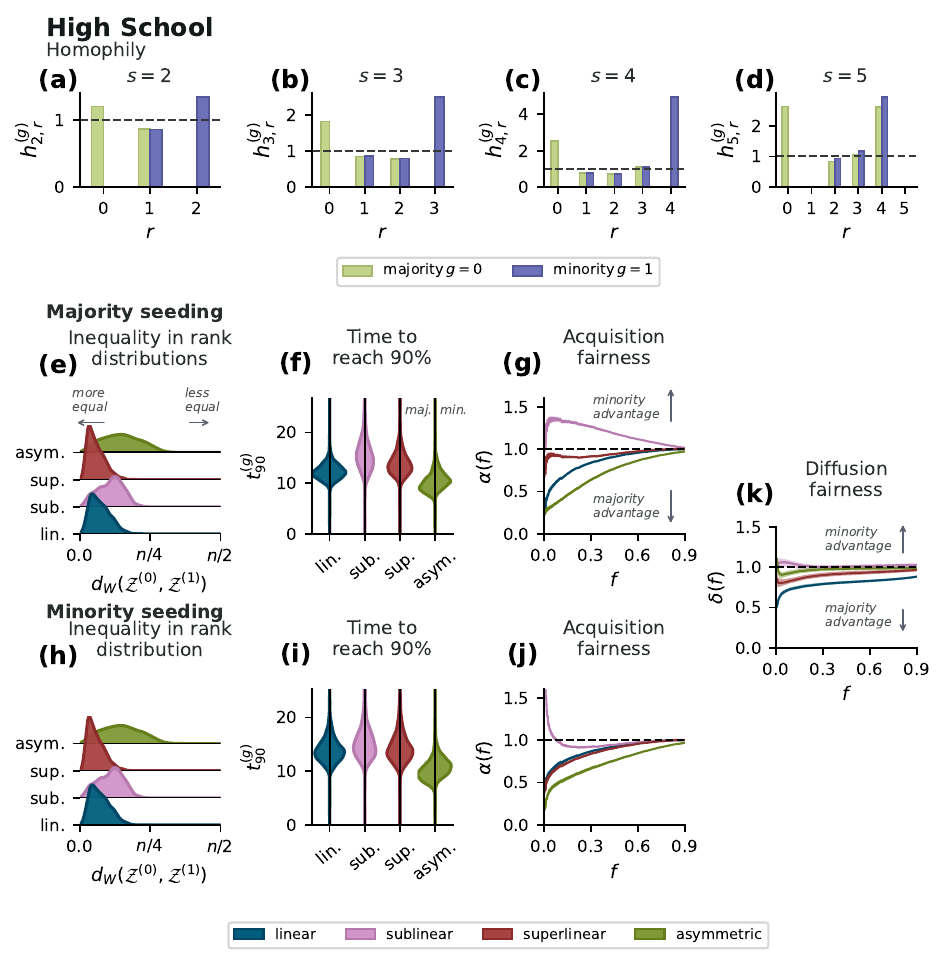}
\caption{\textbf{Homophily patterns and information diffusion inequality in the High School hypergraph.} \textbf{(a)-(d)} Hyperedge homophily $h_{s,r}^{(g)}$ is shown for hyperedge sizes $s\in\{2,3,4,5\}$ for majority ($g=0$, green) and minority ($g=1$, purple) groups. The dashed line shows the expected prevalence under random mixing; values above indicate over-representation. Information inequality outcomes are reported under both majority seeding (\textbf{(e)–(g)}) and minority seeding (\textbf{(h)–(j)}): \textbf{(e),(h)} Wasserstein distances $d_W(\mathcal{Z}_0,\mathcal{Z}_1)$, \textbf{(f),(i)} violin plots of the time $t^{(g)}_{90}$ to inform 90\% of nodes (majority left, minority right), and \textbf{(g),(j)} acquisition fairness $\alpha(f)$. 
\textbf{(k)} Shows diffusion fairness $\delta(f)$, where values above the dashed line indicate a minority advantage. Panels (e)–(k) average results over $n_\mathrm{hg}=10^3$ simulations for linear (blue), sublinear (pink), superlinear (red), and asymmetric (green) contagion. Confidence intervals in (g),(j),(k) are estimated from $100$ bootstrap samples. The High School hypergraph shows a strong majority advantage under asymmetric contagion, while sublinear contagion reduces this advantage and can even produce a minority advantage.}
\label{fig:SI_highschool}
\end{figure*}

\paragraph{High School}
As discussed in Section~\ref{sec:results:real_mixed_homophily}, the High School hypergraph is homophilous for both the majority and minority class, but there are no instances of all-minority hyperedges of size $s=5$. Under majority seeding, this leads to a sizable majority advantage under asymmetric transmission. We posit that sublinear transmission deprioritizes spread through the larger hyperedges, meaning the majority group is unable to take advantage of the size $5$ hyperedges leading to a minority advantage. We see qualitatively the same pattern under minority seeding (Fig.~\ref{fig:SI_highschool}(h)-(j)).

\begin{figure*}[tb]
\centering
\includegraphics[width=0.9\textwidth]{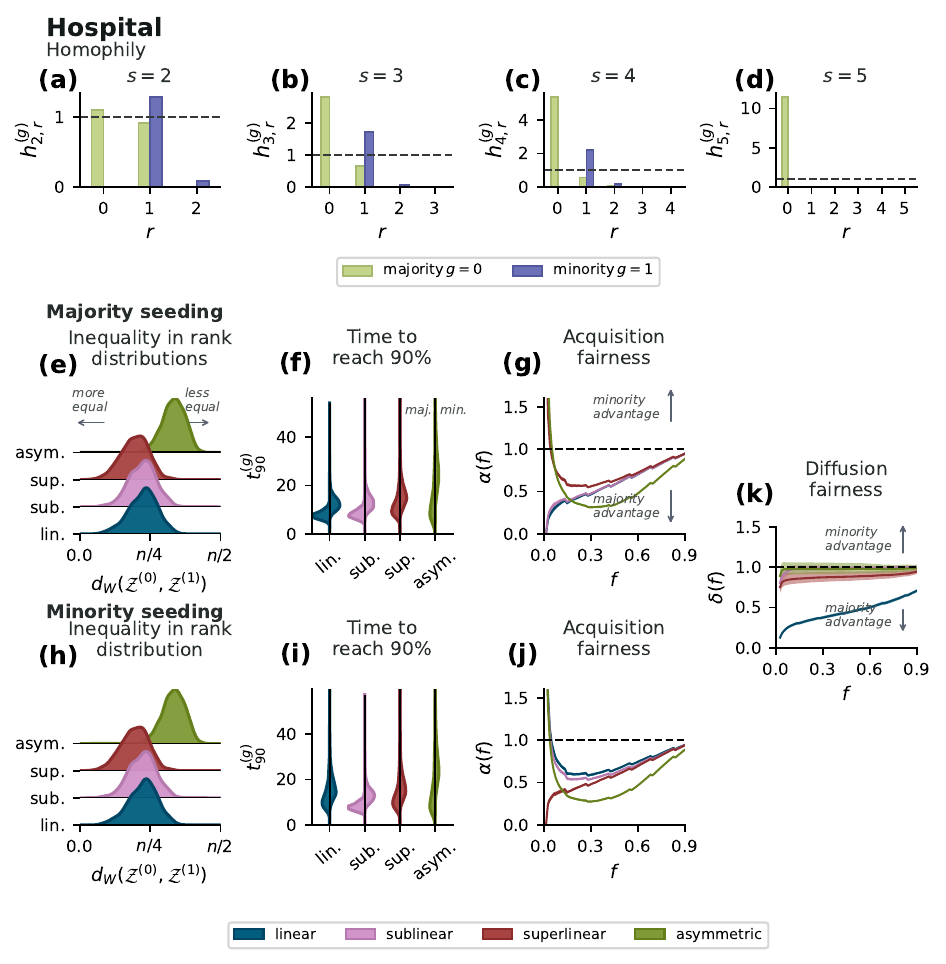}
\caption{\textbf{Homophily patterns and information diffusion inequality in the Hospital hypergraph.} \textbf{(a)-(d)} Hyperedge homophily $h_{s,r}^{(g)}$ is shown for hyperedge sizes $s\in\{2,3,4,5\}$ for majority ($g=0$, green) and minority ($g=1$, purple) groups. The dashed line shows the expected prevalence under random mixing; values above indicate over-representation. Information inequality outcomes are reported under both majority seeding (\textbf{(e)–(g)}) and minority seeding (\textbf{(h)–(j)}): \textbf{(e),(h)} Wasserstein distances $d_W(\mathcal{Z}_0,\mathcal{Z}_1)$, \textbf{(f),(i)} violin plots of the time $t^{(g)}_{90}$ to inform 90\% of nodes (majority left, minority right), and \textbf{(g),(j)} acquisition fairness $\alpha(f)$. 
\textbf{(k)} Shows diffusion fairness $\delta(f)$, where values above the dashed line indicate a minority advantage. Panels (e)–(k) average results over $n_\mathrm{hg}=10^3$ simulations for linear (blue), sublinear (pink), superlinear (red), and asymmetric (green) contagion. Confidence intervals in (g),(j),(k) are estimated from $100$ bootstrap samples. The Hospital hypergraph shows extreme inequalities under asymmetric contagion approaching the theoretical maximum of $n/2$, with majority nodes infected well before minority nodes; initial minority advantages in $\alpha(f)$ under some regimes quickly give way to a pronounced majority advantage.}
\label{fig:SI_hospital}
\end{figure*}

\paragraph{Hospital}
As discussed in Section~\ref{sec:results:real_mixed_homophily}, the Hospital hypergraph has strong homophily for the majority group (hospital staff) only, whereas the minority group (patients) almost exclusively participates in mixed edges. This leads to strong inequality approaching the theoretical maximum in the asymmetric case, suggesting that almost all of the majority nodes are infected before any of the minority nodes. We see qualitatively the same results under minority seeding (Fig.~\ref{fig:SI_hospital}). The only observable difference is in acquisition fairness. With majority seeding, the minority starts with an advantage under superlinear and asymmetric contagion which quickly shifts to a stark majority advantage (Fig.~\ref{fig:SI_hospital}(g)). With minority seeding, the minority starts with an advantage under normal, sublinear, and asymmetric contagion which quickly shifts to a stark majority advantage (Fig.~\ref{fig:SI_hospital}(j)). It is possible that seeding the minority now leads to an initial minority advantage in the linear and sublinear cases since minority nodes are the first to be infected and the spreading dynamics make it more likely that the infected majority nodes infect other minority nodes through pairwise or small group interactions.

\begin{figure*}[tb]
\centering
\includegraphics[width=0.9\textwidth]{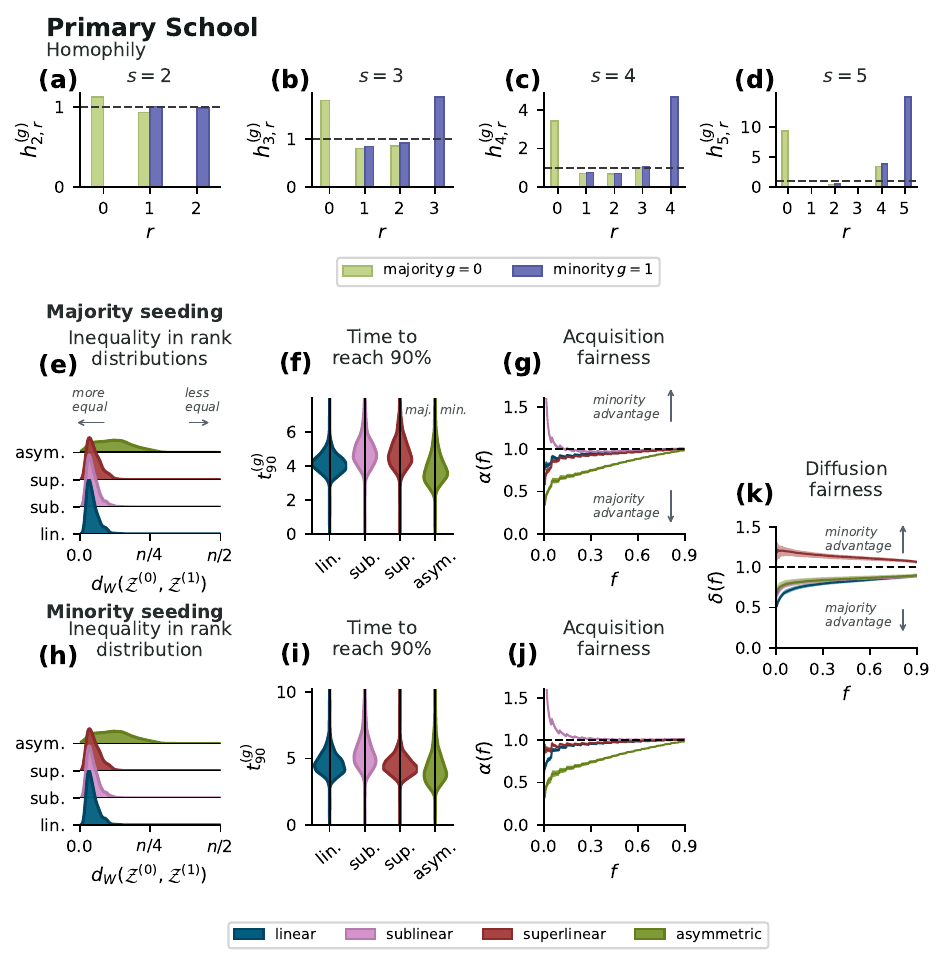}
\caption{\textbf{Homophily patterns and information diffusion inequality in the Primary School hypergraph.} \textbf{(a)-(d)} Hyperedge homophily $h_{s,r}^{(g)}$ is shown for hyperedge sizes $s\in\{2,3,4,5\}$ for majority ($g=0$, green) and minority ($g=1$, purple) groups. The dashed line shows the expected prevalence under random mixing; values above indicate over-representation. Information inequality outcomes are reported under both majority seeding (\textbf{(e)–(g)}) and minority seeding (\textbf{(h)–(j)}): \textbf{(e),(h)} Wasserstein distances $d_W(\mathcal{Z}_0,\mathcal{Z}_1)$, \textbf{(f),(i)} violin plots of the time $t^{(g)}_{90}$ to inform 90\% of nodes (majority left, minority right), and \textbf{(g),(j)} acquisition fairness $\alpha(f)$. 
\textbf{(k)} Shows diffusion fairness $\delta(f)$, where values above the dashed line indicate a minority advantage. Panels (e)–(k) average results over $n_\mathrm{hg}=10^3$ simulations for linear (blue), sublinear (pink), superlinear (red), and asymmetric (green) contagion. Confidence intervals in (g),(j),(k) are estimated from $100$ bootstrap samples. The Primary School hypergraph shows modest majority advantages under most contagion regimes, with a minority advantage in diffusion fairness emerging under superlinear contagion that is likely linked to stronger minority homophily in larger hyperedges.}
\label{fig:SI_primaryschool}
\end{figure*}

\paragraph{Primary School}
The Primary School homophily pattern closely remsembles the basic homophilous pattern explored in Section~\ref{sec:results:synthetic_emd} (Fig.~\ref{fig:SI_primaryschool}(a)-(d)). Under majority seeding, there is minimal inequality with a modest majority advantage in the linear and superlinear cases, while there is a similar modest advantage for the minority in the sublinear case. In the asymmetric case, there is a slightly more pronounced majority advantage (Fig.~\ref{fig:SI_primaryschool}(e)-(g)). We see qualitatively the same trends under minority seeding (Fig.~\ref{fig:SI_primaryschool}(h)-(j)). These results follow our synthetic homophilous results, but are less pronounced. It is possible that the magnitude of the inequality is smaller due to a more modest class imbalance, i.e., about $54\%$ majority and $46\%$ minority for the Primary School hypergraph compared to a $75\%$ majority in the synthetic experiments.

Regarding diffusion fairness, the minority have a slight advantage in the superlinear case, while the majority remain advantaged under other contagion dynamics (Fig.~\ref{fig:SI_primaryschool}(k)). While the homophily pattern is a standard homophilous pattern, the minority group is more homophilous in larger hyperedge sizes. In hyperedges of size $5$, there are both more all-minority hyperedges and more four-minority and one-majority node hyperedges than expected. This high degree of homophily in size $5$ hyperedges might explain why the minority has an advantage in information diffusion when the spread is routed more often through larger hyperedges (i.e., superlinear contagion).

\begin{figure*}[tb]
\centering
\includegraphics[width=0.9\textwidth]{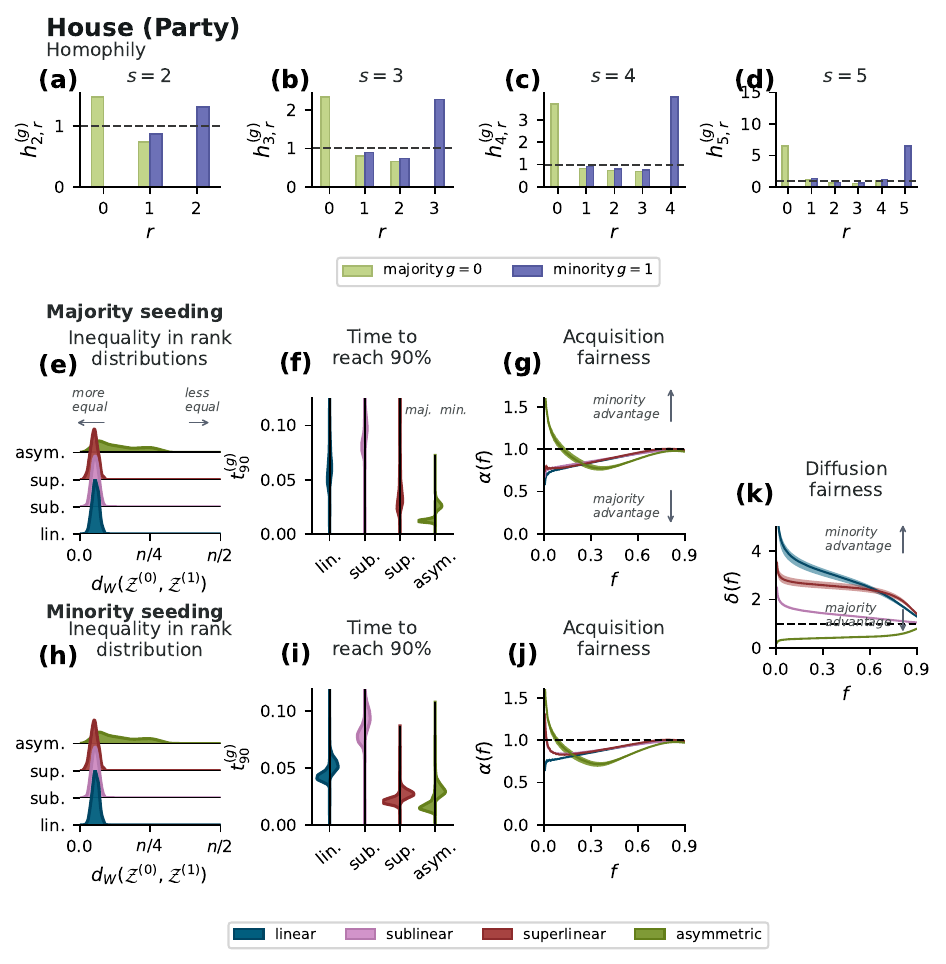}
\caption{\textbf{Homophily patterns and information diffusion inequality in the House hypergraph with party labels.} \textbf{(a)-(d)} Hyperedge homophily $h_{s,r}^{(g)}$ is shown for hyperedge sizes $s\in\{2,3,4,5\}$ for majority ($g=0$, green) and minority ($g=1$, purple) groups. The dashed line shows the expected prevalence under random mixing; values above indicate over-representation. Information inequality outcomes are reported under both majority seeding (\textbf{(e)–(g)}) and minority seeding (\textbf{(h)–(j)}): \textbf{(e),(h)} Wasserstein distances $d_W(\mathcal{Z}_0,\mathcal{Z}_1)$, \textbf{(f),(i)} violin plots of the time $t^{(g)}_{90}$ to inform 90\% of nodes (majority left, minority right), and \textbf{(g),(j)} acquisition fairness $\alpha(f)$. 
\textbf{(k)} Shows diffusion fairness $\delta(f)$, where values above the dashed line indicate a minority advantage. Panels (e)–(k) average results over $n_\mathrm{hg}=10^3$ simulations for linear (blue), sublinear (pink), superlinear (red), and asymmetric (green) contagion. Confidence intervals in (g),(j),(k) are estimated from $100$ bootstrap samples. The House (party) hypergraph shows homophily-driven majority advantages across contagion regimes, broadly consistent with synthetic predictions.}
\label{fig:SI_houseparty}
\end{figure*}

\paragraph{House (by party)}
The House hypergraph is relatively balanced ($53\%$ Democrat and $47\%$ Republican) and exhibits a homophilous pattern for hyperedge sizes $s \in \{2, 3, 4, 5\}$ when labeled by party (Fig.~\ref{fig:SI_houseparty}(a)-(d)). There is a modest majority advantage in the linear, sublinear, and superlinear cases, and a slightly more pronounced majority advantage in the asymmetric case for both majority and minority seeding (majority: Fig.~\ref{fig:SI_houseparty}(e)-(g); minority: Fig.~\ref{fig:SI_houseparty}(h)-(j)). These patterns are consistent with our synthetic results.

Interestingly, there is a strong minority advantage according to diffusion fairness in the linear, sublinear, and superlinear cases (Fig.~\ref{fig:SI_houseparty}(k)). Given that the hypergraph is approximately balanced, it is difficult to interpret a minority or majority advantage. However, we note that the House dataset collects bill co-sponsorship over time and our hypergraph construction collapses the temporal dependency into a single static hypergraph. It is feasible that ignoring time introduces structural properties, such as temporally-correlated node clusters, which could cause this effect. We leave further study of this phenomenon to future work.

\begin{figure*}[tb]
\centering
\includegraphics[width=0.9\textwidth]{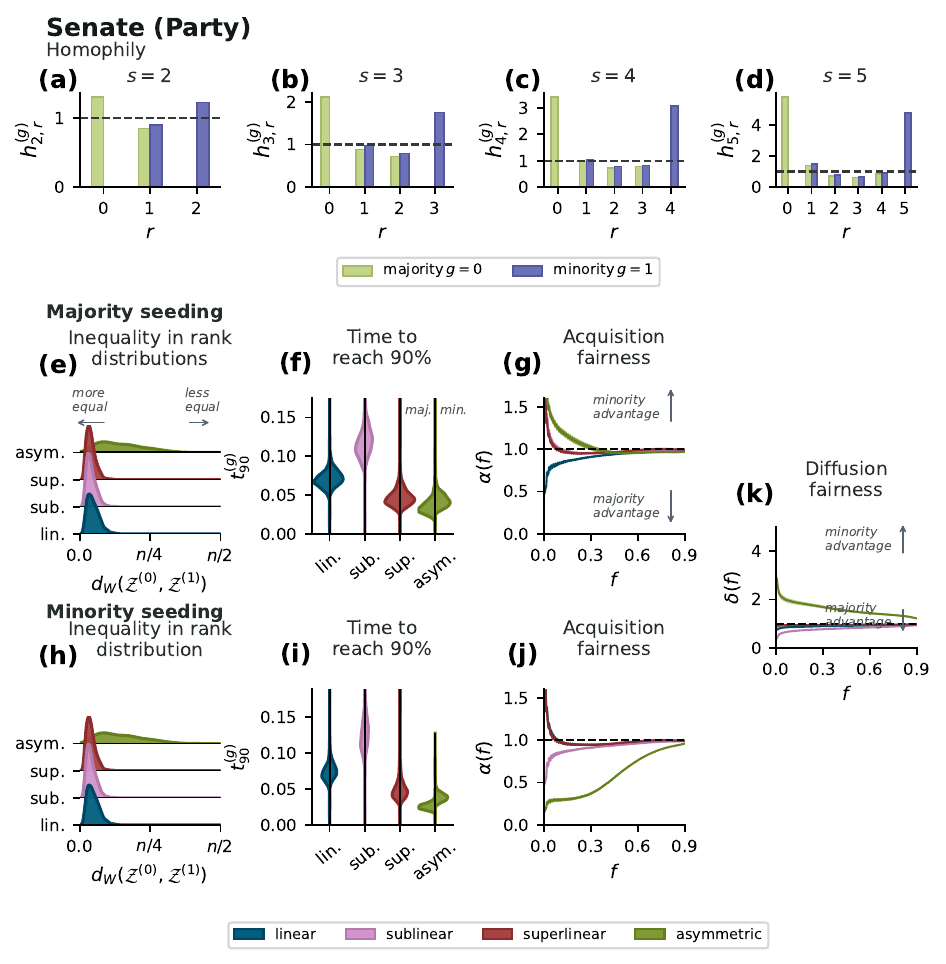}
\caption{\textbf{Homophily patterns and information diffusion inequality in the Senate hypergraph with party labels.} \textbf{(a)-(d)} Hyperedge homophily $h_{s,r}^{(g)}$ is shown for hyperedge sizes $s\in\{2,3,4,5\}$ for majority ($g=0$, green) and minority ($g=1$, purple) groups. The dashed line shows the expected prevalence under random mixing; values above indicate over-representation. Information inequality outcomes are reported under both majority seeding (\textbf{(e)–(g)}) and minority seeding (\textbf{(h)–(j)}): \textbf{(e),(h)} Wasserstein distances $d_W(\mathcal{Z}_0,\mathcal{Z}_1)$, \textbf{(f),(i)} violin plots of the time $t^{(g)}_{90}$ to inform 90\% of nodes (majority left, minority right), and \textbf{(g),(j)} acquisition fairness $\alpha(f)$. 
\textbf{(k)} Shows diffusion fairness $\delta(f)$, where values above the dashed line indicate a minority advantage. Panels (e)–(k) average results over $n_\mathrm{hg}=10^3$ simulations for linear (blue), sublinear (pink), superlinear (red), and asymmetric (green) contagion. Confidence intervals in (g),(j),(k) are estimated from $100$ bootstrap samples. The Senate hypergraph shows a homophily-driven majority advantages across contagion regimes, but diffusion fairness remains near parity, likely reflecting the near balance between parties.}
\label{fig:SI_senateparty}
\end{figure*}

\paragraph{Senate (by party)}
Like the House hypergraph, the Senate hypergraph is almost completely balanced ($51\%$ Democrat and $49\%$ Republican) with a homophilous pattern for hyperedge sizes $s \in \{2, 3, 4, 5\}$ when labeled by party (Fig.~\ref{fig:SI_senateparty}(a)-(d)). The information access inequality follows the same patterns as the House hypergraph (majority seeding: Fig.~\ref{fig:SI_senateparty}(e)-(g); minority seeding: Fig.~\ref{fig:SI_senateparty}(h)-(j)). However, we do see the opposite inequality in diffusion fairness as in the House case (Fig.~\ref{fig:SI_senateparty}(k)). This is likely because the hypergraph is almost completely balanced.

\begin{figure*}[tb]
\centering
\includegraphics[width=0.9\textwidth]{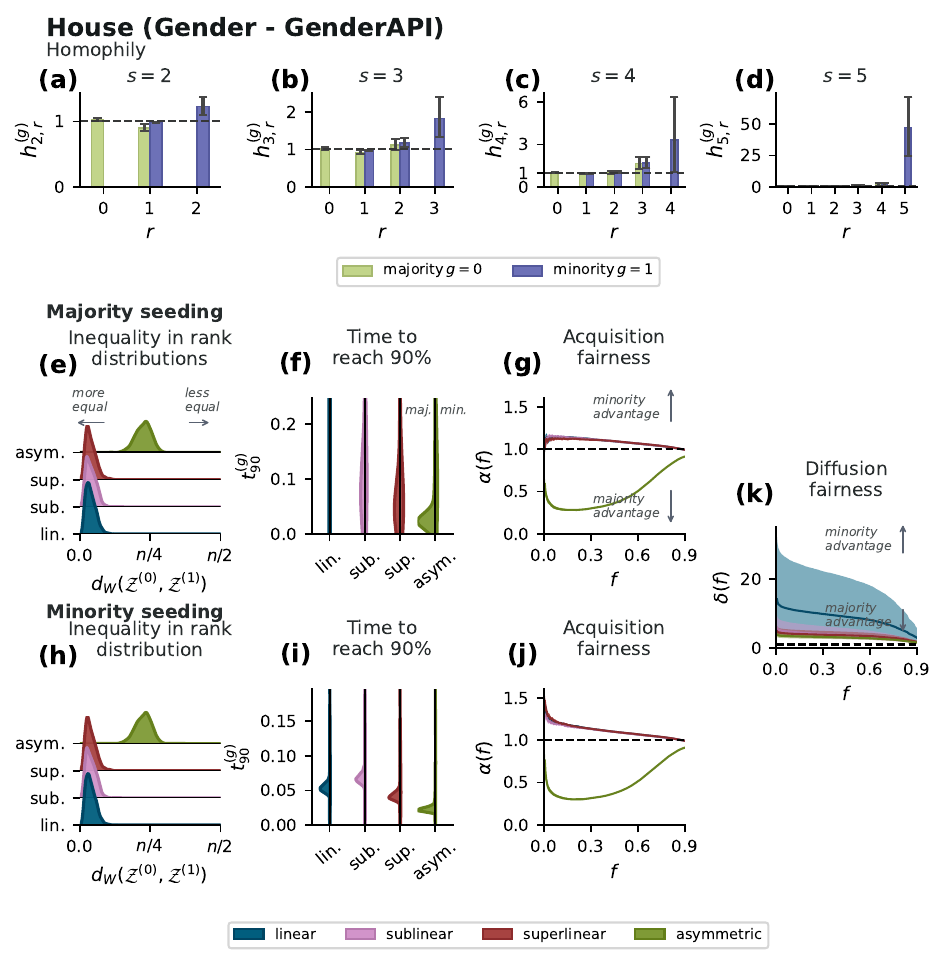}
\caption{\textbf{Homophily patterns and information diffusion inequality in the House hypergraph with gender labels (\texttt{GenderAPI)}.} \textbf{(a)-(d)} Hyperedge homophily $h_{s,r}^{(g)}$ is shown for hyperedge sizes $s\in\{2,3,4,5\}$ for majority ($g=0$, green) and minority ($g=1$, purple) groups. The dashed line shows the expected prevalence under random mixing; values above indicate over-representation. Information inequality outcomes are reported under both majority seeding (\textbf{(e)–(g)}) and minority seeding (\textbf{(h)–(j)}): \textbf{(e),(h)} Wasserstein distances $d_W(\mathcal{Z}_0,\mathcal{Z}_1)$, \textbf{(f),(i)} violin plots of the time $t^{(g)}_{90}$ to inform 90\% of nodes (majority left, minority right), and \textbf{(g),(j)} acquisition fairness $\alpha(f)$. 
\textbf{(k)} Shows diffusion fairness $\delta(f)$, where values above the dashed line indicate a minority advantage. Panels (e)–(k) average results over $n_\mathrm{hg}=10^3$ simulations for linear (blue), sublinear (pink), superlinear (red), and asymmetric (green) contagion. Confidence intervals in (g),(j),(k) are estimated from $100$ bootstrap samples. The House (\texttt{GenderAPI}) hypergraph shows modest minority advantages under linear, sublinear, and superlinear contagion, but a pronounced majority advantage under asymmetric contagion, with diffusion fairness favoring the minority.}
\label{fig:SI_housegender_genderapi}
\end{figure*}

\begin{figure*}[tb]
\centering
\includegraphics[width=0.9\textwidth]{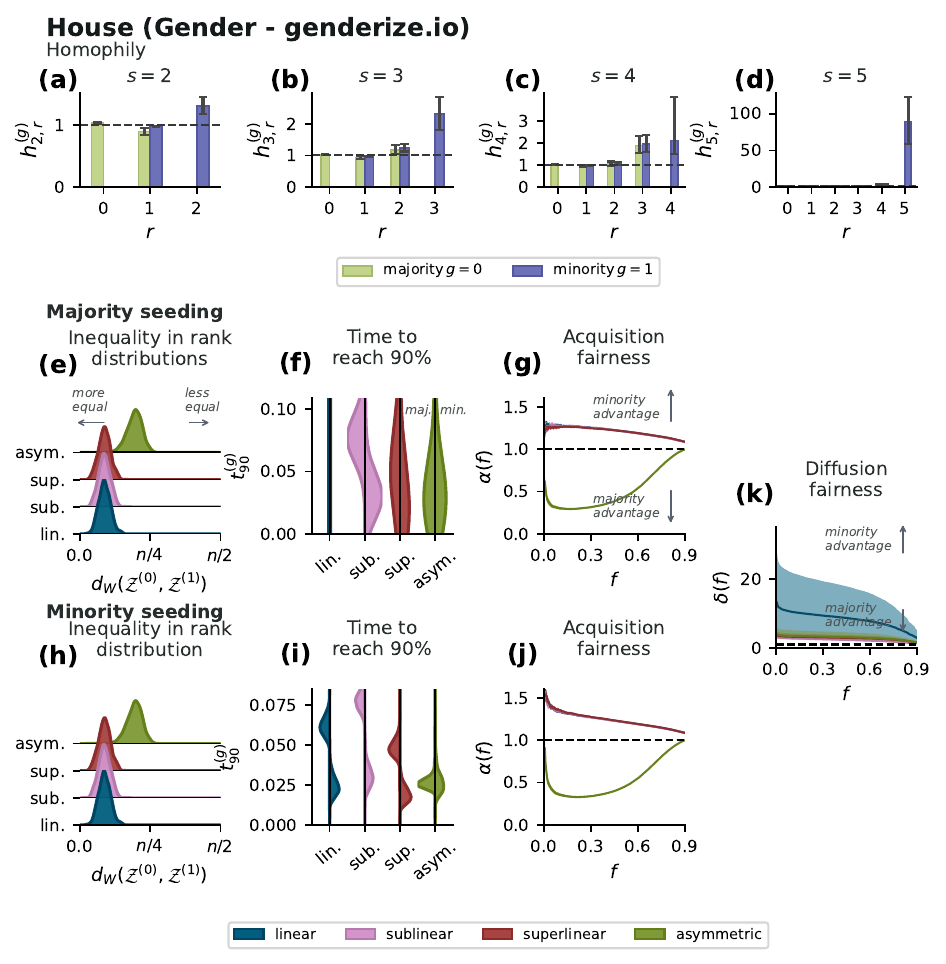}
\caption{\textbf{Homophily patterns and information diffusion inequality in the House hypergraph with gender labels (\texttt{genderize.io)}.} \textbf{(a)-(d)} Hyperedge homophily $h_{s,r}^{(g)}$ is shown for hyperedge sizes $s\in\{2,3,4,5\}$ for majority ($g=0$, green) and minority ($g=1$, purple) groups. The dashed line shows the expected prevalence under random mixing; values above indicate over-representation. Information inequality outcomes are reported under both majority seeding (\textbf{(e)–(g)}) and minority seeding (\textbf{(h)–(j)}): \textbf{(e),(h)} Wasserstein distances $d_W(\mathcal{Z}_0,\mathcal{Z}_1)$, \textbf{(f),(i)} violin plots of the time $t^{(g)}_{90}$ to inform 90\% of nodes (majority left, minority right), and \textbf{(g),(j)} acquisition fairness $\alpha(f)$. 
\textbf{(k)} Shows diffusion fairness $\delta(f)$, where values above the dashed line indicate a minority advantage. Panels (e)–(k) average results over $n_\mathrm{hg}=10^3$ simulations for linear (blue), sublinear (pink), superlinear (red), and asymmetric (green) contagion. Confidence intervals in (g),(j),(k) are estimated from $100$ bootstrap samples. Results are qualitatively consistent with those based on \texttt{GenderAPI} labels (Fig.~\ref{fig:SI_housegender_genderapi}).}
\label{fig:SI_housegender_genderizeio}
\end{figure*}

\paragraph{House (by gender)}
When labeled by gender, the House hypergraph has a class imbalance of about $87\%$ men and $13\%$ women when labeled with \texttt{GenderAPI}, and about $89\%$ men and $11\%$ women with \texttt{genderize.io}, averaged across $n_{hg}=1000$ samples. For small hyperedge sizes ($s\in{2,3,4,5}$), men appear approximately neutral, with the number of all-male edges close to expectation, while women exhibit homophily, with more all-female edges than expected (\texttt{GenderAPI}: Fig.~\ref{fig:SI_housegender_genderapi}(a)–(d); \texttt{genderize.io}: Fig.~\ref{fig:SI_housegender_genderizeio}(a)–(d)). The maximum hyperedge size in this dataset is large ($s_\mathrm{max}=399$), and it is not clear whether these patterns persist at larger hyperedge sizes. Given the combinatorial growth in possible group compositions, analyzing homophily across all hyperedge sizes remains methodologically challenging, and we leave more detailed investigation of this structure to future work.

With respect to information access inequality, we observe a modest minority (female) advantage under linear, sublinear, and superlinear contagion, and a pronounced majority (male) advantage under asymmetric contagion (\texttt{GenderAPI}: Fig.~\ref{fig:SI_housegender_genderapi}(e)–(g),(h)-(j); \texttt{genderize.io}: Fig.~\ref{fig:SI_housegender_genderizeio}(e)–(g),(h)-(j)). Diffusion fairness, by contrast, indicates a minority advantage, particularly under linear contagion, though with substantial variance (\texttt{GenderAPI}: Fig.~\ref{fig:SI_housegender_genderapi}(k); \texttt{genderize.io}: Fig.~\ref{fig:SI_housegender_genderizeio}(k)). While these findings suggest interactions between homophily patterns and contagion dynamics, such as homophily potentially leading to larger inequality in the asymmetric case, the complexity of large hyperedges and other structural features of the House hypergraph make it difficult to draw strong causal conclusions. A more systematic characterization of these properties is an important direction for future research.

\begin{figure*}[tb]
\centering
\includegraphics[width=0.9\textwidth]{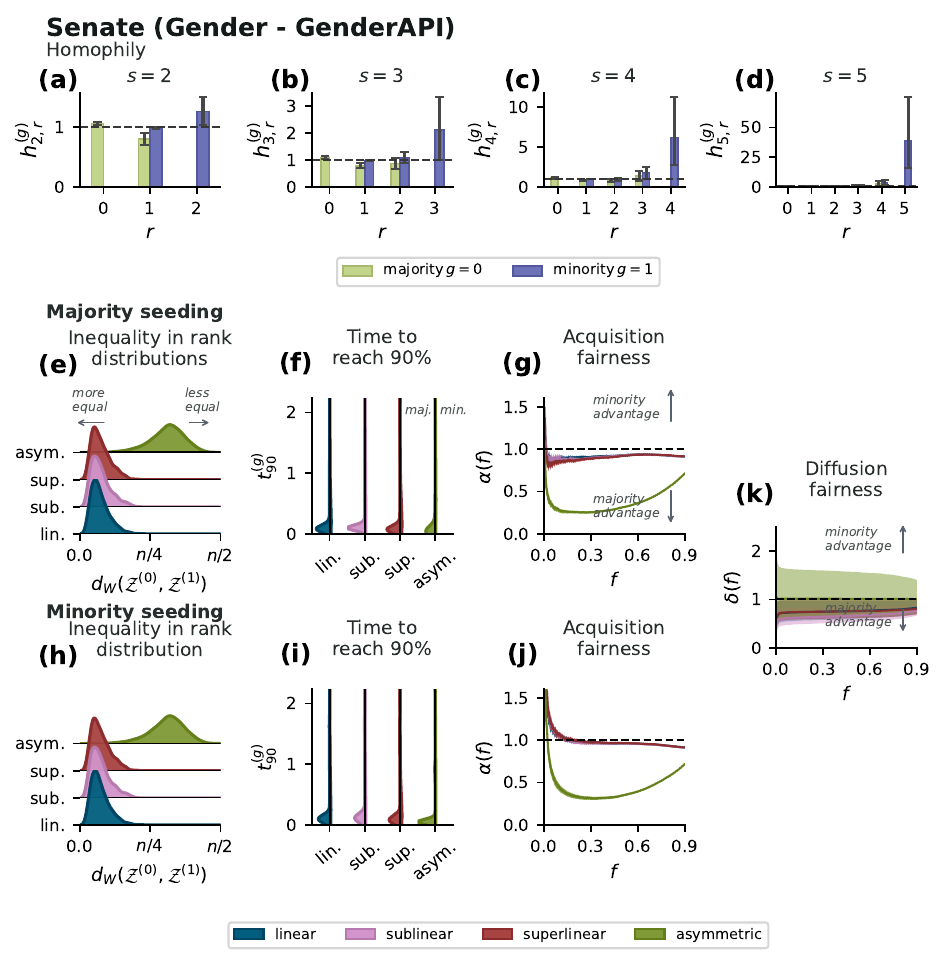}
\caption{\textbf{Homophily patterns and information diffusion inequality in the Senate hypergraph with gender labels (\texttt{GenderAPI)}.} \textbf{(a)-(d)} Hyperedge homophily $h_{s,r}^{(g)}$ is shown for hyperedge sizes $s\in\{2,3,4,5\}$ for majority ($g=0$, green) and minority ($g=1$, purple) groups. The dashed line shows the expected prevalence under random mixing; values above indicate over-representation. Information inequality outcomes are reported under both majority seeding (\textbf{(e)–(g)}) and minority seeding (\textbf{(h)–(j)}): \textbf{(e),(h)} Wasserstein distances $d_W(\mathcal{Z}_0,\mathcal{Z}_1)$, \textbf{(f),(i)} violin plots of the time $t^{(g)}_{90}$ to inform 90\% of nodes (majority left, minority right), and \textbf{(g),(j)} acquisition fairness $\alpha(f)$. 
\textbf{(k)} Shows diffusion fairness $\delta(f)$, where values above the dashed line indicate a minority advantage. Panels (e)–(k) average results over $n_\mathrm{hg}=10^3$ simulations for linear (blue), sublinear (pink), superlinear (red), and asymmetric (green) contagion. Confidence intervals in (g),(j),(k) are estimated from $100$ bootstrap samples. The Senate hypergraph shows modest majority advantages across most contagion regimes, a stronger advantage under asymmetric contagion, and diffusion fairness that likewise favors the majority.}
\label{fig:SI_senategender_genderapi}
\end{figure*}

\begin{figure*}[tb]
\centering
\includegraphics[width=0.9\textwidth]{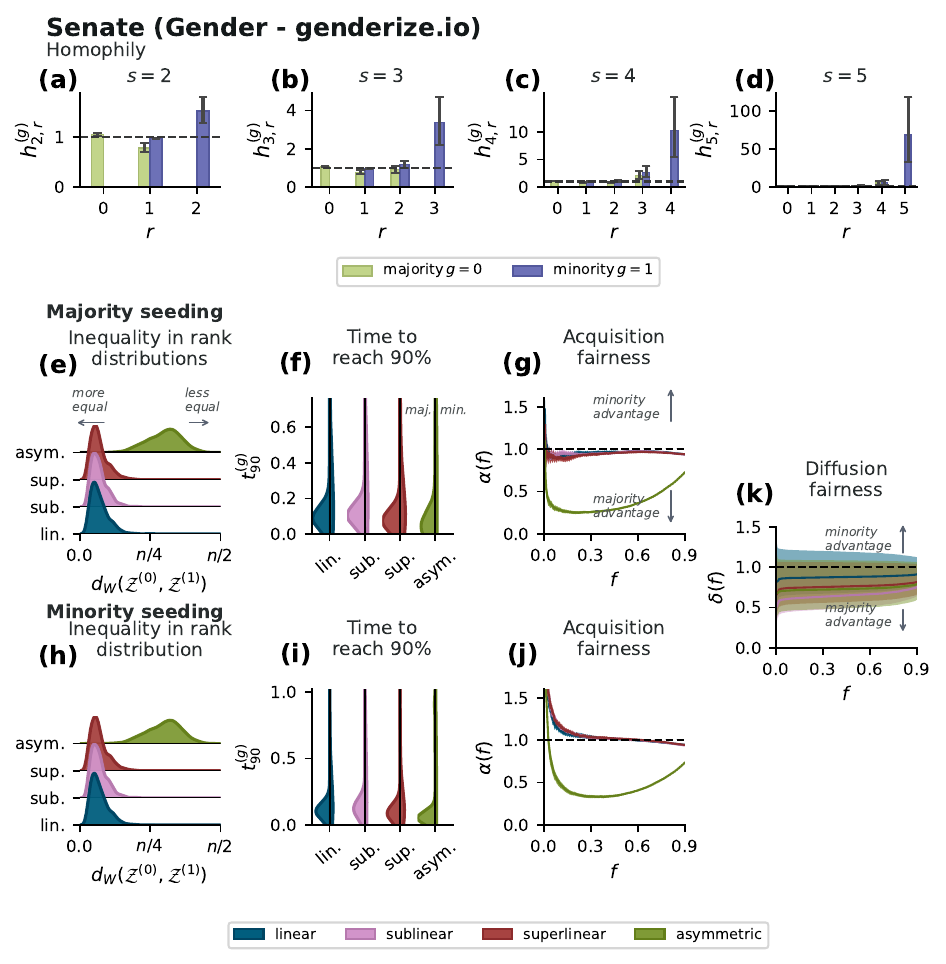}
\caption{\textbf{Homophily patterns and information diffusion inequality in the Senate hypergraph with gender labels (\texttt{genderize.io)}.} \textbf{(a)-(d)} Hyperedge homophily $h_{s,r}^{(g)}$ is shown for hyperedge sizes $s\in\{2,3,4,5\}$ for majority ($g=0$, green) and minority ($g=1$, purple) groups. The dashed line shows the expected prevalence under random mixing; values above indicate over-representation. Information inequality outcomes are reported under both majority seeding (\textbf{(e)–(g)}) and minority seeding (\textbf{(h)–(j)}): \textbf{(e),(h)} Wasserstein distances $d_W(\mathcal{Z}_0,\mathcal{Z}_1)$, \textbf{(f),(i)} violin plots of the time $t^{(g)}_{90}$ to inform 90\% of nodes (majority left, minority right), and \textbf{(g),(j)} acquisition fairness $\alpha(f)$. 
\textbf{(k)} Shows diffusion fairness $\delta(f)$, where values above the dashed line indicate a minority advantage. Panels (e)–(k) average results over $n_\mathrm{hg}=10^3$ simulations for linear (blue), sublinear (pink), superlinear (red), and asymmetric (green) contagion. Confidence intervals in (g),(j),(k) are estimated from $100$ bootstrap samples. Results are qualitatively consistent with those based on \texttt{GenderAPI} labels (Fig.~\ref{fig:SI_senategender_genderapi}).}
\label{fig:SI_senategender_genderizeio}
\end{figure*}

\paragraph{Senate (by gender)}
Like the House hypergraph, the Senate hypergraph shows extreme gender imbalance (\texttt{GenderAPI}: $88\%$ male, $12\%$ female on average; \texttt{genderize.io}: $90\%$ male, $10\%$ female on average). For small hyperedge sizes, men appear roughly neutral, while women form all-female edges more often than expected (\texttt{GenderAPI}: Fig.~\ref{fig:SI_senategender_genderapi}(a)–(d); \texttt{genderize.io}: Fig.~\ref{fig:SI_senategender_genderizeio}(a)–(d)). Although the maximum hyperedge size is smaller than in the House case ($s_\mathrm{max}=99$), it remains large enough relative to the total hypergraph size to complicate analysis.

In terms of inequality, we observe a modest majority advantage under linear, sublinear, and superlinear contagion, and a more pronounced majority advantage under asymmetric contagion (\texttt{GenderAPI}: Fig.~\ref{fig:SI_senategender_genderapi}(e)–(g),(h)-(j); \texttt{genderize.io}: Fig.~\ref{fig:SI_senategender_genderizeio}(e)–(g),(h)-(j)). Diffusion fairness likewise indicates a majority advantage across contagion regimes, though with substantial variance (\texttt{GenderAPI}: Fig.~\ref{fig:SI_senategender_genderapi}(k); \texttt{genderize.io}: Fig.~\ref{fig:SI_senategender_genderizeio}(k)). Given the interplay of homophily patterns and other features, a more complete characterization is left for future work.

\begin{figure*}[tb]
\centering
\includegraphics[width=0.9\textwidth]{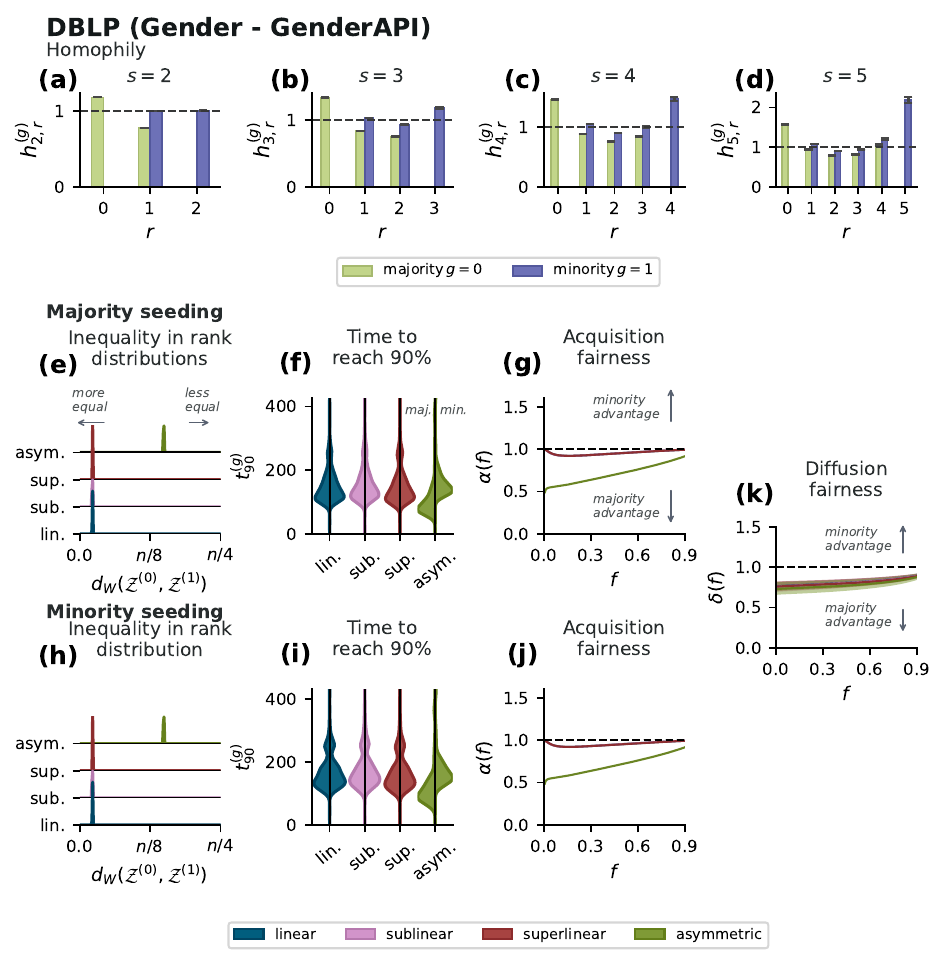}
\caption{\textbf{Homophily patterns and information diffusion inequality in the DBLP hypergraph (\texttt{GenderAPI}).} \textbf{(a)-(d)} Hyperedge homophily $h_{s,r}^{(g)}$ is shown for hyperedge sizes $s\in\{2,3,4,5\}$ for majority ($g=0$, green) and minority ($g=1$, purple) groups. The dashed line shows the expected prevalence under random mixing; values above indicate over-representation. Information inequality outcomes are reported under both majority seeding (\textbf{(e)–(g)}) and minority seeding (\textbf{(h)–(j)}): \textbf{(e),(h)} Wasserstein distances $d_W(\mathcal{Z}_0,\mathcal{Z}_1)$, \textbf{(f),(i)} violin plots of the time $t^{(g)}_{90}$ to inform 90\% of nodes (majority left, minority right), and \textbf{(g),(j)} acquisition fairness $\alpha(f)$. 
\textbf{(k)} Shows diffusion fairness $\delta(f)$, where values above the dashed line indicate a minority advantage. Panels (e)–(k) average results over $n_\mathrm{hg}=10^3$ simulations for linear (blue), sublinear (pink), superlinear (red), and asymmetric (green) contagion. Confidence intervals in (g),(j),(k) are estimated from $100$ bootstrap samples. The DBLP hypergraph shows minimal inequality across contagion regimes, with only slight majority advantages; local imbalances may exist within subfields, but they appear to wash out at the global scale.}
\label{fig:SI_dblp_genderapi}
\end{figure*}

\begin{figure*}[tb]
\centering
\includegraphics[width=0.9\textwidth]{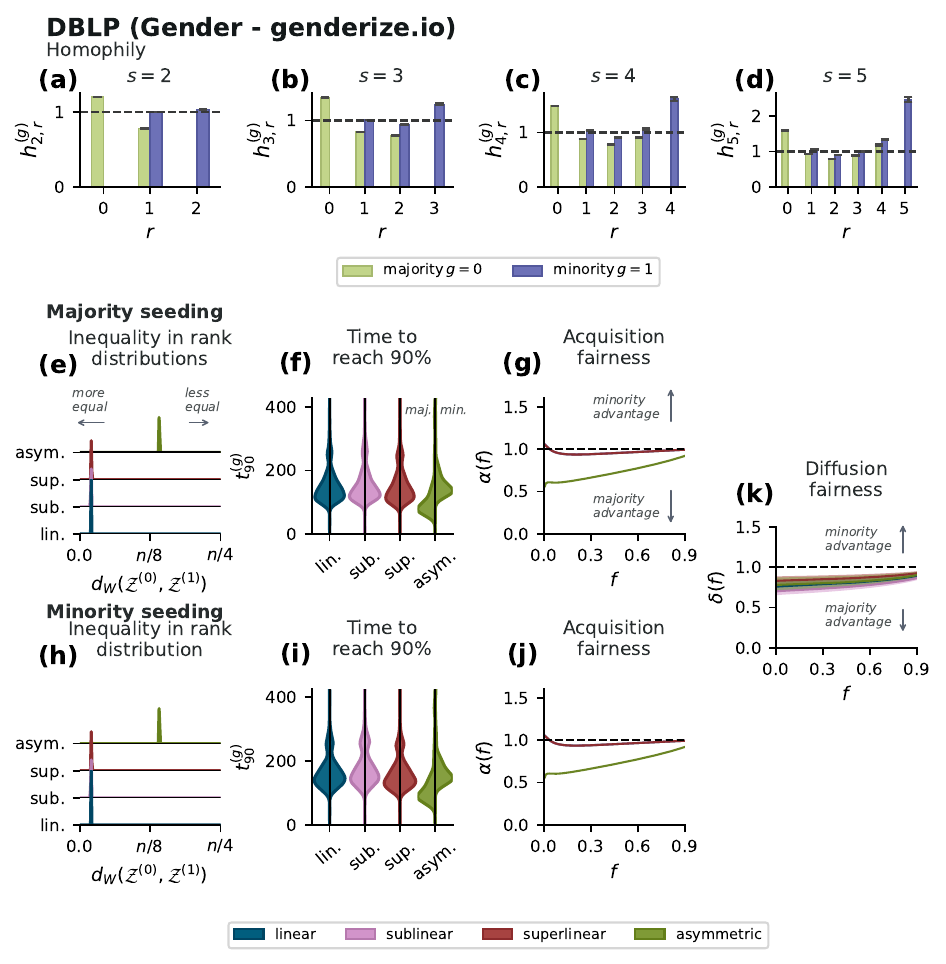}
\caption{\textbf{Homophily patterns and information diffusion inequality in the DBLP hypergraph (\texttt{genderize.io}).} \textbf{(a)-(d)} Hyperedge homophily $h_{s,r}^{(g)}$ is shown for hyperedge sizes $s\in\{2,3,4,5\}$ for majority ($g=0$, green) and minority ($g=1$, purple) groups. The dashed line shows the expected prevalence under random mixing; values above indicate over-representation. Information inequality outcomes are reported under both majority seeding (\textbf{(e)–(g)}) and minority seeding (\textbf{(h)–(j)}): \textbf{(e),(h)} Wasserstein distances $d_W(\mathcal{Z}_0,\mathcal{Z}_1)$, \textbf{(f),(i)} violin plots of the time $t^{(g)}_{90}$ to inform 90\% of nodes (majority left, minority right), and \textbf{(g),(j)} acquisition fairness $\alpha(f)$. 
\textbf{(k)} Shows diffusion fairness $\delta(f)$, where values above the dashed line indicate a minority advantage. Panels (e)–(k) average results over $n_\mathrm{hg}=10^3$ simulations for linear (blue), sublinear (pink), superlinear (red), and asymmetric (green) contagion. Confidence intervals in (g),(j),(k) are estimated from $100$ bootstrap samples. Results are qualitatively consistent with those based on \texttt{GenderAPI} labels (Fig.~\ref{fig:SI_dblp_genderapi}).}
\label{fig:SI_dblp_genderizeio}
\end{figure*}

\paragraph{DBLP}
The DBLP hypergraph has a sizable class imbalance (\texttt{GenderAPI}: $71\%$ male, $29\%$ female on average; \texttt{genderize.io}: $70\%$ male, $30\%$ female on average) and exhibits a standard homophilous pattern for hyperedge sizes $s \in {2,3,4,5}$ (\texttt{GenderAPI}: Fig.~\ref{fig:SI_dblp_genderapi}(a)–(d); \texttt{genderize.io}: Fig.~\ref{fig:SI_dblp_genderizeio}(a)–(d)). However, the maximum hyperedge size of $s_\mathrm{max}=450$ introduces additional challenges for analyzing homophily at larger scales.

Inequality is minimal under linear, sublinear, and superlinear contagion, with only a slight majority advantage in the asymmetric case (\texttt{GenderAPI}: Fig.~\ref{fig:SI_dblp_genderapi}(e)–(g),(h)-(j); \texttt{genderize.io}: Fig.~\ref{fig:SI_dblp_genderizeio}(e)–(g),(h)-(j)). Diffusion fairness likewise shows a minor majority advantage (\texttt{GenderAPI}: Fig.~\ref{fig:SI_dblp_genderapi}(k); \texttt{genderize.io}: Fig.~\ref{fig:SI_dblp_genderizeio}(k)). There are also no apparent differences in inequality between minority and majority seeding, likely due to the relatively small seed budget and the large size of the hypergraph; once roughly $1\%$ of nodes are infected (about $34000$ authors), any seeding effects are likely to wash out at the resolution shown in the plots. One possible explanation for the limited inequality observed, despite homophily, is the dense and modular structure of the network. Within subfields, women may co-author more often with each other than with men in the same subfield, which could produce local inequalities. However, because subfields are infected at different times, these effects may wash out at the global scale, making information access appear more balanced overall. A full characterization of these patterns and their implications for inequality is left for future work.

\begin{figure*}[tb]
\centering
\includegraphics[width=0.9\textwidth]{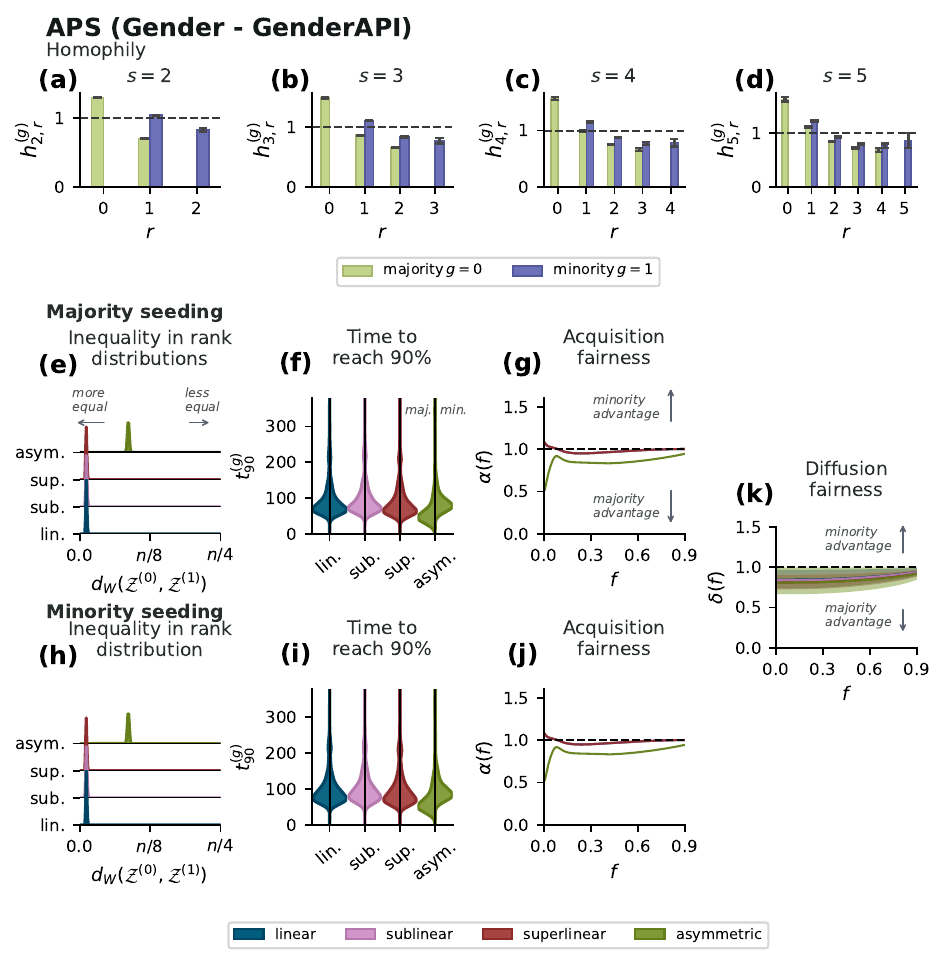}
\caption{\textbf{Homophily patterns and information diffusion inequality in the APS hypergraph (\texttt{GenderAPI}).} \textbf{(a)-(d)} Hyperedge homophily $h_{s,r}^{(g)}$ is shown for hyperedge sizes $s\in\{2,3,4,5\}$ for majority ($g=0$, green) and minority ($g=1$, purple) groups. The dashed line shows the expected prevalence under random mixing; values above indicate over-representation. Information inequality outcomes are reported under both majority seeding (\textbf{(e)–(g)}) and minority seeding (\textbf{(h)–(j)}): \textbf{(e),(h)} Wasserstein distances $d_W(\mathcal{Z}_0,\mathcal{Z}_1)$, \textbf{(f),(i)} violin plots of the time $t^{(g)}_{90}$ to inform 90\% of nodes (majority left, minority right), and \textbf{(g),(j)} acquisition fairness $\alpha(f)$. 
\textbf{(k)} Shows diffusion fairness $\delta(f)$, where values above the dashed line indicate a minority advantage. Panels (e)–(k) average results over $n_\mathrm{hg}=10^3$ simulations for linear (blue), sublinear (pink), superlinear (red), and asymmetric (green) contagion. Confidence intervals in (g),(j),(k) are estimated from $100$ bootstrap samples. The APS hypergraph shows minimal inequality across contagion regimes, with only slight majority advantages; local imbalances may exist within subfields, but they appear to wash out at the global scale.}
\label{fig:SI_aps_genderapi}
\end{figure*}

\begin{figure*}[tb]
\centering
\includegraphics[width=0.9\textwidth]{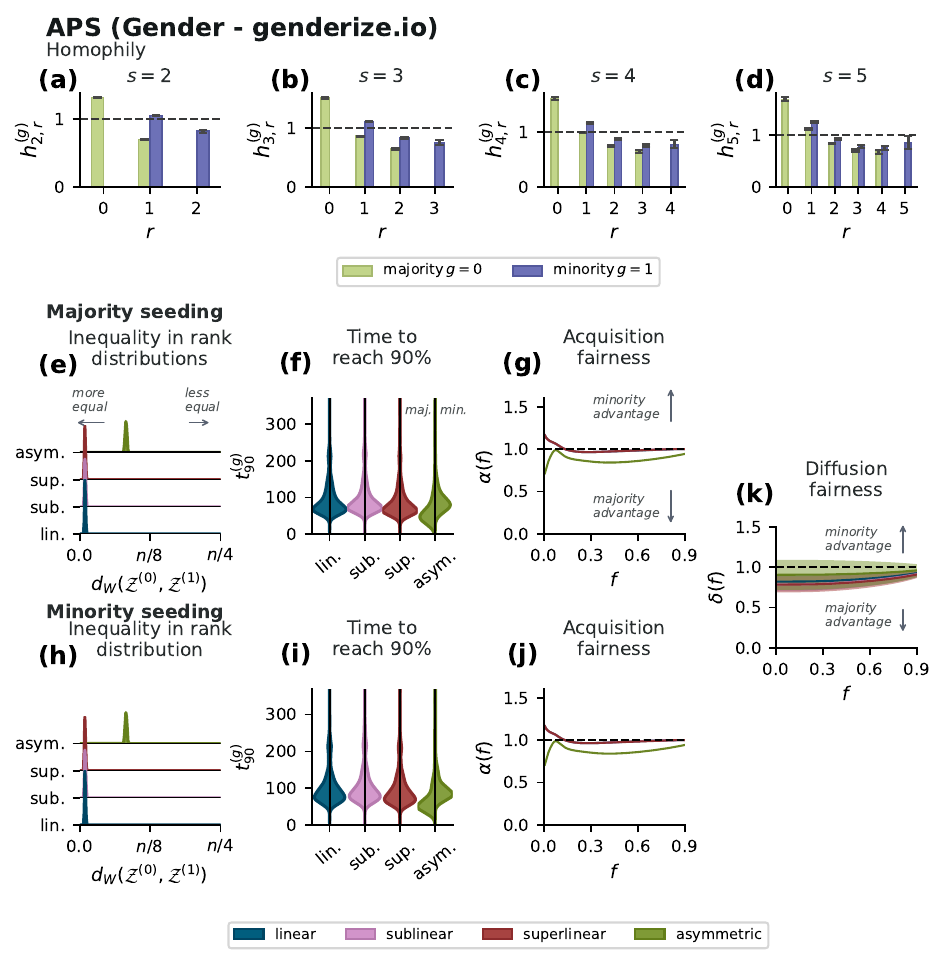}
\caption{\textbf{Homophily patterns and information diffusion inequality in the APS hypergraph (\texttt{genderize.io}).} \textbf{(a)-(d)} Hyperedge homophily $h_{s,r}^{(g)}$ is shown for hyperedge sizes $s\in\{2,3,4,5\}$ for majority ($g=0$, green) and minority ($g=1$, purple) groups. The dashed line shows the expected prevalence under random mixing; values above indicate over-representation. Information inequality outcomes are reported under both majority seeding (\textbf{(e)–(g)}) and minority seeding (\textbf{(h)–(j)}): \textbf{(e),(h)} Wasserstein distances $d_W(\mathcal{Z}_0,\mathcal{Z}_1)$, \textbf{(f),(i)} violin plots of the time $t^{(g)}_{90}$ to inform 90\% of nodes (majority left, minority right), and \textbf{(g),(j)} acquisition fairness $\alpha(f)$. 
\textbf{(k)} Shows diffusion fairness $\delta(f)$, where values above the dashed line indicate a minority advantage. Panels (e)–(k) average results over $n_\mathrm{hg}=10^3$ simulations for linear (blue), sublinear (pink), superlinear (red), and asymmetric (green) contagion. Confidence intervals in (g),(j),(k) are estimated from $100$ bootstrap samples. Results are qualitatively consistent with those based on \texttt{GenderAPI} labels (Fig.~\ref{fig:SI_aps_genderapi}).}
\label{fig:SI_aps_genderizeio}
\end{figure*}

\paragraph{APS}
The APS hypergraph has a modest class imbalance (\texttt{GenderAPI}: $66\%$ male, $34\%$ female on average; \texttt{genderize.io}: $66\%$ male, $34\%$ female on average). It also has a one-sided homophily pattern in its smaller edge sizes, i.e., $s \in \{2, 3, 4, 5\}$, where majority nodes are forming more all-majority edges than expected while minority nodes are forming fewer all-minority edges than expected (\texttt{GenderAPI}: Fig.~\ref{fig:SI_aps_genderapi}(a)–(d); \texttt{genderize.io}: Fig.~\ref{fig:SI_aps_genderizeio}(a)–(d)). The APS hypergraph also contains much larger hyperedges than any other dataset, with a maximum size of $s_\mathrm{max}=1666$, making it unclear whether these patterns persist at larger scales.

As with the DBLP hypergraph, there is minimal inequality in the linear, sublinear, and superlinear cases and only a modest majority advantage in the asymmetric case (\texttt{GenderAPI}: Fig.~\ref{fig:SI_aps_genderapi}(e)–(g),(h)-(j); \texttt{genderize.io}: Fig.~\ref{fig:SI_aps_genderizeio}(e)–(g),(h)-(j)). Diffusion fairness likewise shows a minor majority advantage (\texttt{GenderAPI}: Fig.~\ref{fig:SI_aps_genderapi}(k); \texttt{genderize.io}: Fig.~\ref{fig:SI_aps_genderizeio}(k)). Again like with DBLP, there are no apparent differences in inequality between minority and majority seeding since any seeding effects are likely to wash out at the resolution shown in the plots. This pattern may reflect information spreading within subfields, where local inequalities might wash out at the global scale. It could also stem from more heterophilous interactions in larger hyperedges, or from overestimation of women in the dataset due to probabilistic gender sampling (Supplementary Note 6). A fuller characterization of these factors and their role in shaping inequality is left for future work.

%
%

%
%

%
%

%
%

%
%

%
%

%
%

%
%

%
%

%
%

%
%

%
%

%
%

\end{appendices}

\end{document}